\documentclass[pra,amsmath,amssymb,twocolumn, longbibliography, superscriptaddress]{revtex4-2}
\usepackage{graphicx}
\usepackage{dcolumn}
\usepackage{bm}
\usepackage[utf8]{inputenc}
\usepackage[english]{babel}
\usepackage{amsfonts}
\usepackage{hyperref}
\usepackage{physics}
\usepackage{epstopdf}
\usepackage{mathtools}
\usepackage[capitalise]{cleveref}
\usepackage{float}
\usepackage[space]{grffile}
\usepackage{color}
\usepackage[normalem]{ulem}
\usepackage{subfigure}
\usepackage{natbib}
\usepackage{enumitem}
\usepackage{bbm}
\usepackage{multirow}

\binoppenalty=\maxdimen
\relpenalty=\maxdimen

\addto\captionsenglish{}
\DeclareMathAlphabet\mathbfcal{OMS}{cmsy}{b}{n}

\def \hrho{\hat{\rho}}
\def \ha{\hat{a}}
\def \hc{\hat{c}}
\def \nth{\bar{n}_{\rm th}}

\def \hL{\mathbfcal{\hat{L}}}
\def \hV{\mathbfcal{\hat{V}}}

\def \hacl{\boldsymbol{\hat{a}}_{\rm cl}}
\def \haq{\boldsymbol{\hat{a}}_{\rm q}}
\def \hcone{\boldsymbol{\hat{c}}_1}
\def \hctwo{\boldsymbol{\hat{c}}_2}
\def \hone{\boldsymbol{\hat{1}}}
\def \halphacl{\hat{\alpha}_{\rm cl}}
\def \halphaq{\hat{\alpha}_{\rm q}}

\def \hetaq{\hat{\eta}_{\rm q}}

\def \aclj{a_{{\rm cl}, j}}
\def \aqj{a_{{\rm q}, j}}

\newcommand{\hacln}[1]{\boldsymbol{\hat{a}}_{\text{cl}, #1}}
\newcommand{\haqn}[1]{\boldsymbol{\hat{a}}_{\text{q}, #1}}

\newcommand{\hconen}[1]{\boldsymbol{\hat{c}}_{1, #1}}
\newcommand{\hctwon}[1]{\boldsymbol{\hat{c}}_{2, #1}}
\newcommand{\mat}[1]{\mathbb{#1}} 
\newcommand{\ssb}[1]{\mat{S}_{\rm ss}}

\DeclareSymbolFont{bbold}{U}{bbold}{m}{n}
\DeclareSymbolFontAlphabet{\mathbbold}{bbold}

\begin{document}
	

	\title{Third quantization of open quantum systems: new dissipative symmetries and connections to phase-space and Keldysh field theory formulations}

    \author{Alexander McDonald }
    \affiliation{Pritzker School of Molecular Engineering, University of Chicago, Chicago, IL 60637, USA}
    \affiliation{Department of Physics, University of Chicago, Chicago, IL 60637, USA}
    \affiliation{Institut Quantique \& Département de Physique, Université de Sherbrooke, Sherbrooke, Québec, J1K 2R1, Canada}

    \author{Aashish A. Clerk}
    \affiliation{Pritzker School of Molecular Engineering, University of Chicago, Chicago, IL 60637, USA}

	\begin{abstract}
The connections between standard theoretical tools used to study open quantum systems can sometimes seem opaque.  Whether it is a Lindblad master equation, the equation of motion for the Wigner function or a dissipative Keldysh action, features evident in one formalism are often masked in another. Here, we reformulate the technique of third quantization in a way that explicitly connects all three methods.  We first show that our formulation reveals a fundamental dissipative symmetry present in all quadratic bosonic or fermionic Lindbladians. This symmetry can then be used  to easily diagonalize these models, and provides a intuitive way to demonstrate the separation of dissipation and fluctations in linear systems. For bosons, we then show that the Wigner function and the characteristic function can be thought of as ``wavefunctions" of the density matrix in the eigenbasis of the third-quantized superoperators we introduce. The field-theory representation of the time-evolution operator in this basis is then the Keldysh path integral. To highlight the utility of our approach, we apply our version of third quantization to a dissipative non-linear oscillator, and use it to obtain new exact results.  
	\end{abstract}

	\maketitle
    \section{Introduction}

    Open quantum systems are generically much more difficult to characterize than their closed-system counterparts. Quadratic systems of bosons or fermions coupled to Markovian baths are arguably an exception: computing average values and correlation functions is straightforward using either the master equation, Keldysh field theory or the Heisenberg-Langevin equations. However, it is often useful to understand the full structure of the eigenvalues and eigenvectors of the Lindbladian. 
    This information can help answer questions that are difficult to address using Keldysh or Langevin equations, e.g.~the full time evolution of an arbitrarily complicated non-Gaussian initial state. In a set of seminal works,  Prosen \cite{Prosen_Fermions_2008} and Prosen and Seligman \cite{Prosen_Seligman_Bosons_2010} introduced the technique of third quantization, which allows one to obtain the spectral decomposition of quadratic multi-mode Lindbladians in a manner that is analogous to diagonalizing a second-quantized quadratic Hamiltonian. For all its virtues and widespread use in the literature \cite{Lieu_Ten_Fold_Lindblad_PRL, Shu_Boundary_Mode_arXiv, Prosen_Random_Liouvillian_Fermions, Dissipation_Flat_Bands_PRL, Prosen_Integrable_Spin_PRL, PT_Transition_PRA, Bergholtz_Exact_PRR, Buca_Multistability_Comm_Physics, Non-Hermitian_Kitaev_PRB}, the method as presented does not provide an intuitive picture of open quantum system dynamics. Further, it is not a priori clear how it is related to more standard methods. A more physically-transparent formulation of third quantization which addresses these issues could thus help make it an even more powerful tool.

    In this work we present an alternate reformulation of third quantization which satisfies both criteria simultaneously. This is achieved by introducing a new set of canonical superoperators, defined in Eq.~(\ref{eq:Bosons_Sup_Def}) for bosons and Eq.~(\ref{eq:Fermions_Sup_Def}) for fermions. In this basis, one directly identifies the two pieces of data that determine any quadratic Lindbladian: a non-Hermitian Hamiltonian which acts as an effective dynamical matrix, and a second matrix which encodes the noise. This separation of dissipation and fluctuations further reveals a fundamental symmetry of all such models. We show that this symmetry can be used to effectively gauge away the fluctuation terms in the master equation using a novel similarity transformation implemented at the superoperator level. This provides a simple quantum-oriented way to demonstrate that \textit{the eigenvalues of the Lindbladian are independent of noise in linear systems}. 
    
\begin{figure}[t]
		\centering
		\includegraphics [width=0.475\textwidth]{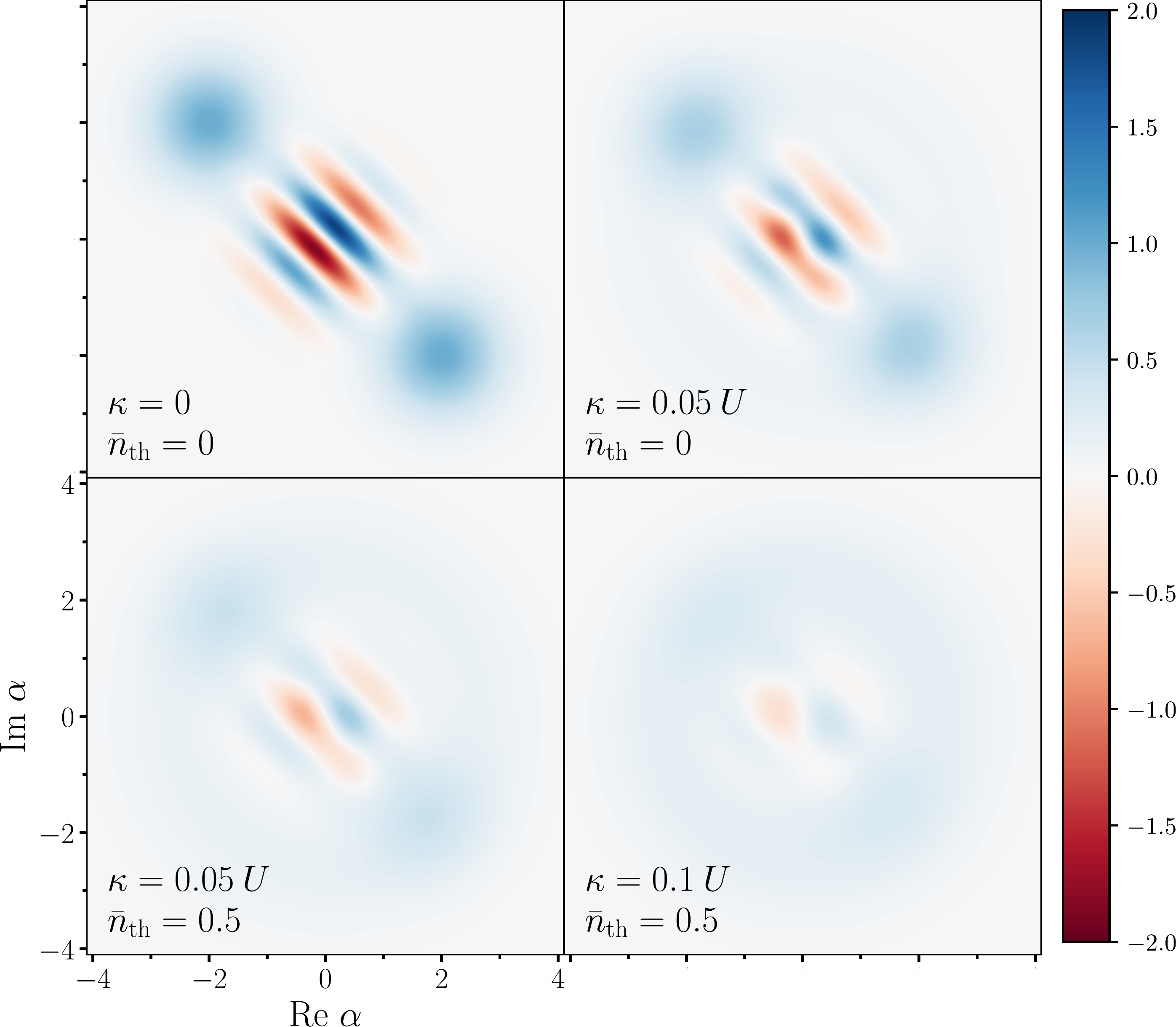}
		\caption{Exact Wigner function of a dissipative non-linear cavity mode at time $t = \pi U$ (with $U$ the Kerr constant), where in each case the initial state is a coherent state $\hrho(0) = |\alpha \rangle \langle \alpha|$, $\alpha = \sqrt{2}(1+i)$.  Evolution is generated by the master equation Eq.~(\ref{eq:NonLinear_Oscillator_EOM}). With no dissipation $\kappa = \nth = 0$, the resulting state is a superposition of coherent states \cite{Yurke_Cat}. The formation of such a state is however very sensitive to any amount of dissipation. As explained in Sec.~\ref{sec:Kerr}, third quantization allows  us to obtain the exact evolution of the Wigner function for an arbitrary initial state.  
		}
		\label{fig:Wigner_Function}
	    \end{figure}

    The superoperators we introduce also make the connection between the Lindbladian and the Keldysh formalism evident. We make this notion exact by formulating a path integral representation of third quantization over a finite time contour. For bosons, we go further by demonstrating that third quantization is naturally linked to phase-space representations of the density matrix. Namely, we show that the Wigner function and its characteristic function can be thought of as wavefunctions in the basis of what is essentially the third-quantized equivalent of position and momentum eigenkets. We use this insight to study a dissipative interacting model, a thermally-damped Kerr cavity. Our approach lets us 
    analytically describe the time-evolution of cavity's Wigner function, starting from an {\it arbitrary} initial state (see Fig.~\ref{fig:Wigner_Function} for an example).  To the best of our knowledge, this result has not been previously presented, and highlights the power of our approach.

  The paper is organized as follows. In Sec.~\ref{sec:Harmonic_Oscillator} we work out in detail the diagonalization procedure of the Lindbladian describing a thermally-damped harmonic oscillator using our formulation of third quantization. We then discuss how our approach is naturally connected to both phase-space and Keldysh formalisms of open quantum systems in Secs.~\ref{sec:Single_Osc_Phase_Space} and \ref{sec:Keldysh_Single_Osc} respectively. The insight gained from this example allows for a straightforward generalization to arbitrary quadratic bosonic Lindbladians as shown in Sec.~\ref{sec:Quadratic_Bosons}. Despite the change in statistics, a similar approach is presented for the fermionic version of this problem as detailed in Sec.~\ref{sec:Quadratic_Fermions}. We finish by applying our phase-space formulation of third quantization to the dissipative Kerr oscillator problem in Sec.~\ref{sec:Kerr}. 

Before we proceed, let us highlight the differences between this work and the work of Prosen and Prosen and Seligman. The main dissimilarity is our starting point: we use a distinct set of superoperators to express the third-quantized Lindbladian, see Eq.~(\ref{eq:Bosons_Sup_Def}) and Eq.~(\ref{eq:Fermions_Sup_Def}). Although a seemingly-minor difference, this alternate choice of basis makes the symmetry we use to gauge away fluctuations apparent. Our superoperators also provide a natural connection between third-quantization, Keldysh field theory and phase-space representations of open quantum systems, which has not been considered in previous works \cite{Prosen_Fermions_2008, Prosen_Seligman_Bosons_2010, Barthel_Thomas_Third_Quant_2022, Englert_Damping_Basis_1993, Honda_Spectral_Resolution_2010, Kamanev_arXiv_Field_Theory}. 
 
  \section{Diagonalizing a Liouvillian using third quantization }\label{sec:Harmonic_Oscillator}
\subsection{Setup}
While the diagonalization method we discuss can be applied to arbitrary multi-mode bosonic quadratic Lindbladians, for clarity we discuss these ideas in the simplest possible setting: a harmonic oscillator coupled to a thermal Markovian bath. 
Our starting point is the equation of motion for the density matrix (with $\hbar = 1$ throughout)
    \begin{align} \nonumber 
		i \partial_t \hrho
		&
		=
		\omega_0[\ha^\dagger \ha, \hrho ] + i \kappa(\nth+1) \mathcal{D}[\ha]\hrho
		\\ \label{eq:Linear_Oscillator_EOM}
		&
		+
		i \kappa \nth \mathcal{D}[\ha^\dagger]\hrho
		\equiv 
		\mathcal{L} \hrho
	\end{align}
	where $\ha$ is the bosonic annihilation operator of the oscillator, $\omega_0$ is its frequency, $\kappa$ its decay rate and $\nth$ is the thermal occupation of the bath. Here, we have defined the usual dissipator as $\mathcal{D}[\hat{X}]\hrho = \hat{X} \hrho \hat{X}^\dagger - \{\hat{X}^\dagger \hat{X},\hrho\}/2$ and $\mathcal{L}$ as the Liouvillian superoperator. Our goal is to diagonalize $\mathcal{L}$. It is convenient to think of operators as being elements of a Hilbert space $\hrho \to |\hrho \rangle \rangle$ with the usual Hilbert-Schmidt inner product \cite{Victor_Thesis}
	\begin{align}\label{eq:Inner_Prod}
	  \langle \langle \hat{X} | \hat{Y} \rangle \rangle
	  \equiv
	  \Tr\left( \hat{X}^\dagger \hat{Y} \right)
	\end{align}
	for arbitrary operators $\hat{X}$ and $\hat{Y}$\footnote{Technically, for the infinite-dimensional spaces under consideration, both $\hat{X}$ and $\hat{Y}$ must be trace class to be part of this Hilbert space of operators. This is cumbersome, since operators of interest such as $\ha$ and $\ha^\dagger$ do not satisfy this property. We do not concern ourselves with these technical details. We will thus use the language and notation that one uses for a normal Hilbert space.}. 
 
 Using this notation, superoperators become operators acting on this new space and we thus write them with hats. To differentiate them from operators acting on the Hilbert space of wavefunctions, we write all superoperators using a bold typeface $\mathcal{L}  \to \hL.$ The inner product Eq.~(\ref{eq:Inner_Prod}) then allows us to define the adjoint of any superoperator. For instance the adjoint Liouvillian, which controls the time evolution of observables, is in standard second quantized form given by
	\begin{align}
	    \mathcal{L}^\dagger \hat{Y}
	    =
	    \omega_0[\ha^\dagger \ha, \hat{Y}]
	    - i \kappa(\nth+1)\mathcal{D}^\dagger[\ha]\hat{Y}
	    - i \kappa \nth \mathcal{D}^\dagger[\ha^\dagger]\hat{Y}
	\end{align}
    with $\mathcal{D}^\dagger[\hat{X}] \hat{Y} = \hat{X}^\dagger \hat{Y} \hat{X}-\{\hat{X}^\dagger \hat{X},\hat{Y}\}/2$ the adjoint dissipator. As the Linbladian is non-Hermitian $\hL \neq \hL^\dagger$, to diagonalize it we must find both its right and left eigenvectors. That is, we seek operators and complex numbers which satisfy
    \begin{subequations}
    \begin{align}
        \hL |\hat{r}_{\mu, \nu}\rangle \rangle
        =
        | \mathcal{L} \hat{r}_{\mu, \nu}\rangle \rangle
        &=
        E_{\mu, \nu}
        |\hat{r}_{\mu, \nu}\rangle \rangle
        \\
        \langle \langle \hat{l}_{\mu, \nu}|\hL
        =
        \langle \langle \mathcal{L}^\dagger \hat{l}_{\mu, \nu}|
        &=
        E_{\mu, \nu} \langle \langle \hat{l}_{\mu, \nu}|
    \end{align}
    \end{subequations}
    where, with some foresight, we have labelled the eigenvectors by two non-negative integers $\mu, \nu \geq 0$. 
    
    Although the physical interpretation of the right and left eigenvectors may not be straightforward, the eigenvalues have a simple intuitive meaning. The real part of $E_{\mu, \nu}$ plays the role of energy differences between eigenstates, whereas the imaginary component dictates the rate of decay. For the linear system of interest however, just as in the classical case,  the characteristic oscillation and decay rates are independent of temperature. This simply reflects the fact that a linear system's response to noise is always independent of the noise itself, something that is clear e.g. by using Langevin equations to describe our system. We thus conclude that the eigenvalues of $\hL$ must not depend on $\nth$. 
    A crucial question though still remains:  how do we rigorously identify these  fluctuations and reach this conclusion directly from the master equation?

\subsection{Classical and Quantum Superoperators}
 We argue that this can be straightforwardly done by introducing appropriate third-quantized superoperators defined by
 \begin{subequations}
\begin{align}\label{eq:Def_acl_aq}
	\hacl | \hrho \rangle \rangle
	&\equiv \frac{1}{\sqrt{2}}|\{\ha,\hrho\}\rangle \rangle,
	\hspace{0.3cm}
	&&\haq |\hrho \rangle \rangle 
	\equiv 
	\frac{1}{\sqrt{2}}|[\ha,\hrho] \rangle \rangle
	\\ \label{eq:Def_acldag_aqdag}
	\hacl^\dagger |\hrho\rangle \rangle 
	&\equiv 
	\frac{1}{\sqrt{2}}|\{\ha^\dagger,\hrho\}\rangle \rangle,
	\hspace{0.3cm}
	&&\haq^\dagger |\hrho\rangle \rangle 
	\equiv 
	\frac{1}{\sqrt{2}}|[\ha^\dagger,\hrho]\rangle \rangle
\end{align}
\end{subequations}
with which we refer to as classical and quantum superoperators. The naming convention is in analogy with Keldysh field theory, which is justified by writing the Lindbladian in this basis of superoperators
\begin{align}\nonumber
	\hL
	=&
	(\omega_0 - i \frac{\kappa}{2}) \haq^\dagger \hacl
	+
	(\omega_0 + i \frac{\kappa}{2}) \haq \hacl^\dagger
	\\ \label{eq:L_Linear_Oscillator}
	&
	-i \kappa(2 \nth+1) \haq^\dagger \haq.
\end{align}
This form directly mimics the structure of the Keldysh action for this same model \cite{Kamenev_Book_2011, Diehl_Keldysh_Review_2016}; we make this connection more explicit in Sec~\ref{sec:Keldysh_Single_Osc}. Just as in the field theory description, the coefficients of $\haq^\dagger \hacl$ and $\haq \hacl^\dagger$ serve as an effective non-Hermitian Hamiltonian, whereas the term $\haq^\dagger \haq$ should be thought of as the fluctuations that must accompany the dissipation \cite{RMP_Clerk}. Although $\hacl^\dagger$ and $\haq^\dagger$ are the conjugate of $\hacl$ and $\haq$, they are not the creation operators of true bosonic modes since $[\hacl, \hacl^\dagger] = [\haq, \haq^\dagger] = 0$. Thus, we cannot interpret $\hacl^\dagger \hacl$ and $\haq^\dagger \haq $ as number operators. 

Rather, despite being non-Hermitian, we should think of $\haq^\dagger \hacl$ and $- \haq \hacl^\dagger$ as operators whose eigenvalues correspond to a number of quanta in a mode. This follows since the only non-vanishing commutation relation are
\begin{align}\label{eq:Commutator}
	[\hacl, \haq^\dagger] = [\hacl^\dagger, -\haq  ] = \hone,
\end{align}
with $\hone$ the identity superoperator. Given their status as number operators, to determine the eigenvectors of $\haq^\dagger \hacl$ and $- \haq \hacl^\dagger$ we must first find the ``vacuums" of $\hacl, \hacl^\dagger$ and $\haq, \haq^\dagger$.  These are simply the parity and identity operator 
\begin{subequations}
\begin{align}\label{eq:Classical_Vacuum}
    \hat{0}_{\rm cl}
    \equiv
    2 e^{i \pi \ha^\dagger \ha}
    &\rightarrow
    \hacl
    |\hat{0}_{\rm cl}\rangle \rangle
    =
    \hacl^\dagger
    |\hat{0}_{\rm cl} \rangle \rangle
    =
    0
    \\\label{eq:Quantum_Vacuum}
    \hat{0}_{\rm q}
    \equiv
    \hat{1}
    &\rightarrow
    \langle \langle \hat{0}_{\rm q}| \haq 
    =
    \langle \langle \hat{0}_{\rm q}| \haq^\dagger 
    =
    0
\end{align}
\end{subequations}
which follows since $e^{i \pi \ha^\dagger \ha}$ and $\hat{1}$ respectively anti-commute and commute with both $\ha$ and $\ha^\dagger$. Using Schur's lemma, these vacuums are unique. In complete analogy with a simple harmonic oscillator, we can obtain the simultaneous eigenvectors of these number operators by repeatedly applying the appropriate creation operators on the vacuums. Thus, we find that
\begin{subequations}
    \begin{align} \label{eqs:RandLEigenvectors_1}
        |\hat{r}'_{\mu, \nu} \rangle \rangle
        &=
        \frac{1}{\sqrt{\mu! \nu!}}(\haq^\dagger)^{\mu}(-\haq)^\nu|\hat{0}_{\rm cl}\rangle \rangle
        \\ \label{eqs:RandLEigenvectors_2}
        \langle \langle \hat{l}'_{\mu, \nu}
        |,
        &=
        \frac{1}{\sqrt{\mu! \nu!}}\langle \langle \hat{0}_{\rm q}|(\hacl)^\mu (\hacl^\dagger)^\nu
    \end{align}
\end{subequations}
 are the right and left eigenvectors of $\haq^\dagger \hacl$ and $\haq \hacl^\dagger$ with eigenvalue $\mu$ and $-\nu$ respectively. 

The above eigenvectors are unfortunately not eigenvectors of the Liouvillian: while the first two terms in Eq.~(\ref{eq:L_Linear_Oscillator}) are proportional to our generalized number operators, the last term is not.  We next show that there is a simply way to deal with this.

\subsection{Gauging away the noise}
Recall that we have already reasoned that fluctuations in this linear system should never affect the eigenvalues of $\hL$, only its eigenvectors. We should thus be able to remove the temperature-dependent noise term $-i \kappa (2\nth+1)\haq^\dagger \haq$ entirely using a similarity transformation. This can indeed be achieved by  defining
\begin{align}\label{eq:V_def}
    \hV 
    \equiv
    e^{-(2\nth+1)\haq^\dagger \haq},
\end{align}
and using the commutation relations Eq.~(\ref{eq:Commutator}) along with the Baker-Campbell-Hausdorff identity to verify that
\begin{align}\label{eq:L_no_fluctuations}
    \hV^{-1}\hL \hV
    =
    (\omega_0-i\frac{\kappa}{2})\haq^\dagger \hacl
    +
    (\omega_0+i\frac{\kappa}{2})\haq \hacl^\dagger.
\end{align}
Since this superoperator and $\hL$ are isospectral, we confirm that as promised, the temperature does not effect the eigenvalues of the Lindbladian. 
Further, the transformed Lindbladian is now just a sum of our generalized number operators, and hence is diagonalized by the eigenvectors introduced in Eqs.~(\ref{eqs:RandLEigenvectors_1})-(\ref{eqs:RandLEigenvectors_2}).  Note that the transformed Lindbladian is not in Lindblad form due to the presence of a negative rate
\begin{align}\label{eq:V_Frame}
    \mathcal{V}^{-1}
    \mathcal{L}
    \mathcal{V}
    \hrho
    =
    \omega_0
    [\ha^\dagger \ha, \hrho]
    +
    i
    \frac{\kappa}{2}
    \left(
    \mathcal{D}[\ha]\hrho
    -
    \mathcal{D}[\ha^\dagger]\hrho
    \right),
\end{align}
which is however irrelevant for diagonalization purposes \footnote{One should be careful when working with the inverse of $\hV$, since as we we show in Sec.~\ref{sec:Single_Osc_Phase_Space} it is unbounded. One can always write down expression which involve only $\hV$ however, e.g. $\hL \hV |\hrho' \rangle \rangle = \hV\left[(\omega_0-i \frac{\kappa}{2})\haq^\dagger \hacl + (\omega_0+i \frac{\kappa}{2})\haq \hacl^\dagger)\right] |\hrho' \rangle \rangle $  }.

We have thus made a non-trivial mapping at the level of master equations from a noisy damped harmonic oscillator to its fluctuation-free equivalent.  Our ability to remove the noise term exactly is a consequence of the usual separation of dynamics and fluctuations in linear systems, yet here there are quantum consequences, seeing as how the eigenmodes of the Liouvillian have a Fock state structure. 

We pause to note that one could imagine adding a quadratic term to our Liouvillian that would break our symmetry, i.e.~a term proportional to  $\hacl^\dagger \hacl$. The similarity transformation $e^{-(2\nth+1)\haq^\dagger \haq}$ would no longer bring $\hL$ to a diagonal form, since the former commutes with the quantum superoperators but not the classical ones. Such a term is however forbidden by the trace-preserving nature of the dynamics $\Tr(\mathcal{L} \hrho) = \langle \langle \hat{0}_q| \hL | \hrho \rangle \rangle = 0 $, which a classical-classical term would violate. At a quantum level, being able to eliminate the noise can be understood as a special kind of symmetry; the combination of linearity and conservation of probability. 

This structure and an analogous transformation to Eq.~(\ref{eq:V_def}) will also be present in the multi-mode bosonic case considered in Sec.~\ref{sec:Quadratic_Bosons}. A comparable symmetry, transformation and conclusion will also be presented for fermionic master equations in Sec.~\ref{sec:Quadratic_Fermions}. In non-linear quantum systems, which are of course also probability-conserving, fluctuations can impact the dynamics as we show explicitly in Sec.~\ref{sec:Kerr}

\subsection{Eigenvalues, eigenvectors and non-Hermitian quasiparticles}
With Eq.~(\ref{eq:L_no_fluctuations}) already in diagonal form, the similarity transformation allows us to simply relate the eigenvectors $\hat{r}_{\mu, \nu}, \hat{l}_{\mu, \nu}$ of $\hL$ to $\hat{r}'_{\mu, \nu}, \hat{l}'_{\mu, \nu}$, those of $\haq^\dagger \hacl$ and $\haq \hacl^\dagger$. They are given by
\begin{subequations}\label{eq:r_l_Single_Oscillator}
\begin{align}\nonumber
    |\hat{r}_{\mu, \nu}\rangle \rangle 
    &=
    e^{-(2\nth+1)\haq^\dagger \haq}
     |\hat{r}'_{\mu, \nu}\rangle \rangle 
    \\ \label{eq:Right_Eigen_L_Harmonic}
    &=
    \frac{1}{\sqrt{\mu! \nu!}} 
    (\haq^\dagger)^{\mu} (-\haq)^\nu | \hrho_{\rm ss} \rangle \rangle,
    \\ \nonumber
     \langle \langle \hat{l}_{\mu,\nu} |
     &=
     \langle \langle \hat{l}'_{\mu,\nu} |
     e^{(2\nth+1)\haq^\dagger \haq}
    \\\label{eq:Left_Eigen_L_Harmonic}
     &=
     \frac{1}{\sqrt{\mu! \nu!}} 
     \langle \langle{\hat{0}_{\rm q}}|
     (\hacl(\nth))^\mu
     (\hacl^\dagger(\nth))^\nu
\end{align}
\end{subequations}
with corresponding eigenvalues 
\begin{align}
    E_{\mu, \nu}
    =
    (\omega_0-i\frac{\kappa}{2})\mu
    -
    (\omega_0+i \frac{\kappa}{2})\nu.
\end{align}
Here we have identified $\hrho_{\rm ss}$ as the unique zero-eigenvalue eigenvector of $\hL$, i.e.~the steady state of dissipative dynamics. We have also defined
\begin{subequations}
\begin{align}\label{eq:Def_acl_nth}
    \hacl(\nth)
    \equiv
    \hV
    \hacl
    \hV^{-1}
    =
    \hacl+(2\nth+1) \haq
    \\\label{eq:Def_acl_dag_nth}
    \hacl^\dagger(\nth)
    \equiv
    \hV
    \hacl^\dagger
    \hV^{-1}
    =
    \hacl^\dagger-(2\nth+1) \haq^\dagger
\end{align}
\end{subequations}
where note that since $\hV$ is not unitary, $\hacl^\dagger(\nth)$ is not the conjugate of $\hacl(\nth)$ despite the notation. This is however of no concern, seeing as the similarity transformations preserve all commutation relations, e.g.
\begin{align}
    [\hacl(\nth), \haq^\dagger]
    =
    [\hacl^\dagger(\nth), -\haq]
    =
    \hone.
\end{align}
We can thus identify $\hacl(\nth), \haq$ and $\hacl^\dagger(\nth), \haq$ as the correct independent ``non-Hermitian quasiparticle" superoperators which create and destroy excitations on top of the their vacuum with a definite energy and decay rate. Although the vacuum of the quantum superoperators is the identity, the right vacuum of these new classical superoperators is the steady state
\begin{align}\label{eq:New_Classical_Vacuum}
    \hacl(\nth)|\hrho_{\rm ss}\rangle \rangle
    =
    \hacl^\dagger(\nth)|\hrho_{\rm ss} \rangle \rangle
    =
    0,
\end{align}
which is of course explicitly temperature-dependent. 

The superoperators $\hacl(\nth), \hacl^\dagger(\nth)$ can also be interpreted as having displaced the classical superoperators by fluctuations. This can be made more precise by noting that $\hV$ does not affect average values of the creation or annihilation operators $\sqrt{2}\langle \ha(t) \rangle =  \langle \langle \hat{0}_q | \hacl |\hrho(t)\rangle \rangle = \langle \langle \hat{0}_q | \hacl(\nth)| \hrho(t) \rangle \rangle $ $= \sqrt{2}e^{-i\omega_0 t-\frac{\kappa}{2}t} \langle \ha(0) \rangle$, consistent with the zero-mean nature of the noise. Higher-order correlation functions like average particle number however are affected by fluctuations, which is why the original noise-free classical superoperators cannot be used to build up the left eigenvectors. For instance 
\begin{align}\nonumber
    \langle \langle \hat{0}_q | 
    \hacl^\dagger(\nth) & \hacl(\nth) 
    |\hrho(t)\rangle\rangle
    \\ \label{eq:Left_Eigen_Time_Evolution}
    &=
    e^{-\kappa t}
    \left[
    \langle
    \{\ha(0), \ha^\dagger(0) \}
    \rangle
    -(2\nth+1)
    \right]
\end{align}
has the simple exponential decay characteristic of a left eigenvector whereas its noise-free equivalent
$\langle \langle \hat{0}_q | \hacl^\dagger \hacl |\hrho(t)\rangle\rangle = \langle \{\ha(t), \ha^\dagger(t)\} \rangle = e^{-\kappa t}\langle \{\ha(0), \ha^\dagger(0)\} \rangle$ $+ (2\nth+1)(1-e^{-\kappa t})  $ does not.

\subsection{Remaining unsatisfactory issues}
This completes the diagonalization of a single thermally-damped oscillator (something that can be easily generalized to the multi-mode case, as we show below). There are however two unsatisfactory features of the procedure presented above. First, the similarity transformation $\hV$ at this stage does not seem to possess a simple physical interpretation. For instance, it is well-known that the steady state is a Gaussian state, with an average particle number of $\nth$.  In contrast, our procedure simply asserts that it is equivalent to $e^{-(2\nth+1)\haq^\dagger \haq}|\hat{0}_{\rm cl} \rangle \rangle$.  Is there a direct way to see the connection here,  that is
\begin{align}\nonumber
\hrho_{\rm ss}
&=
    \frac{1}{\nth+1}
    \sum_{n=0}^{\infty}
    \left(
    \frac{\nth}{\nth+1}
    \right)^{n}
    |n\rangle \langle n |
    \\
    &\stackrel{?}{=}
    2
    e^{-(\nth+\frac{1}{2})
      [\ha^\dagger, [\ha, \cdot ]]}
     e^{i \pi \ha^\dagger \ha}.
\end{align}
It is at this stage not clear how we should think about this result or even explicitly demonstrate this equality. Second, the right eigenvectors are given by nested commutators of $\ha^\dagger, \ha$ and the steady state; their explicit second-quantized form is tedious to obtain and largely uninformative. Take for example the right eigenvector describing population decay with a characteristic decay rate of $2\kappa$, 
\begin{align}
    \hat{r}_{2, 2}
    =
    \frac{1}{8\nth^2(\nth+1)^2}
    \left(
    \hat{n}^2
    -(4\nth+1)\hat{n}
    +2\nth^2
    \right)
    \hrho_{\rm ss}
\end{align}
with $\hat{n} \equiv \ha^\dagger \ha$. At first glance, there is no obvious intuitive way to think about such an eigenvector, let alone a simple connection to a 2-excitation Fock state.  

\section{Equivalence between third-quantization and phase-space formalism}\label{sec:Single_Osc_Phase_Space}
While seemingly devoid of physical content, we shall now demonstrate that both the similarity transformation $\hV$ in Eq.~(\ref{eq:V_def}) and eigenvectors $\hat{r}_{\mu, \nu}$, $\hat{l}_{\mu, \nu}$ of Eq.~(\ref{eq:r_l_Single_Oscillator}) have transparent phase-space representations. The continuous nature of phase space arises by asking a natural question: what are the eigenvectors and eigenvalues of the classical and quantum superoperators? As we now show, the answer to this question leads to the main results of this section: the similarity transformation $\hV$ is equivalent to a phase-space convolution with a Gaussian of width $2\nth+1$ and the eigenvectors of $\hL$ have a functional form equivalent to that of a Fock state. 
\begin{table}
\centering
\resizebox{\columnwidth}{!}{
	\begin{tabular}{|r|c|c|c|}
				\cline{2-4}
				\multicolumn{1}{c|}{}
				& \multicolumn{1}{c|}{$|\hrho \rangle \rangle$}
				& \multicolumn{1}{c|}{$W_{\hrho}(\alpha)$}
				& \multicolumn{1}{c|}{$\Lambda_{\hrho}(\eta)$} \\ \cline{1-4} 
				$\hacl$ & $\frac{1}{\sqrt{2}}|\{\ha, \hrho\} \rangle \rangle$ & $\alpha W_{\hrho}(\alpha)$ & $\partial_{\eta^*} \Lambda_{\hrho}(\eta)$ \\
				$\haq$ & $\frac{1}{\sqrt{2}}|[\ha, \hrho] \rangle \rangle$ & $\partial_{\alpha^*} W_{\hrho}(\alpha)$ & $\eta \Lambda_{\rho}(\eta)$\\
				$\hacl^\dagger$ & $\frac{1}{\sqrt{2}}|\{\ha^\dagger, \hrho\} \rangle \rangle$ & $\alpha^* W_{\hrho}(\alpha)$ & $-\partial_{\eta} \Lambda_{\hrho}(\eta)$  \\
				$\haq^\dagger$ & $\frac{1}{\sqrt{2}}|[\ha^\dagger, \hrho] \rangle \rangle$ & $-\partial_{\alpha} W_{\hrho}(\alpha)$ & $\eta^* \Lambda_{\rho}(\eta)$
				\\
				\hline
			\end{tabular}%
}
  \caption{\label{tb:Quantization_Rules} The action of third-quantized creation and annihilation superoperators on the density matrix $|\hrho \rangle \rangle$, Wigner function $W_{\hrho}(\alpha)$ or characteristic function $\Lambda_{\hrho}(\eta)$. These rules can be used to obtain an equivalent set of equations of motion, see Eq.~(\ref{eq:Master_Eq}) for the single linear oscillator case.}
\end{table}

\subsection{Classical and quantum eigenvectors}
Thankfully, the eigenvectors of $\hacl, \hacl^\dagger$ and $\haq, \haq^\dagger$ are already familiar from quantum optics. For any complex number $\alpha$ and $\eta$ let us define corresponding displaced parity and displacement operators \cite{Displaced_Squeezed_Ops_1994_PRA} respectively
\begin{subequations}
    \begin{align}\label{eq:alphacl_def}
        \halphacl
        &
        \equiv
        2 \hat{D}(\sqrt{2}\alpha)
        e^{i \pi \ha^\dagger \ha},
        \\ \label{eq:alphaq_def}
        \hetaq
        &\equiv
        \hat{D}(\sqrt{2}\eta)
    \end{align}
\end{subequations}
with $\hat{D}(\alpha) \equiv e^{\alpha \ha^\dagger - \alpha^* \ha}$. Using the defining property of the displacement operator $ \ha \hat{D}(\alpha) = \hat{D}(\alpha)(\ha + \alpha) $ and parity operator $ \ha e^{i \pi \ha^\dagger \ha} = -e^{i\pi \ha^\dagger \ha} \ha $, we arrive at
\begin{subequations}
    \begin{align} \label{eq:acl_eigen_property}
        \hacl |\halphacl \rangle \rangle &= \alpha|\halphacl \rangle \rangle,
        \: \: 
        \hacl^\dagger |\halphacl \rangle \rangle
        = \alpha^* |\halphacl \rangle \rangle,
        \\\label{eq:aq_eigen_property}
        \haq |\hetaq \rangle \rangle &= \eta |\hetaq \rangle \rangle,
        \: \:\: \: \:
        \haq^\dagger |\hetaq \rangle \rangle = \eta^* |\hetaq \rangle \rangle.
    \end{align}
\end{subequations}
     The spectrum of each of our third-quantization superoperators is thus the whole complex plane. 
     
     The corresponding eigenvectors are reminiscent of standard single-mode coherent states, an analogy we can make even stronger by writing
     \begin{subequations}
    \begin{align}\label{eq:Displaced_cl}
        |\halphacl \rangle \rangle
        &=
        e^{\alpha \haq^\dagger
        -
        \alpha^* \haq}
        |
        \hat{0}_{\rm cl}
        \rangle \rangle,
        \\ \label{eq:Displaced_q}
        | \hetaq \rangle \rangle
        &=
        e^{\eta \hacl^\dagger
        -
        \eta^* \hacl}
        |
        \hat{0}_{\rm q}
        \rangle \rangle
    \end{align}
    \end{subequations}
    in the same way that coherent states are displaced vacuum. Unlike coherent states however, which are not eigenvectors of $\ha^\dagger$, $\hacl$ and $\hacl^\dagger$ share the same eigenvectors, as do $\haq$ and $\haq^\dagger$. This a consequence of $[\hacl, \hacl^\dagger] = [\haq, \haq^\dagger] = 0$. For the same reason, $\halphacl$ and $\hetaq$ are orthogonal to other eigenstates of the same kind:
    \begin{subequations}
      \begin{align}
        \langle \langle \hat{\alpha}_{\rm cl} | \hat{\beta}_{\rm cl} \rangle \rangle
        &=
        \Tr
        \left(
        \hat{\alpha}_{\rm cl }^\dagger 
        \hat{\beta}_{\rm cl}
        \right)
        =
        2\pi
        \delta^2(\alpha - \beta),
        \\
        \langle \langle \hat{\eta}_{\rm q} | \hat{\xi}_{\rm q} \rangle \rangle
        &=
        \Tr
        \left(
        \hat{\eta}_{\rm q }^\dagger 
        \hat{\xi}_{\rm q}
        \right)
        =
        \frac{\pi}{2}
        \delta^2(\eta - \xi)
    \end{align}
    \end{subequations}
    where the trace was computed using coherent states. In this sense, these operators are more reminiscent to position and momentum eigenkets. 
   Just like their first-quantized counterparts, they can then be used to form a resolution of the identity
    \begin{align}\label{eq:Res_Identity}
        \hone
        =
        \int \frac{d^2 \alpha}{2 \pi}
        |\hat{\alpha}_{\rm cl} \rangle \rangle
        \langle \langle \hat{\alpha}_{\rm cl} |
        =
        \int \frac{2 d^2\eta}{\pi}
        | \hat{\eta}_{\rm q} \rangle \rangle
        \langle \langle \hat{\eta}_{\rm q} |
    \end{align}
   where $d^2\beta \equiv d\text{Re}\beta  \, d \text{Im} \beta$ for any complex $\beta$. It follows that we can write the density matrix as
    \begin{align}\label{eq:rho_classical_quantum_basis}
        | \hrho \rangle \rangle
        =
        \int \frac{d^2 \alpha}{2\pi}
        W_{\hrho}(\alpha)
        | \hat{\alpha}_{\rm cl} \rangle \rangle
        =
        \int \frac{2 d^2 \eta}{\pi}
        \Lambda_{\hrho}(\eta)
        | \hat{\eta}_{\rm q} \rangle \rangle
    \end{align}
    with 
    \begin{subequations}
    \begin{align}\nonumber
        W_{\hrho}(\alpha)
        &\equiv
        \langle \langle \halphacl | \hrho \rangle \rangle
        =
        2
        \Tr 
      \left(
      e^{i \pi \hat{a}^\dagger \hat{a}}
      \hat{D}^{\dagger}(\sqrt{2} \alpha)
      \hrho
      \right)
      \\ \label{eq:W_alphacl}
      & =
      2 
      \Tr 
      \left(
      e^{i \pi \hat{a}^\dagger \hat{a}}
      \hat{D}^{\dagger}(\frac{\alpha}{\sqrt{2}})
      \hrho
      \hat{D}(\frac{\alpha}{\sqrt{2}}),
      \right)
      \\ \label{eq:W_alphaq}
       \Lambda_{\hrho}(\eta)
         &\equiv
        \langle \langle \hetaq | \hrho \rangle \rangle
        =
        \Tr 
      \left(
      \hat{D}^{\dagger}(\sqrt{2} \eta)
      \hrho
      \right).
    \end{align}
    \end{subequations}
    Remarkably, these ``wavefunctions" of the density matrix in these bases are well-known objects: $W_{\hrho}(\alpha)$ and $\Lambda_{\hrho}(\eta)$ are precisely the Wigner function of $\hrho$ and its characteristic function respectively \cite{Royer_1977_Wigner_Disaplced, gardiner_zoller_2010}.

    To recover the result that $W_{\hrho}(\alpha)$ and $\Lambda_{\hrho}(\eta)$ are Fourier transforms of each other, we compute the overlap
    \begin{align}\label{eq:alhpacl_alphaq_overlap}
        \langle \langle \hat{\alpha}_{\rm cl} | \hat{\eta}_{\rm q} \rangle \rangle
        =
        \Tr
        \left(
        \hat{\alpha}_{\rm cl }^\dagger 
        \hat{\eta}_{\rm q}
        \right)
        =
        e^{\eta \alpha^*-\eta^* \alpha},
    \end{align}
    use the definitions Eqs.~(\ref{eq:W_alphacl})-(\ref{eq:W_alphaq}) and the resolution of the identify Eq.~(\ref{eq:Res_Identity})  to obtain
    \begin{subequations}
    \begin{align}\label{eq:Fourier_Transform_W}
        W_{\hrho}(\alpha)
        &=
        \int \frac{2 d^2\eta}{\pi}
        e^{\eta \alpha^*-\eta^* \alpha}
        \Lambda_{\hrho}(\eta),
        \\
        \Lambda_{\hrho}(\eta)
        &=
        \int \frac{d^2\alpha}{2\pi}
        e^{\eta^* \alpha-\eta \alpha^*}
        W_{\hrho}(\alpha).
    \end{align}
    \end{subequations}
   Much like position and momentum eigenkets corresponds to a perfectly spatially-localized or delocalized particle in standard quantum mechanics, the eigenkets $|\halphacl \rangle \rangle$ and $|\hetaq \rangle \rangle$ can be thought of as states in phase space which are perfectly localized at $\alpha$ or delocalized with wavevector $\eta$. That $\alpha$ and $\eta$ are Fourier transform pairs  is also consistent with $\hacl, \haq^\dagger$ and $\haq, \hacl^\dagger$ being conjugate to one another. This can be further demonstrated by using Eqs.~(\ref{eq:Displaced_cl})-(\ref{eq:Displaced_q}) to write
   \begin{subequations}
\begin{align}\label{eq:acl_aqdag_derivative}
         \langle \langle \hetaq |\hacl 
        &=
        \partial_{\eta^*} 
         \langle \langle \hetaq |, 
        \: \: \:
        \langle \langle \halphacl| \haq^\dagger 
        =
        -\partial_{\alpha} 
       \langle \langle \halphacl| 
        \\ \label{eq:aq_acldag_derivative}
        \langle \langle \halphacl | \haq
        &=
        \partial_{\alpha^*} 
        \langle \langle \halphaq |, 
        \: \: \:
        \langle \langle \hetaq | \hacl^\dagger 
        =
        -\partial_{\eta} 
        \langle \langle \hetaq |.
    \end{align}
    \end{subequations}
   Given any third-quantized Lindbladian, we can then use Eqs.~(\ref{eq:W_alphacl})-(\ref{eq:W_alphaq}) and Eqs.~(\ref{eq:acl_aqdag_derivative})-(\ref{eq:aq_acldag_derivative}) to readily find the equation of motion for either the Wigner or the characteristic function. These rules are summarized in Table~\ref{tb:Quantization_Rules}.  For example, we can recover the standard equivalent equations of motion \cite{gardiner_zoller_2010}
   \begin{widetext}
   \begin{subequations}\label{eq:Master_Eq}
   \begin{align}
       i \partial_t |\hrho \rangle \rangle
       &=
       \left[
       	(\omega_0 - i \frac{\kappa}{2}) \haq^\dagger \hacl
	    +
    	(\omega_0 + i \frac{\kappa}{2}) \haq \hacl^\dagger
	    -i \kappa(2 \nth+1) \haq^\dagger \haq
	    \right]
	    |\hrho\rangle \rangle
	    \\ \nonumber
	     &\hspace{4.5cm}\updownarrow
	    \\ \label{eq:FP_Wigner}
	    i \partial_t W_{\hrho}(\alpha)
	    &=
	   \left[
       	-(\omega_0 - i \frac{\kappa}{2})\partial_{\alpha} \alpha
	    +
    	(\omega_0 + i \frac{\kappa}{2}) \partial_{\alpha^*} \alpha^*
	    +i \kappa(2 \nth+1) \partial_{\alpha}\partial_{\alpha^*}
	    \right] 
	    W_{\hrho}(\alpha)
	    \\\nonumber
	    &\hspace{4.5cm}\updownarrow
	    \\ \label{eq:W_alphaq_EOM}
	    i \partial_t \Lambda_{\hrho}(\eta)
	    &=
	    \left[
       	(\omega_0 - i \frac{\kappa}{2}) \eta^* \partial_{\eta^*}
	    -
    	(\omega_0 + i \frac{\kappa}{2}) \eta\partial_{\eta}
	    -i \kappa(2 \nth+1) \eta^* \eta
	    \right]
	    \Lambda_{\hrho}(\eta).
   \end{align}
   \end{subequations}
   \end{widetext}
  From now on, we will write the classical and quantum superoperators and their action on these eigenvectors as equivalent, e.g. $\hacl = \alpha = \partial_{\eta^*}$. 

\subsection{Interpretation of $\hV$ and phase-space eigenvectors of $\hL$}
It is worth stressing that by using the resolution of the identity Eq.~(\ref{eq:Res_Identity}), one can obtain a phase-space representation of an arbitrary operator $W_{\hat{X}}(\alpha) \equiv \langle \langle \halphacl | \hat{X} \rangle \rangle = 2\Tr(e^{i \pi \ha^\dagger \ha}\hat{D}(\sqrt{2}\alpha)\hat{X})$. This is nothing but the Wigner-Weyl transform \cite{Quantization_Phase_Space_Treatise}, yet here it appears much more naturally using the language of standard quantum mechanics. This allows us to easily describe how the similarity transformation $\hV$ which eliminated the noise acts on an arbitrary density matrix, something which would be extremely challenging using standard formulations and tools of quantum mechanics in phase space, e.g. phase-point operators \cite{Fano_1957_RMP, Phase_Point_Wooter_1987}.


Being diagonal in the quantum basis, $\hV = e^{-(2\nth+1)\haq^\dagger \haq}$ simply acts by multiplication on the characteristic function $\Lambda_{\mathcal{V} \hrho}(\eta) = e^{-(2\nth+1)|\eta|^2} \Lambda_{\hrho}(\eta)$. We can then obtain the Wigner function by Fourier transforming using Eq.~(\ref{eq:Fourier_Transform_W}) 
 \begin{align}\label{eq:V_Interpret}
     W_{\mathcal{V} \hrho}(\alpha)
     =
     \int \frac{d^2 \beta}{(2\nth+1)\pi}
     e^{-\frac{|\alpha-\beta|^2}{2\nth+1}}
     W_{\hrho}(\beta).
 \end{align}
 As promised, $\hV$ is phase-space convolution with a Gaussian of width $(2\nth+1)$, spreading out and delocalizing any distribution. This is also consistent with the similarity transformation $e^{-(2\nth+1)\haq^\dagger \haq }$ removing the noise, or equivalently the diffusion term in the Fokker-Planck equation Eq.~(\ref{eq:FP_Wigner}), which has an identical delocalization effect. Note that if $\nth = 0$, then this is precisely the Glauber-Sudarshan $P$-function of the oscillator, which is insensitive to vacuum noise \cite{gardiner_zoller_2010}.

Turning our attention to the right eigenvectors of $\hL$, we find that they have a compact form in phase space
\begin{align} \nonumber
    W_{\hat{r}_{\mu, \nu}}(\alpha)
     &\equiv
     \frac{1}{\sqrt{\mu!\nu!}}
     \langle \langle \halphacl|
     (\haq^\dagger)^\mu
     (-\haq)^\nu
     |\hrho_{\rm ss} \rangle \rangle
     \\ \label{eq:Right_Eigen_Phase_Space}
     &=
     \frac{(-1)^{\mu+\nu}}{\sqrt{\mu!\nu!}}
    \partial_{\alpha}^\mu 
     \partial_{\alpha^*}^\nu
     W_{\hrho_{\rm ss}}(\alpha).
\end{align}
Focusing on the $\mu = \nu$ case for simplicity and comparing with the Wigner function of Fock state projectors $|\mu \rangle \langle \mu |$ we obtain
\begin{subequations}
 \begin{align}\label{eq:Pop_Eigen}
     W_{\hat{r}_{\mu, \mu}}(\alpha)
     &=
     \frac{2(-1)^\mu e^{-\frac{|\alpha|^2}{2\nth+1}}}{(2\nth+1)^{\mu+1}}
     L_{\mu}
     \left(
     \frac{|\alpha|^2}{2\nth+1}
     \right),
     \\
     W_{|\mu\rangle \langle \mu|}(\alpha)
     &=
     2(-1)^\mu
     e^{-|\alpha|^2}
     L_{\mu}
     (
     2|\alpha|^2
     )
 \end{align}
 \end{subequations}
 where $L_{\mu}$ is the $\mu$-th Laguerre polynomial. Unlike the explicit second-quantized form, which from Eq.~(\ref{eq:Right_Eigen_L_Harmonic}) requires computing nested commuators of $\ha, \ha^\dagger$ and $\hrho_{\rm ss}$, in this representation $\hat{r}_{\mu, \nu}$ bear a striking similarity to Fock states. Although they can be made almost identical by re-scaling the phase-space variable $W_{\hat{r}_{\mu, \mu}}(\alpha) \to W_{\hat{r}_{\mu, \mu}}(\sqrt{2\nth+1}\alpha$), which would make the Gaussian factors match, the argument of the Laguerre polynomial would still differ by a factor of $2$. This property also holds when comparing arbitrary right eigenmodes $\hat{r}_{\mu, \nu}$ and outer products of Fock states $|\mu \rangle \langle \nu|$ which we show in Appendix~\ref{app_section:Phase_space_right_left_eigenmodes}. There we also give the phase-space representations of the left eigenvectors.

 Using the equivalent representations of the master equation Eq.~(\ref{eq:Master_Eq})), we conclude that the phase-space representations of $|\hat{r}_{\mu, \nu} \rangle \rangle$ and $\langle \langle \hat{l}_{\mu, \nu} |$ are also right and left eigenvectors of the classical Fokker-Planck differential operator. The existence of quantized Laguerre-polynomial eigenfunctions is a well-known property of this classical damped oscillator problem \cite{Fokker_Planck_Textbook}, but the corresponding quantum analog arguably is not. The quantized nature of these modes in a classical or quantum setting is easily understood for this linear problem: any observable should only oscillate and decay at integer multiples of the fundamental frequency and decay rate respectively. For instance, $ \langle (\ha^\dagger)^{\nu} \ha^\mu \rangle $ and its classical counterpart includes the harmonics $\omega_0(\mu-\nu)$ and decay $\frac{\kappa}{2}(\mu+\nu)$.

 Finally, note that these eigenvectors are still correct even when there is no dissipation $\nth \to 0, \kappa \to 0 $.  This might seem surprising at first, as in this limit  $W_{\hat{r}_{\mu, \mu}}(\alpha) \neq W_{|\mu\rangle\langle \mu|}(\alpha)$. This discrepancy is easily understood.  Without decay, each eigenvalue $E_{\mu,\nu} = \omega_0(\mu-\nu)$ is infinitely degenerate, which is most obvious by writing $\hL = \omega_0(\haq^\dagger  \hacl + \haq \hacl^\dagger) = \omega_0(-\partial_{\alpha} \alpha + \partial_{\alpha^*} \alpha^*) = \omega_0 i\partial_{\phi}$ where $\phi = \arg \alpha$. There is thus no unique way to diagonalize $\hL$; in this limit our approach picks out one linearly-independent set of eigenvectors out of the infinitely-many available ones. 

\section{Correspondence between third quantization and Keldysh field theory}\label{sec:Keldysh_Single_Osc}
Despite the Lindbladian Eq.~(\ref{eq:L_Linear_Oscillator}) being reminiscent of the dissipative Keldysh action of the same model, the connection between these approaches we have been suggesting are at this stage not rigorous.  In this section, we make the correspondence concrete by showing that Keldysh field theory emerges directly from the structure of the third-quantized Lindbladian expressed in the basis of classical and quantum superoperators. 
Our approach here complements standard derivations of the Keldysh action for open systems \cite{Diehl_Keldysh_Review_2016, Kamenev_Book_2011}.  Unlike these treatments, our focus is not the derivation of a partition function for the calculation of correlation functions, but rather on obtaining the time evolution of an arbitrary initial state.  In this sense, this section introduces an alternate formulation of the Feynman-Vernon influence functional approach  \cite{Feynman_Vernon_1962} and subsequent closely-related descriptions \cite{Star_product_PRD, Marinov_1991}.


\subsection{Path integral for a thermally-damped harmonic oscillator}
The Keldysh path integral emerges here by asking how the Wigner or characteristic function evolves in time. 
Indeed, since $e^{-i \hL t}$ is the time evolution superoperator, we can use the resolution of the identity Eq.~(\ref{eq:Res_Identity}) formed using the classical and quantum operators to write
\begin{align} \nonumber
    |\hrho(t)\rangle \rangle
    &=
    e^{-i \hL t}
    \int\frac{d^2\alpha}{2\pi}
    W_{\hrho(0)}(\alpha)
    |\halphacl \rangle \rangle \\
    &
    =
    \int 
    \frac{2 d^2 \eta}{\pi}
    \Lambda_{\hrho(t)}(\eta)
    | \hetaq \rangle \rangle 
\end{align}
where 
\begin{subequations}
\begin{align}\label{eq:Prop_Integral_Definition}
    \Lambda_{\hrho(t)}(\eta)
    &=
    \int\frac{d^2\alpha}{2\pi}
    K(\eta,\alpha;t)
    W_{\hrho(0)}(\alpha),
    \\ \label{eq:K_Def}
    K(\eta, \alpha; t)
    &\equiv
    \langle \langle \hetaq |
    e^{-i \hL t}
    | \halphacl \rangle \rangle.
\end{align}
\end{subequations}
The matrix element $K(\eta, \alpha;t)$, also referred to as the kernel, thus serves as the mixed phase-space propagator, and  in principle contains the same information as the spectral decomposition of $\hL$.

Using a resolution of the identity
\begin{align}\label{eq:Resolution_Identity_cl_q_j}
    \hone 
    =
    \int
    \frac{d^2 \aclj d^2 \aqj}{\pi^2}
    e^{\aqj \aclj^*-\aqj^* \aclj}
    |\ha_{{\rm cl}, j} \rangle \rangle 
    \langle \langle \ha_{{\rm q}, j} |,
\end{align}
which follows directly from Eq.~(\ref{eq:Res_Identity}) and Eq.~(\ref{eq:alhpacl_alphaq_overlap}), along with the defining property of the classical and quantum eigenvectors, we can obtain a path-integral description of Eq.~(\ref{eq:K_Def}) using the familiar prescription \cite{Altland_Simons_2010}. The central object will be a functional very similar to the standard Keldysh action of a thermally-damped harmonic oscillator. There are however important differences that emerge. We emphasize these here, leaving the technical and mostly-familiar details in Appendix~\ref{app_sec:Propegator_Harmonic_Oscillator}.

Once these details have been taken into account, we are left with a path integral representation of the kernel
\begin{align}\nonumber
    &K(\eta, \alpha; t)
    =
    \\ \label{eq:Continuum_Limit}
    &\int \mathcal{D}[a_{\rm cl}, a_{\rm q}]
    e^{i S_0[a_{\rm cl}, a_{\rm q}]
    +
    \int_{0}^t
    dt' 
    \left(
    \mathbbm{a}^\dagger(t')
    \mat{J}(t')
    -
    \mat{J}^\dagger(t')
    \mathbbm{a}(t')
    \right)
    }
\end{align}
where $\mathcal{D}[a_{\rm cl}, a_{\rm q}]$ is the usual functional integration measure, the action takes the form
  \begin{align}\nonumber
    &S_0[a_{\rm cl}, a_{\rm q}]
    =
    \\ \label{eq:S_Finite_Contour}
    &
    \int_{0}^{t}
    \int_{0}^{t}
    dt'
    dt''
    \begin{pmatrix}
    a^*_{\rm cl}(t') & a^*_{\rm q}(t')
    \end{pmatrix}
    \mat{G}^{-1}(t',t'')
    \begin{pmatrix}
    a_{\rm cl}(t'') \\ a_{\rm q}(t'')
    \end{pmatrix},
\end{align}
and the inverse of the matrix Green's function reads
\begin{align}
    \mat{G}^{-1}(t',t'')
    =
    \delta(t'-t'')
    \begin{pmatrix}
    0 & i\partial_{t'} - (\omega_0+i\frac{\kappa}{2}) \\
    i\partial_{t'} - (\omega_0-i\frac{\kappa}{2}) & i \kappa(2\nth+1)
    \end{pmatrix}.
\end{align}
Throughout this work, we will denote vectors and matrices with a
blackboard bold font to distinguish them from operators
and superoperators. Here we have defined the row vectors $\mathbbm{a}^\dagger(t') \equiv \left(a^*_{\rm{cl}}(t'), a^*_{\rm{q}}(t')\right)$ and the source term
\begin{align}\label{eq:Source_Term}
    \mat{J}^\dagger(t')
    \equiv
    \left(
    -\eta^* \delta(t'-t)
    ,
    \alpha^* \delta(t')
    \right).
\end{align}
Note that in most path integrals, from Eq.~(\ref{eq:K_Def}), one would normally impose boundary conditions on the fields such as $a_{\rm cl}(0) = \alpha$, $a_{\rm q}(t) = \eta$; this is incorrect here as it implies that $\mat{G}^{-1}$ has a zero-eigenvalue eigenvector. We stress that one must instead keep track of the boundary term through the source $\mat{J}$ and work directly the with the discrete representation of the Green's function as we do in Appendix \ref{app_sec:Propegator_Harmonic_Oscillator}.

There we show that, although $\mat{G}^{-1}$ is symbolically the same regardless of the length of the time contour, the matrix Green's function
\begin{align}\label{eq:G_Matrix_Single_Osc}
    \mat{G}(t',t'')
    \equiv
    \begin{pmatrix}
    G^{K}(t',t'') & G^R(t',t'') \\
    G^A(t',t'') & 0
    \end{pmatrix}
\end{align}
is sensitive to such changes. While the retarded $G^R(t',t'')$ and advanced $G^A(t',t'')$ Green's function remains the same regardless of the contour, the Keldysh Green's $G^K(t',t'')$ function does not 
\begin{subequations}
\begin{align}
    &G^R(t',t'') 
    = 
    -i \Theta(t'-t'')e^{-i(\omega_0-i\frac{\kappa}{2})(t'-t'')},
    \\
    &G^A(t',t'')
    =
    (G^R(t'',t'))^*,
    \\\nonumber 
    &G^K(t',t'')
    =
    -i\kappa(2\nth+1)
    \int_0^t
    dT
    G^R(t',T)G^A(T,t'')
    \\
    &=
     -i(2\nth+1) e^{-i \omega_0(t'-t'')}
     \left(e^{-\frac{\kappa}{2}|t'-t''|}-e^{-\frac{\kappa}{2}(t'+t'')}
     \right)
\end{align}
\end{subequations}
with $\Theta(t)$ the Heaviside step function. Crucially, the discrete-time path integral makes it clear that the correct choice of the Heaviside step function at zero is $\Theta(0) = 1$.

The path integral can be computed in the usual manner when one has a quadratic action and a source term. Making the displacement
\begin{align}
    \mathbbm{a}(t') \to \mathbbm{a}(t')+i\int_0^t dt'' \mat{G}(t',t'')\mat{J}(t'')
\end{align}
and, using $\int \mathcal{D}[a_{\rm cl}, a_{\rm q}] e^{i S_0[a_{\rm cl}, a_{\rm q}]} = \det(i \mat{G})= 1$ as in the usual Keldysh formalism, we obtain
\begin{align} \nonumber
    &K(\eta, \alpha; t)
    =
    e^{-i
    \int_0^t 
    dt' dt''
    \mat{J}^\dagger(t')
    \mat{G}(t',t'')
    \mat{J}(t'')
    }
    \\ \label{eq:K_Solution}
    &
    =
    e^{
    -(2\nth+1)(1-e^{-\kappa t})|\eta|^2
    +
    e^{-\frac{\kappa}{2}t}
    \left(
    \eta^* \alpha e^{-i \omega_0 t}
    -
    \rm{c.c.}
    \right)
    }.
\end{align}
As expected due to the quadratic nature of the Lindbladian, the propagator is a Gaussian function of $\alpha$ and $\eta$. Equation~(\ref{eq:K_Solution}) can independently be verified to be correct by noting that it must satisfy Eq.~(\ref{eq:W_alphaq_EOM}) along with the initial condition $K(\eta, \alpha; 0) = e^{\eta^* \alpha-\rm{c.c.}}$. 

\subsection{Correlation functions and timer evolution of the Wigner function}
For the sake of completeness, in this subsection we demonstrate how, with knowledge of $K(\eta, \alpha;t)$ and the initial Wigner function $W_{\hrho(0)}(\alpha)$, one can straightforwardly obtain both arbitrary correlation functions and the time-evolved Wigner function $W_{\hrho(t)}(\alpha)$. Since the characteristic function $\Lambda_{\hrho(t)}(\eta)$ is the generator of symmetrically-ordered  correlation functions, we can simply use Eq.~(\ref{eq:Prop_Integral_Definition}) to obtain
\begin{align}\nonumber
    &(\sqrt{2})^{\mu + \nu}
    \langle \{\ha^{\mu} (\ha^\dagger)^{\nu} \}_{\rm sym}  \rangle
    =
    (\partial_{\eta^*})^\mu 
    (-\partial_\eta)^\nu
    \Lambda_{\hrho(t)}(\eta)|_{\eta = \eta^* = 0}
    \\
    &
    =
    \int\frac{d^2\alpha}{2\pi}
    \Big[
    (\partial_{\eta^*})^\mu 
    (-\partial_\eta)^\nu
    K(\eta,\alpha;t)_{\eta = \eta^* = 0}
    \Big]
    W_{\hrho(0)}(\alpha),
\end{align}
which is valid for any Lindbladian, and not simply the damped oscillator considered here. 

Using the resolution of the identity Eq.~(\ref{eq:Res_Identity}) we can obtain the classical-classical phase-space propagator 
\begin{align}\label{eq:Prop_Cl_Cl_One}
    \Xi(\beta, \alpha; t)
    \equiv
    \langle \langle \hat{\beta}_{\rm cl} | 
    e^{-i \hL t}
    | \hat{\alpha}_{\rm cl} \rangle \rangle
    =
    \int \frac{2 d^2 \eta}{\pi}
    e^{\eta \beta^*-\eta^* \beta}
    K(\eta, \alpha; t)    
\end{align}
which we can then use to write
\begin{align}\label{eq:Prop_Cl_Cl_Two}
    W_{\hrho(t)}(\beta)
    =
    \int\frac{d^2\alpha}{2\pi}
    \Xi(\beta, \alpha; t)
    W_{\hrho(0)}(\alpha).
\end{align}
Once again this is valid for an arbitrary Lindbladian, but can be done here exactly due to the Gaussian nature of the problem.

\subsection{Path integral for arbitrary Lindbladians}
Although in this example $\hL$ was quadratic in creation and annihilation superoperators, it is evident that a path integral representation of the propagator can be obtained for arbitrary Lindbladians. Replacing all $\haq, \haq^\dagger$ and $\hacl, \hacl^\dagger$ by their respective fields in the action however is only valid once all quantum superoperators have been placed to the left of the classical ones. This is due to the form of the resolution of the identity Eq.~(\ref{eq:Res_Identity}). This convention is different than the one used in the standard coherent-state path integral, where one must ensure that the Lindlabidan is normal-ordered \cite{Kamenev_Book_2011, Diehl_Keldysh_Review_2016}. No differences emerge in this linear setting, but the two different prescriptions will produce two different Keldysh actions for an interacting problem. We expand on this point in Sec.~\ref{sec:Kerr} when we study a dissipative Kerr oscillator. 

\section{Quadratic Bosonic Lindbladians}\label{sec:Quadratic_Bosons}
   \subsection{Setup}
    We now generalize the previous results to an arbitrary quadratic multi-mode bosonic Lindbladian. The model now under consideration is a Lindblad master equation of the form
    \begin{widetext}
    \begin{align} \nonumber
        &i \partial_t \hrho 
         = 
        \sum_{n,m}
        \mat{H}_{nm} [\ha_n^\dagger \ha_m, \hrho]
        +
        \frac{1}{2}
        \sum_{n,m}
        [\mat{K}_{nm}\ha_n^\dagger \ha_m^\dagger + {\rm{h.c.}}
        ,
        \hrho]
        +
        i 
        \sum_b
        \mathcal{D}
        \left[
        \sum_m
        \left(
        l_{bm} \ha_m
        +
        p^*_{bm}\ha_m^\dagger 
        \right)
        \right] \hrho
        \\ \nonumber 
        & =
        \sum_{n,m}
        \mat{H}_{nm} [\ha_n^\dagger \ha_m, \hrho]
        +
        i \sum_{n,m} 
        \mat{L}_{nm}
        \left( 
        \ha_m \hrho \ha_n^\dagger
        -
        \frac{1}{2}
        \{\ha_n^\dagger \ha_m, \hrho\}
        \right)
        +
        i
        \sum_{n,m}
        \mat{P}_{nm}
        \left( 
        \ha^\dagger_n \hrho \ha_m
        -
        \frac{1}{2}
        \{\ha_m \ha_n^\dagger, \hrho\}
        \right)
        \\ \label{eq:L_U(1)_sym_bosons}
        &
        +
        \frac{1}{2}
        \sum_{n,m}
        [\mat{K}_{nm}\ha_n^\dagger \ha_m^\dagger + {\rm h.c.}
        ,
        \hrho]
        +
        i \sum_{n,m}
        \mat{C}_{nm}
        \left(
        \ha_m^\dagger \hrho \ha_n^\dagger
        -
        \frac{1}{2}
        \{
        \ha_n^\dagger \ha_m^\dagger, \hrho 
        \}
        \right)
        +
        i \sum_{n,m}
        \mat{C}^*_{nm}
        \left(
        \ha_n \hrho \ha_m
        -
        \frac{1}{2}
        \{
        \ha_m \ha_n, \hrho 
        \}
        \right)
        \equiv
        \mathcal{L} \hrho
    \end{align}
    \end{widetext}
    where $n$ and $m$ label independent bosonic modes $[\ha_m, \ha_n^\dagger] = \delta_{n,m}$. Further $\mat{H}_{nm}$ and $\mat{K}_{nm}$ are, respectively,  arbitrary Hermitian and complex symmetric matrices which describes the coherent particle-conserving and non-conserving process of the isolated system. The coefficients $l_{bm}$, $p^*_{bm}$ characterize the coupling to a set of dissipative baths indexed by $b$ and, consequently, the positive semi-definite Hermitan matrices $\mat{L}_{nm} = \sum_{b} l^*_{bn} l_{bm}$ and $\mat{P}_{nm} = \sum_b p^{*}_{bn} p_{bm}$ capture the phase-insensitive loss and pumping respectively. The complex matrix $\mat{C}_{nm} = \sum_{b}l^*_{bn} p^*_{bm}$ describes coherences between these processes. Its phase-sensitivity indicates that not all quadratures are equivalent. 

    \subsection{Multi-mode classical and quantum eigenvectors}
    Mimicking the procedure of the single-oscillator case, for each mode $m$ we define a set of classical and quantum creation and annihilation superoperators
    \begin{subequations}\label{eq:Bosons_Sup_Def}
    \begin{align}\label{eq:Def_aclm_aqm}
	\hacln{m} | \hrho \rangle \rangle
	&\equiv \frac{1}{\sqrt{2}}|\{\ha_m,\hrho\}\rangle \rangle,
	&&\haqn{m} |\hrho \rangle \rangle
	\equiv 
	\frac{1}{\sqrt{2}}|[\ha_m,\hrho] \rangle \rangle,
	\\ \label{eq:Def_acldagm_aqdagm}
	\hacln{m}^\dagger |\hrho\rangle \rangle 
	&\equiv 
	\frac{1}{\sqrt{2}}|\{\ha^\dagger_m,\hrho\}\rangle \rangle,
	&&\haqn{m}^\dagger |\hrho\rangle \rangle 
	\equiv 
	\frac{1}{\sqrt{2}}|[\ha_m^\dagger,\hrho]\rangle \rangle
\end{align}
\end{subequations}
whose only non-vanishing commutators are  
\begin{align}\label{eq:Boson_MM_Commutators}
    [
    \hacln{m}, \haqn{n}^\dagger
    ]
    =
    [
    \hacln{n}^\dagger,
    -
    \haqn{m}
    ]
    =
    \delta_{nm}
    \hone.
\end{align}
A tedious but straightforward calculation 
gives the third-quantization representation of the Lindbladian 
\begin{widetext}
\begin{align}\nonumber
    \hL
    &=
    \sum_{n,m}
    \left[
    (\mat{H}_{\rm eff})_{nm}
    \haqn{n}^\dagger \hacln{m}
    +
    (\mat{H}^\dagger_{\rm eff})_{nm}
    \haqn{m} \hacln{n}^\dagger
    \right]
    -i
    \sum_{n,m}
    \mat{N}_{nm}
    \haqn{n}^\dagger \haqn{m}
    \\ \label{eq:L_Boson_Third_Quantized_Many_Modes}
    &+
    \sum_{n,m}
    \left[
    (\mat{K}_{\rm eff})_{nm} \haqn{n}^\dagger \hacln{m}^\dagger 
    +
    (\mat{K}^\dagger_{\rm eff})_{nm} \haqn{m}\hacln{n}
    \right]
    -
    i\sum_{n,m}
    \left[
    \mat{Q}_{nm} \haqn{n}^\dagger \haqn{m}^\dagger 
    +
    \mat{Q}_{nm}^*\haqn{n}\haqn{m}
    \right].
\end{align}
\end{widetext}
Here we have defined the effective Hamiltonian and two-photon pumping matrices
\begin{align}
    \mat{H}_{\rm eff}
    \equiv
    \mat{H}-\frac{i}{2}\left(\mat{L}-\mat{G}\right),
    \hspace{0.5cm}
    \mat{K}_{\rm eff}
    \equiv
    \mat{K}-\frac{i}{2}\left(\mat{C}-\mat{C}^{T} \right)
\end{align}
in addition to the phase-insensitive and phase-sensitive noise matrices 
\begin{align}
    \mat{N}
    \equiv
    \mat{L}+\mat{G},
    \hspace{0.5cm}
    \mat{Q}
    \equiv
    \frac{1}{2}
    (
    \mat{C}+\mat{C}^{T}
    ).
\end{align}
The form of $\hL$ directly parallels the single-mode example. The quantum-classical terms encodes the dynamics, the quantum-quantum terms the noise, and there is no classical-classical term due to conservation of probability. 

The presence of the matrices $\mat{K}$ and $\mat{C}$ breaks a global $U(1)$ symmetry $\ha_m \to e^{i \theta }\ha_m, \ha_n^\dagger \to e^{-i \theta}\ha_n^\dagger$ and, just like in the closed-system case, implies that we have to work with a Nambu structure. For clarity, we first set $\mat{K} = \mat{C} = 0$ and avoid the additional technical details that comes with doubling the size of the matrix one has to diagonalize. At the end of this section we briefly discuss how the presence of these terms changes the diagonalziation procedure. 

\subsection{Correlation matrix}
To justify referring to $\mat{H}_{\rm eff}$ and $\mat{N}$ as the dynamical and noise matrices respecticely, let us compute the equations of motion for the symmetrically-ordered correlation matrix
\begin{align}
    \mat{S}_{mn}(t)
    \equiv
    \langle \{\ha_m(t), \ha_n^\dagger(t) \} \rangle
    =
    \langle \langle\hat{0}_{\rm q}|\hacln{m}\hacln{n}^\dagger |\hrho(t)\rangle \rangle
\end{align}
which is given by
\begin{align}\label{eq:S_EOM_bosons}
    i \partial_t \mat{S}(t)
    =
    \mat{H}_{\rm eff}
    \mat{S}(t)
    -
    \mat{S}(t)
    \mat{H}_{\rm eff}^\dagger
    +
    i \mat{N}.
\end{align}
As promised, $\mat{H}_{\rm eff}$ serves as the dynamical matrix which determines the fluctuation-free evolution of the system. It is also the dynamical matrix that would appear in the equation of motion for the average values of $\ha_m(t)$. The noise matrix $\mat{N}$ only affects the correlations and occupation of the modes. Using the Heisenberg-Langevin equations, it can be shown that $\mat{N}$ corresponds to the correlation matrix of a set of fluctuating noisy forces acting on the oscillators. Further, note that the dissipation cannot be larger than the noise $i(\mat{H}_{\rm eff}- \mat{H}^\dagger_{\rm eff}) \leq \mat{N}$, which ensures that the Heisenberg uncertainty principle is satisfied \cite{RMP_Clerk}. Finally,
we remind the reader that the identification of noise is always contingent on the ordering prescription used to define the corelation matrix.  For example, the equations of motion for the normal-ordered covariance matrix takes the same form as that of $\mat{S}$ but with modified noise terms:  $\mat{N} \to \mat{P}$.

Let us assume that all eigenvalues of $\mat{H}_{\rm eff}$ have negative imaginary part, ensuring the existence of a unique steady state.  The formal solution to Eq.~(\ref{eq:S_EOM_bosons}) then reads
\begin{align} \nonumber
    \mat{S}(t)
    &
    =
    e^{-i \mat{H}_{\rm eff}}
    \mat{S}(0)
    e^{i \mat{H}_{\rm eff}^\dagger t}
    +
    \int_{0}^t
    dt' 
    e^{-i \mat{H}_{\rm eff}(t-t')}
    \mat{N}
    e^{i \mat{H}^\dagger_{\rm eff}(t-t')}
    \\ \label{eq:S_Solution_Bosons}
    &
    =
    e^{-i \mat{H}_{\rm eff}t}
    \Bigl(
    \mat{S}(0)
    - 
    \mat{S}_{\rm ss}
    \Bigr)
     e^{i \mat{H}_{\rm eff}^\dagger t}
    +
    \mat{S}_{\rm ss}.
\end{align}
where the steady state correlation matrix can be written as 
\begin{subequations}
\begin{align}\label{eq:S_ss_time}
\mat{S}_{\rm ss}
&
=
\int_{0}^\infty
dt
\:
e^{-i \mat{H}_{\rm eff} t}
\mat{N}
e^{i \mat{H}_{\rm eff}^\dagger t}
\\\label{eq:S_ss_frequency}
&
=
\int_{-\infty}^\infty
\frac{d \omega}{2\pi}
\mat{G}^{R}(\omega)
\:
\mat{N}
\:
\mat{G}^A(\omega)
\\ \label{eq:S_ss_eigen}
&
=
-i
\sum_{\sigma, \tau}
|\Psi^R_\sigma \rangle
\frac{
\langle \Psi^L_\sigma |
\mat{N}
|\Psi^L_\tau \rangle
}
{
E_\sigma-E_\tau^*
}
\langle \Psi^R_\tau |.
\end{align}
\end{subequations}
Here,
\begin{align}
    \mat{G}^R(\omega)
    \equiv
    \frac{1}{\omega \mathbbm{1}-\mat{H}_{\rm eff}}
    =
    (\mat{G}^A(\omega))^\dagger
\end{align}
are the frequency-space retarded and advanced Green's function respectively whereas $\sigma $ and $\tau$ label the bi-orthonormal eigenvectors and eigenvalues of $\mat{H}_{\rm eff}$ which satisfy
\begin{subequations}
\begin{align}
    \mat{H}_{\rm eff} | \Psi^R_\sigma \rangle
    =
    E_{\sigma} | \Psi^R_\sigma\rangle,
    \: 
    &
    \:
    \langle \Psi^L_\sigma | \mat{H}_{\rm eff}
    =
    \langle \Psi^L_\sigma | E_{\sigma},
    \\
    \langle \Psi^L_{\sigma} | \Psi^R_\tau \rangle 
    &=
    \delta_{\sigma \tau}.
\end{align}
\end{subequations}
 It is worth stressing that while Eq.~(\ref{eq:S_ss_eigen}) assumes $\mat{H}_{\rm eff}$ can be diagonalized, Eqs.~(\ref{eq:S_ss_time})-(\ref{eq:S_ss_frequency}) are always valid. Thus, even when the dynamical matrix $\mat{H}_{\rm eff}$ is defective or equivalently at an exceptional point \cite{Bergholtz_Kunst_RMP, Ueda_NH_Review}, the steady state is well defined.

 \subsection{Gauging away noise, eigenvectors and eigenvalues}
Returning to the question of how to diagonalize $\hL$, recall that in the single-mode case 
linearity and the lack of a classical-classical term is ultimately what let us easily diagonalize the Lindbladian.  
As we now show, a similar approach works in the multi-mode case. We can again find a multi-mode similarity transformation that effectively gauges away the fluctuations, leaving a transformed Liouvillian that can be written in terms of generalized number operators.  The generalization of $e^{-(2\nth+1) \haq^\dagger \haq}$ to several oscillators is simply
\begin{align}
    \hV 
    \equiv
    \exp
    \left(
    -\sum_{n,m} (\mat{S}_{\rm ss})_{nm} \haqn{n}^\dagger \haqn{m}
    \right). 
\end{align}
The Liouvillian in the new frame defined by $\hV$ is, using the Baker-Cambell-Hausdorff identity and Eq.~(\ref{eq:Boson_MM_Commutators}),
\begin{widetext}
\begin{align}\nonumber
    \hV^{-1}\hL\hV
    &=
    \sum_{n,m}
    \left[
    (\mat{H}_{\rm eff})_{nm}
    \haqn{n}^\dagger \hacln{m}
    +
    (\mat{H}^\dagger_{\rm eff})_{nm}
    \haqn{m} \hacln{n}^\dagger
    \right]
    -
    \sum_{n,m}
    \left( 
    \mat{H}_{\rm eff} \mat{S}_{\rm ss}
    -
    \mat{S}_{\rm ss}\mat{H}_{\rm eff}^\dagger
    +
    i
    \mat{N}
    \right)_{nm}
    \haqn{n}^\dagger \haqn{m}
    \\
    &=
    \sum_{n,m}
    \left[
    (\mat{H}_{\rm eff})_{nm}
    \haqn{n}^\dagger \hacln{m}
    +
    (\mat{H}^\dagger_{\rm eff})_{nm}
    \haqn{m} \hacln{n}^\dagger
    \right]
\end{align}
\end{widetext}
where we have used $ \mat{H}_{\rm eff} \mat{S}_{\rm ss} - \mat{S}_{\rm ss}\mat{H}_{\rm eff}^\dagger + i\mat{N} = 0$ which follows from Eq.~(\ref{eq:S_EOM_bosons}). In this new basis, we are then left with two commuting third-quantized non-Hermitian Hamiltonians. Using the same eigenvectors and eigenvalues we used to diagonalize $\mat{H}_{\rm eff}$, these two terms can easily be brought to diagonal form (i.e.~written as a sum of commuting generalized number operators).  Moving back to the original frame, we finally arrive at
\begin{align}
    \hL 
    =
    \sum_{\sigma}
    \left[
    E_{\sigma}
    \haq^\dagger(\sigma) \hacl(\sigma)
    +
    E_{\sigma}^*
    \haq(\sigma)\hacl^\dagger(\sigma)
    \right]
\end{align}
where the relevant quasiparticle superoperators are
\begin{subequations}
\begin{align}\label{eq:aqdag_acl_alpha}
\hacl(\sigma)
&\equiv
\sum_{m}
    (\Psi^L_m(\sigma))^*
    \left(
    \hacln{m}
    +
    \sum_{n}
    (\mat{S}_{\rm ss})_{m n}
    \haqn{n}
    \right),
    \\ 
    \haq^\dagger(\sigma)
    &\equiv
    \sum_{m}
    \Psi^R_m(\sigma)
    \haqn{m}^\dagger,
    \\ 
    \hacl^\dagger(\sigma)
    &
=
\sum_{m}
    \Psi^L_m(\sigma)
    \left(
    \hacln{m}^\dagger
    -
    \sum_{n}
    (\mat{S}_{\rm ss})_{nm}
    \haqn{n}^\dagger
    \right),
    \\ \label{eq:aq_acldag_alpha}
     \haq(\sigma)
    &\equiv
    \sum_{m}
    (\Psi^R_m(\sigma))^*
    \haqn{m}
\end{align}
\end{subequations}
with $\Psi^R_m(\sigma) \equiv \langle m | \Psi^R_\sigma \rangle$ and $(\Psi^L_m(\sigma))^* \equiv \langle \Psi^L_\sigma |m \rangle$.  Although $\hacl^\dagger(\sigma)$ is not the conjugate of $\hacl(\sigma)$, this is irrelevant since, using the bi-orthonormality of the eigenvectors of $\mat{H}_{\rm eff}$, the only non-vanishing commutators amongst the set Eqs.~(\ref{eq:aqdag_acl_alpha})-(\ref{eq:aq_acldag_alpha}) are 
\begin{align}\label{eq:Comm_Relation}
    [\hacl(\sigma), \haq^\dagger(\sigma')]
    =
     [\hacl^\dagger(\sigma),-\haq(\sigma')]
    =
    \delta_{\sigma \sigma'}
    \hone.
\end{align}
We can therefore think of $\haq^\dagger(\sigma)\hacl(\sigma)$ and $-\haq(\sigma)\hacl^\dagger(\sigma)$ as non-Hermitian number operators. The right and left vacuums of both number operators is simply the steady state and identity operator respectively
\begin{subequations}
\begin{align}
    \hacl(\sigma)|\hat{\rho}_{\rm ss}\rangle \rangle = \hacl^{\dagger}(\sigma)|\hat{\rho}_{\rm ss}\rangle \rangle = 0
    \\
    \langle \langle \hat{0}_{\rm q} | \haq(\sigma) = \langle \langle \hat{0}_{\rm q} | \haq^\dagger (\sigma)= 0
\end{align}
\end{subequations}
with $\hat{0}_{\rm q} \equiv \hat{1}$. The steady state $| \hat{\rho}_{\rm ss}\rangle \rangle = \hV|\hat{0}_{\rm cl} \rangle \rangle$ is Gaussian with two-point correlation matrix determined by $\mat{S}_{\rm ss}$. Here $\hat{0}_{\rm cl} = 2^{M}e^{i \pi \hat{N}}$ is the total parity operator with $\hat{N} = \sum_m \ha^\dagger_m \ha_m$, which serves as the joint vacuum of the classical superoperators, and $M$ is the total number of modes. 

In complete analogy to a second-quantized quadratic Hamiltonian, we can now write the eigenvectors and eigenvalues of $\hL$ as
\begin{subequations}
\begin{align}
    | \hat{r}_{\vec{\mu}, \vec{\nu}} \rangle \rangle
     &=
     \prod_{\sigma}
    \frac{1}{\sqrt{\mu_\sigma! \nu_\sigma!}}
    \left[\haq^\dagger(\sigma)\right]^{\mu_\sigma}
    \left[-\haq(\sigma)\right]^{\nu_\sigma}
     |\hrho_{\rm ss}\rangle \rangle
     \\
     \langle \langle \hat{l}_{\vec{\mu}, \vec{\nu}} |
     &=
     \langle \langle \hat{0}_q |
     \prod_\sigma
     \frac{1}{\sqrt{\mu_\sigma! \nu_\sigma!}}
     \left[\boldsymbol{\hat{a}}_{\rm cl}(\sigma)\right]^{\mu_\sigma}
     [\boldsymbol{\hat{a}}^\dagger_{\rm cl}(\sigma)]^{\nu_\sigma}
\end{align}
\end{subequations}
with corresponding eigenvalues\\
\begin{align}
    E_{\vec{\mu}, \vec{\nu}}
    =
    \sum_{\sigma}
    \left(
    E_{\sigma} \mu_{\sigma}
    -
    E_{\sigma}^* \nu_{\sigma}
    \right).
\end{align}
Here, $\vec{\mu}, \vec{\nu}$ are lists of non-negative integers $\mu_\sigma, \nu_\sigma$ which characterize the number of ``particles" $\mu_{\sigma}$ or ``holes" $\nu_{\sigma}$ in mode $\sigma$. The right and left eigenvectors are normalized such that the bi-orthonormality condition $\langle \langle \hat{l}_{\vec{\mu'}, \vec{\nu'}} | \hat{r}_{\vec{\mu}, \vec{\nu}} \rangle \rangle = \delta_{\vec{\mu}, \vec{\mu}'} \delta_{\vec{\nu}, \vec{\nu}'}$ is satisfied. 

\subsection{Multi-mode Wigner function and path integral}
 The equivalence between third quantization, phase-space representations of the density matrix and Keldysh field theory presented in Secs.~\ref{sec:Harmonic_Oscillator}-\ref{sec:Keldysh_Single_Osc} can be extended to this multi-mode setting in an obvious manner. For any set of complex numbers $\alpha_m$ and  $\eta_m$ the multi-mode displaced parity and displacement operator 
 \begin{subequations}
 \begin{align}
     \vec{\hat{\alpha}}_{\rm cl}
     &\equiv
     2^M
     e^{
     \sum_m 
     \left(
     \alpha_m \ha_m^\dagger 
     -
     \alpha_m^* \ha_m
     \right)
     }
     e^{i \pi \hat{N}}
     \\
     \vec{\hat{\eta}}_{\rm q}
     &\equiv
     e^{
     \sum_m 
     \left(
     \alpha_m \ha_m^\dagger 
     -
     \alpha_m^* \ha_m
     \right)
     }
 \end{align}
 \end{subequations}
are the eigenvectors of the classical and quantum superoperators respectively. The  multi-mode Wigner function and characteristic function are given by 
\begin{subequations}
    \begin{align}\label{eq:W_alphacl_MM}
        W_{\hrho}(\vec{\alpha})
        &\equiv
        \Tr 
      \left(
     \vec{\hat{\alpha}}_{\rm cl}^\dagger
      \hrho
      \right),
      \\ \label{eq:W_alphaq_MM}
       \Lambda_{\hrho}(\vec{\eta})
         &\equiv
        \Tr 
      \left(
      \vec{\hat{\eta}}_{\rm cl}^\dagger
      \hrho
      \right).
    \end{align}
    \end{subequations}
from which we can obtain a simple physical interpretation of the similarity transformation used to eliminate the noise. Using the basis which diagonalizes the Hermitian covariance matrix $\mat{S}_{\rm ss}$, that is the set of orthogonal modes $\ha_{\lambda}$ satisfying $\langle \{\ha_{\lambda}, \ha^\dagger_{\lambda'}\}\rangle_{\rm ss} = (2\bar{n}_\lambda+1)\delta_{\lambda\lambda'}$, we have $\hV = \exp\left(-\sum_{\lambda} (2 \bar{n}_{\lambda}+1) \boldsymbol{\hat{a}}^\dagger_{{\rm q}, \lambda}\boldsymbol{\hat{a}}_{{\rm q}, \lambda}\right)$. Given the result of Sec.~\ref{sec:Single_Osc_Phase_Space} we conclude that $\hV$ is a convolution with a multi-mode Gaussian in phase spaces along a set of orthogonal axes determined by the modes $\ha_{\lambda}$  whose width depends on the average number of quanta in that mode in steady state. 

The equations of motion for $W_{\hrho}(\vec{\alpha})$ and $\Lambda_{\hrho}(\vec{\eta})$ can be easily read off from the third-quantized form of $\hL$ in Eq.~(\ref{eq:L_Boson_Third_Quantized_Many_Modes}) and the multi-mode generalization of the rules in Table~\ref{tb:Quantization_Rules} e.g. $\hacln{m} = \alpha_m = \partial_{\eta^*_m}$. We do not reproduce them here for the sake of compactness. We can also find the propagator $K(\vec{\eta},\vec{\alpha};t) \equiv \langle \langle \hetaq | e^{-i \hL t} | \halphacl \rangle \rangle $ for this differential equation using the Keldysh path integral. The multi-mode matrix Green's function for the finite-time contour of length $t$ is nothing but the generalization of Eq.~(\ref{eq:G_Matrix_Single_Osc})
\begin{widetext}
\begin{align}
    \mat{G}(t',t'')
    &\equiv
    \begin{pmatrix}
        \mat{G}^K(t',t'') & \mat{G}^R(t',t'') \\
        \mat{G}^A(t',t'') & 0
    \end{pmatrix}
    =
    \begin{pmatrix}
        -i\int_0^t dT
        \mat{G}^R(t',T)\: \mat{N} \: \mat{G}^A(T,t'') & -i \Theta(t'-t'') e^{-i \mat{H}_{\rm eff}(t'-t'')}
        \\
        i \Theta(t''-t') e^{i \mat{H}^\dagger_{\rm eff}(t''-t')} & 0
    \end{pmatrix}
\end{align}
\end{widetext}
and consequently the propagator is nothing but the multi-mode generalization of Eq.~(\ref{eq:K_Solution})
\begin{align}\nonumber
    &K(\vec{\eta},\vec{\alpha};t)
    =
    \exp
    \left(
    -i
    \mat{J}_{\rm int}^\dagger 
    \begin{pmatrix}
        \mat{G}^K(t,t) & \mat{G}^R(t,0) \\
        \mat{G}^A(0,t) & 0
    \end{pmatrix}
    \mat{J}_{\rm int}
    \right)
    \\
    &
    =
    \exp
    \left(
    - \mat{J}_{\rm int}^\dagger 
    \begin{pmatrix}
        \int_0^t dt'
        e^{-i \mat{H}_{\rm eff} t'}
        \:
        \mat{N}
        \:
        e^{i \mat{H}^\dagger_{\rm eff} t'}
        &
        e^{-i \mat{H}_{\rm eff} t} \\
        -e^{i\mat{H}^\dagger_{\rm eff} t} & 0
    \end{pmatrix}
    \mat{J}_{\rm int}
    \right)
\end{align}
where here we have defined the integrated source term $\mat{J}_{\rm int}^\dagger \equiv \int_{-\infty}^{\infty}dt' \mat{J}^\dagger(t') $. 

\subsection{Symmetry-breaking terms}\label{sec:Symm_Breaking_Bosons}
In the presence of $U(1)$ symmetry-breaking terms in the Lindbladian Eq.~(\ref{eq:L_Boson_Third_Quantized_Many_Modes}) due to non-zero $\mat{K}$ and $\mat{C}$, the strategy would be analogous to the one presented above. One would now have non-vanishing anomalous steady-state covariance matrix elements $\langle\{ \ha_m, \ha_n \}\rangle_{\rm ss}$, which would now enter the similarity transformation used to remove the noise terms. What remains is a $4 M \times 4 M$ non-Hermitian Hamiltonian. Just as in the closed-system case, one would have to perform an appropriate Bogoliubov transformation when defining third-quantized quasi-particle superoperators. The propagator would also take essentially the same form, the only difference being that the matrix Green's function would be of size $4M \times 4M$.

\section{Quadratic fermiomic Lindbladians}\label{sec:Quadratic_Fermions}
\subsection{Setup}
%
In this section we will diagonalize an arbitrary fermionic Lindbladian of the form 
\begin{widetext}
\begin{align} \nonumber
        i \partial_t \hrho 
        & =
        \sum_{n,m}
        \mat{H}_{nm} [\hc_n^\dagger \hc_m, \hrho]
        +
        \frac{1}{2}\sum_{n,m}
        [\mat{K}_{nm}\hc_n^\dagger \hc_m^\dagger+{\rm h.c.}, \hrho]
        +
        i
        \sum_{b}
        \mathcal{D}
        \left[
        \sum_m
        \left( 
        l_{bm} \hc_m+p^*_{bm}\hc_m^\dagger
        \right)
        \right]
        \\ \nonumber
        &=
        \sum_{n,m}
        \mat{H}_{nm} [\hc_n^\dagger \hc_m, \hrho]
        +
        i \sum_{n,m} 
        \mat{L}_{nm}
        \left( 
        \hc_m \hrho \hc_n^\dagger
        -
        \frac{1}{2}
        \{\hc_n^\dagger \hc_m, \hrho\}
        \right)
        +
        i
        \sum_{n,m}
        \mat{P}_{nm}
        \left( 
        \hc^\dagger_n \hrho \hc_m
        -
        \frac{1}{2}
        \{\hc_m \hc_n^\dagger, \hrho\}
        \right)
        \\ 
        &
        +
        \frac{1}{2}\sum_{n,m}
        [\mat{K}_{nm}\hc_n^\dagger \hc_m^\dagger+{\rm h.c.}, \hrho]
        +
         i \sum_{n,m}
        \mat{C}_{nm}
        \left(
        \hc_m^\dagger \hrho \hc_n^\dagger
        -
        \frac{1}{2}
        \{
        \hc_n^\dagger \hc_m^\dagger, \hrho 
        \}
        \right)
        +
        i \sum_{n,m}
        \mat{C}^*_{nm}
        \left(
        \hc_n \hrho \hc_m
        -
        \frac{1}{2}
        \{
        \hc_m \hc_n, \hrho 
        \}
        \right)
        \equiv
        \mathcal{L} \hrho.
\end{align}
\end{widetext}
where $m$ and $n$ label independent fermionic degrees of freedom $\{\hc_m, \hc_n^\dagger \} = \delta_{nm}$. The matrices $\mat{H}$, $\mat{L}$, $\mat{P}$ and $\mat{C}$ have the same form and interpretation as in the bosonic case studied in Sec.~\ref{sec:Quadratic_Bosons}. The only difference is that the complex matrix $\mat{K}$ now describes the pairing of fermions, and is thus necessarily anti-symmetric. 

\subsection{Creation and annihilation superoperators}
In analogy with the bosonic case Eq.~(\ref{eq:Bosons_Sup_Def}),  we would like to define a set of fermionic superoperators which satisfy canonical anti-commutation relations. The issue however is that superoperators acting on the left and the right of the density matrix commute with one another; this would seem to preclude introducing superoperators with canonical fermionic anti-commutation relations.   The solution is to introduce the parity operator $e^{i \pi \hat{N}}$ in the definition of these modes, where $\hat{N} = \sum_m \hc^\dagger_m \hc_m$ is the total number operator. Defining
\begin{subequations}\label{eq:Fermions_Sup_Def}
\begin{align}\label{eq:Def_c1_c2}
    \hconen{m} | \hrho \rangle \rangle 
    &\equiv 
    \frac{1}{\sqrt{2}}
    |
    \{ 
    \hc_m,
    \hrho e^{i \pi \hat{N}}
    \}
    \rangle \rangle
    ,
    \hctwon{m} | \hrho \rangle \rangle
    \equiv
    \frac{1}{\sqrt{2}}
    |
    [
    \hc_m
    ,
    \hrho e^{i \pi \hat{N}}
    ]
    \rangle \rangle
    \\\label{eq:Def_c1dag_c2dag}
    \hconen{m}^\dagger | \hrho \rangle \rangle,
    &
    \equiv 
    \frac{1}{\sqrt{2}}
    |
    [
    \hc_m^\dagger,
    \hrho e^{i \pi \hat{N}}
    ] 
    \rangle \rangle,
    \hctwon{m}^\dagger | \hrho \rangle \rangle
    \equiv
    \frac{1}{\sqrt{2}}
    |
    \{
    \hc^\dagger_m
    ,
    \hrho e^{i \pi \hat{N}}
    \}
    \rangle \rangle
\end{align}
\end{subequations}
one verifies that $\hconen{m}^\dagger$ and $\hctwon{m}^\dagger$ are the Hermitian conjugates of $\hconen{m}$ and $\hctwon{m}$ as per the definition of the inner product Eq.~(\ref{eq:Inner_Prod}). The only non-vanishing set of anti-commutators are
\begin{align}\label{eq:Anti_Comm}
    \{ 
    \hconen{m}, \hconen{n}^\dagger
    \} 
    =
    \{
    \hctwon{m}, \hctwon{n}^\dagger
    \}
    =
    \delta_{nm}\hone.
\end{align}
Thus, $\hconen{m}$ and $\hctwon{n}$ can be interpreted as genuine annihilation operators of fermionic modes.

From the definitions Eqs~(\ref{eq:Def_c1_c2})-(\ref{eq:Def_c1dag_c2dag}), the total parity and identity operator serve as the unique joint vacuum for one type of superoperator, while simultaneously serving as the completely-filled state of the other, i.e. 
\begin{subequations}\label{eq:Fermion_Vac_Filled_Def}
\begin{align}
    \hat{1}_1, \hat{0}_{2} \equiv e^{i \pi \hat{N}}
    &\rightarrow
    \hconen{m}^\dagger|\hat{1}_1, \hat{0}_{2}\rangle \rangle 
    =
    \hctwon{m}|\hat{1}_1, \hat{0}_{2}\rangle \rangle 
    =
    0,
    \\
    \hat{0}_1, \hat{1}_{2} \equiv \hat{1}
    &\rightarrow
    \hconen{m}|\hat{0}_1, \hat{1}_{2}\rangle \rangle 
    =
    \hctwon{m}^\dagger|\hat{0}_1, \hat{1}_{2}\rangle \rangle 
    =
    0.
\end{align}
\end{subequations}
To be explicit, $|\hat{1}_1, \hat{0}_{2}\rangle \rangle $ defines a state where all type-1 quasiparticle states are filled, and all type-2 quasiparticle states are empty; $|\hat{0}_1, \hat{1}_{2}\rangle \rangle$ is interpreted in a similar fashion.  
Note that these effective states are not normalized to unity but rather $\langle \langle \hat{1}_1, \hat{0}_2|\hat{1}_1, \hat{0}_{2}\rangle \rangle = \langle \langle \hat{0}_1, \hat{1}_2|\hat{0}_1, \hat{1}_{2}\rangle \rangle = 2^M$ where $M$ is the total number of modes.

We stress that these superoperators and states differ from Prosen \cite{Prosen_Fermions_2008}, who instead introduced Majorana superoperators instead of fermionic creation and annihilation superoperators. Our choice allows us to make the connection to the Keldysh formalism, as is apparent when writing the Lindbladian in this basis
\begin{widetext}
\begin{align} \nonumber
    \hL
    &=
    \sum_{n,m}
    \left[
    (\mat{H}_{\rm eff})_{nm}
    \hconen{n}^\dagger \hconen{m}
    -
    (\mat{H}^\dagger_{\rm eff})_{nm}
    \hctwon{m} \hctwon{n}^\dagger
    \right]
    -i
    \sum_{n,m}
    \mat{N}_{nm}
    \hconen{n}^\dagger \hctwon{m}
    \\ \label{eq:L_fermions}
    &+
    \sum_{n,m}
    \left[
    (\mat{K}_{\rm eff})_{nm}\hconen{n}^\dagger \hctwon{n}^\dagger 
    -
    (\mat{K}_{\rm eff})^*_{nm}\hctwon{n}\hconen{m}
    \right]
    -
    i
    \sum_{n,m}
    \left[
    \mat{Q}_{nm}\hconen{n}^\dagger \hconen{m}^\dagger
    -
    \mat{Q}^*_{nm}\hctwon{n}\hctwon{m}
    \right]
\end{align}
\end{widetext}
where for fermions the effective non-Hermitian Hamiltonians and pairing matrices are defined as 
\begin{align}
    \mat{H}_{\rm eff}
    \equiv
    \mat{H}
    -\frac{i}{2}
    \left(
    \mat{L}+\mat{P}
    \right),
    \hspace{0.5cm}
    \mat{K}_{\rm eff}
    \equiv
    \mat{K}
    -\frac{i}{2}
    \left(
    \mat{C}+\mat{C}^T
    \right)
\end{align}
with the noise matrices taking the form 
\begin{align}
    \mat{N} 
    \equiv
    \mat{L}-\mat{P},
    \hspace{0.5cm}
    \mat{Q}
    \equiv
    \frac{1}{2}
    \left(
    \mat{C}-\mat{C}^T
    \right).
\end{align}
Setting $\mat{K} = \mat{Q} = 0 $ for simplicity, we confirm that the nomenclature is justified by defining the anti-symmetrized covariance matrix
\begin{align}
    \mat{A}_{mn}(t)
    \equiv
    \langle
    [
    \hc_m(t), \hc_n^\dagger(t)
    ]
    \rangle
    =
    \langle \langle \hat{0}_1, \hat{1}_2  |
    \hconen{m}\hctwon{n}^\dagger
    |\hrho(t)\rangle \rangle
\end{align}
and computing its equation of motion
\begin{align}\label{eq:S_EOM_fermions}
    i \partial_t \mat{A}(t)
    =
    \mat{H}_{\rm eff}
    \mat{A}(t)
    -
    \mat{A}(t) \mat{H}^\dagger_{\rm eff}
    +
    i \mat{N}.
\end{align}
Note that here the dissipation must be larger than the fluctuations $i(\mat{H}_{\rm eff} - \mat{H}^\dagger_{\rm eff}) \geq \mat{N}$, whereas the opposite is true for bosons. This ensures that the Pauli exclusion principle is satisfied. The solution and physical content of Eq.~(\ref{eq:S_EOM_fermions}) was already discussed in Sec.~\ref{sec:Quadratic_Bosons}. Our only goal here was to stress that each term in the Lindbladian has a physical interpretation; it describes either dynamics or noise. 

\subsection{Gauging away the noise}
Just as in the bosonic case, there are terms in the third-quantized representation of $\hL$ such as $\hctwon{n}^\dagger \hconen{m}$ which are not allowed due to conservation of probability $\langle \langle \hat{1}| \hL = \langle \langle \hat{0}_1, \hat{1}_2|\hL = 0$. We can use this to move to a frame where there are no fluctuations, interpreting the separation of dissipation and noise in this linear system as the underlying symmetry which allows us find such a basis. Defining the anti-symmetrized steady-state covariance matrix  $(\mat{A}_{\rm ss})_{mn} \equiv \langle[\hc_m, \hc^\dagger_n] \rangle_{\rm ss}$ which satisfies $\mat{H}_{\rm eff} \mat{A}_{\rm ss}-\mat{A}_{\rm ss} \mat{H}_{\rm eff}^\dagger + i \mat{N} = 0$ and the superoperator
\begin{align}\label{eq:V_Fermions}
    \hV 
    \equiv
    \exp
    \left(
    -
    \sum_{n,m}
    (\mat{A}_{\rm ss})_{nm}
    \hconen{n}^\dagger \hctwon{m}
    \right)
\end{align}
we can use the Baker-Campbell-Hausdorff formlua and the anti-commutations relations Eq.~(\ref{eq:Anti_Comm}) to obtain
\begin{widetext}
\begin{align} \nonumber
    \hV^{-1} \hL \hV
    &=
    \sum_{n,m}
    \left[
    (\mat{H}_{\rm eff})_{nm}
    \hconen{n}^\dagger \hconen{m}
    -
    (\mat{H}^\dagger_{\rm eff})_{nm}
    \hctwon{m} \hctwon{n}^\dagger
    \right]
    -
    \sum_{n,m}
    \left(
    \mat{H}_{\rm eff} \mat{A}_{\rm ss}-\mat{A}_{\rm ss} \mat{H}_{\rm eff}^\dagger + i \mat{N}
    \right)_{nm}
    \hconen{n}^\dagger \hctwon{m}
    \\
    &=
    \sum_{n,m}
    \left[
    (\mat{H}_{\rm eff})_{nm}
    \hconen{n}^\dagger \hconen{m}
    -
    (\mat{H}^\dagger_{\rm eff})_{nm}
    \hctwon{m} \hctwon{n}^\dagger
    \right].
\end{align}
\end{widetext}
We can now use the eigenvectors and eigenvalues of $\mat{H}_{\rm eff}$ to bring this superoperator to diagonal form (i.e. written as a sum of commuting generalized number operators). Moving back to the original frame, we have
\begin{align}
     \hL
     =
 \sum_{\sigma}
    \left[
    E_{\sigma}
    \hcone^\dagger(\sigma)
    \hcone(\sigma)
    -
    E_{\sigma}^*
    \hctwo(\sigma)
    \hctwo^\dagger(\sigma)
    \right]
\end{align}
where we have defined
\begin{subequations}
\begin{align}
    \hcone(\sigma)
    &\equiv
    \sum_m
    (\Psi^{L}_m(\sigma))^*
    \left(
    \hconen{m}
    +
    \sum_{n}
    (\mat{A}_{\rm ss})_{mn} \hctwon{n}
    \right)
    \\
    \hcone^\dagger(\sigma)
    &\equiv
    \sum_{m}
    \Psi^R_n(\sigma)
    \hconen{m}^\dagger
    \\
    \hctwo^\dagger(\sigma)
    &\equiv
    \sum_m
    \Psi_m^L(\sigma)
    \left(
    \hctwon{m}
    -
    \sum_{n}
    (\mat{A}_{\rm ss})_{nm}
    \hconen{n}^\dagger
    \right)
    \\
    \hctwo^\dagger(\sigma)
    &\equiv
    \sum_m
    (\Psi_m^R(\sigma))^*
    \hctwon{m}.
\end{align}
\end{subequations}
 Although $\hcone^\dagger(\sigma)$ and $\hctwo^\dagger(\sigma)$ are not the conjugate of $\hcone(\sigma)$ and $\hctwo(\sigma)$, the canonical anti-commutation relations are still preserved, e.g., 
\begin{align}\label{eq:Anti_Comm_QuasiP}
    \{ 
    \hcone(\sigma), \hcone^\dagger(\sigma')
    \}
    =
    \{
    \hctwo(\sigma), \hctwo^\dagger(\sigma')
    \}
    =
    \delta_{\sigma \tau} \hone.
\end{align}
However, as a result of the non-unitary nature of $\hV$, the left and right vacuum and completely filled state of these new superoperators are not the same
\begin{subequations}
\begin{align}\label{eq:Fermion_Left_Vac}
    \langle \langle \hat{0}_1, \hat{1}_2 | \hcone^\dagger(\sigma) 
    &=
    \langle \langle \hat{0}_1, \hat{1}_2 |  \hctwo(\sigma)
    =
    0,
    \\ \label{eq:Fermion_Right_Vac}
    \hcone(\sigma) |\hrho_{\rm ss}\rangle \rangle
    &=
    \hctwo^\dagger(\sigma) |\hrho_{\rm ss} \rangle \rangle
    =
    0.
\end{align}
\end{subequations}
where we can write $|\hrho_{\rm ss}\rangle \rangle = 2^{-M} \hV|\hat{0}_1, \hat{1}_2\rangle \rangle$. In direct analogy with a standard Hermitian Hamiltonian, the eigenvectors of $\hL$ can then be written as
\begin{subequations}
\begin{align}
    | \hat{r}_{\vec{\mu}, \vec{\nu}} \rangle \rangle
    &=
    \prod_{\sigma}
    [\hcone^\dagger(\sigma)]^{\mu_\sigma}
    \left[\hctwo(\sigma)\right]^{\nu_\sigma}
     | \hrho_{\rm ss} \rangle \rangle
     \\
     \langle \langle \hat{l}_{\vec{\mu}, \vec{\nu}} |
     &=
     \langle \langle \hat{0}_1, \hat{1}_2 |
     \prod_\sigma
     [\hctwo^\dagger(\sigma)]^{\mu_\sigma}
     [\hcone(\sigma)]^{\nu_\sigma}
\end{align}
\end{subequations}
with corresponding eigenvalues
\begin{align}
E_{\vec{\mu}, \vec{\nu}}
=
\sum_\sigma
\left(
E_\sigma \mu_\sigma
-
E_{\sigma}^* \nu_\sigma
\right).
\end{align}
Here, $\vec{\mu}, \vec{\nu}$ are lists with entries $\mu_\sigma, \nu_\sigma = \{0,1\}$ and the eigenvectors satisfy $\langle \langle \hat{l}_{\vec{\mu'}, \vec{\nu'}} | \hat{r}_{\vec{\mu}, \vec{\nu}} \rangle \rangle = \delta_{\vec{\mu}, \vec{\mu}'} \delta_{\vec{\nu}, \vec{\nu}'}$.

The inclusion of $U(1)$ symmetry-breaking terms adds no formal degree of complexity; one simply has to diagonalize a larger matrix. The underlying physics, that is the separation of dynamics and fluctuations, still allows one to eliminate the noise using an appropriate similarity transformation analogous to Eq.~(\ref{eq:V_Fermions}).

\subsection{Third quantization and Keldysh field theory for fermions }
Just as we did in the bosonic case, we now develop a path integral representation of the dynamics starting from the third-quantized representation of the Lindbladian in Eq.~(\ref{eq:L_fermions}). For simplicity, let us assume that we have a single fermionic mode, such that the third-quantized Lindbladian takes the form
\begin{align}
    \hL 
    =&
    (\epsilon_0 - i \frac{\gamma}{2}) \hcone^\dagger \hcone
    -
    (\epsilon_0+i\frac{\gamma}{2})
    \hctwo \hctwo^\dagger
    \\
    &
    -i \gamma(1-2\bar{n}) \hcone^\dagger \hctwo
\end{align}
where $\gamma$ is the decay rate and $\bar{n} = \langle \hc^\dagger \hc \rangle_{\rm ss}$ the average number of particles in the steady state. Although here the Hamiltonian loss and pumping matrices have been simply replaced by numbers, $\mat{H} \to \epsilon_0$ $\mat{L} \to \gamma(1-\bar{n}), \mat{P} \to \gamma \bar{n} $, the generalization to several modes is straightforward.

The first step in formulating any fermionic path integral is to introduce the fermionic equivalent of coherent states, for which one must work with the usual Grassmann numbers \cite{Altland_Simons_2010}. We thus introduce four Grassmann numbers $\psi_1, \psi_2, \bar{\psi}_1, \bar{\psi}_2$ which anticommute with each other and all superoperators. Recalling that we have identified the identity operator as the vacuum of type-1 quasiparticlces and the occupied state of type-2 quasiparticles, see Eq.~(\ref{eq:Fermion_Vac_Filled_Def}), we then define
\begin{subequations} \label{eq:Fermion_Coh_Def}
\begin{align}
    |\psi\rangle \rangle
    &\equiv
    \frac{
    e^{
    -\psi_1 \boldsymbol{\hat{c}}_1^\dagger
    -\bar{\psi}_2 \boldsymbol{\hat{c}}_2
    }
    }
    {2}
    |
    \hat{0}_1, \hat{1}_2
    \rangle 
    \rangle,
    \\
    \langle \langle \psi |
    &\equiv
    \langle \langle \hat{0}_1, \hat{1}_2 |
    e^{- \boldsymbol{\hat{c}}_1 \bar{\psi}_1
    -
    \boldsymbol{\hat{c}}_2^\dagger \psi_2
    }
\end{align}
\end{subequations}
which satisfy
\begin{subequations}
\begin{align}\label{eq:Femrionic_Coh_State_1}
    \hcone |\psi \rangle \rangle
    &=
    \psi_1 |\psi\rangle \rangle
    ,
    \hspace{0.2 cm}
    \hctwo^\dagger |\psi \rangle \rangle
    =
    \bar{\psi}_2 |\psi\rangle \rangle
    \\ \label{eq:Femrionic_Coh_State_2}
    \langle \langle \psi | \hcone^\dagger  
    &=
    \langle \langle \psi | \bar{\psi}_1
    ,
    \hspace{0.2cm}
    \langle \langle \psi | \hctwo  
    =
    \langle \langle \psi | \psi_2
\end{align}
\end{subequations}
and
\begin{align}
    \langle \langle \psi' | \psi \rangle \rangle
    =
    e^{\bar{\psi}'_1 \psi_1 +\psi'_2 \bar{\psi}_2}.
\end{align}
One can then verify that these coherent states can be used as a resolution of the identity
\begin{align}\label{eq:Res_Identity_Fermions}
 \int 
 d \bar{\psi}_1 d \psi_1
 d \psi_2 d \bar{\psi}_2
 \,
 e^{-\bar{\psi}_1 \psi_1 -\psi_2 \bar{\psi}_2}
 | \psi \rangle \rangle
 \langle \langle \psi |
 =
 \hone
\end{align}
from which one can construst a Grassmann-valued ``wavefunction" which, in analogy with the Wigner function for bosons, we denote $W_{\hrho}(\psi) \equiv \langle \langle \psi | \hrho \rangle \rangle$. It follows that the matrix element 
\begin{align}
K(\psi', \psi; t)
\equiv
    \langle \langle \psi' |
    e^{-i \hL t}
    |\psi \rangle \rangle
\end{align}
contains all information about the dynamics.

A continuous-time version of this propagator can be build using essentially the same procedure presented in Sec.~\ref{sec:Keldysh_Single_Osc}. Leaving all details to Appendix \ref{app_sec:Propegator_Fermions}, we arrive at
\begin{align} \nonumber
    &K(\psi', \psi; t)
    =
    \\ \label{eq:PI_fermions}
    &\int 
    \mathcal{D}[\psi_1, \psi_2]
    e^{i S_0[\psi_1, \psi_2] 
    +
    \bar{\psi}'_1 \psi_1(t) - \bar{\psi}_{2}(t) \psi_2'
    +\bar{\psi}_1(0) \psi_1- \bar{\psi}_2 \psi_2(0)
    }
\end{align}
where $\mathcal{D}[\psi_1, \psi_2]$ is the standard functional measure, and the action is
  \begin{align}\nonumber
    &S_0[\psi_1, \psi_2]
    =
    \\ \label{eq:S_Finite_Contour_Fermions}
    &
    \int_{0}^{t}
    \int_{0}^{t}
    dt'
    dt''
    \begin{pmatrix}
    \bar{\psi}_1(t') & \bar{\psi}_{2}(t')
    \end{pmatrix}
    \mat{G}^{-1}(t',t'')
    \begin{pmatrix}
    \psi_{1}(t'') \\ \psi_{2}(t'')
    \end{pmatrix},
\end{align}
where the matrix Green's function now reads
\begin{align}
    \mat{G}^{-1}(t',t'')
    =
    \delta(t'-t'')
    \begin{pmatrix}
    i\partial_{t'} - (\epsilon_0-i\frac{\gamma}{2}) & i \gamma(1-2\bar{n}) \\
     0 & i\partial_{t'} - (\epsilon_0+i\frac{\gamma}{2})
    \end{pmatrix}.
\end{align}
The matrix Green's function is then
\begin{align}\label{eq:G_Matrix_Fermion}
    \mat{G}(t',t'')
    \equiv
    \begin{pmatrix}
    G^{R}(t',t'') & G^K(t',t'') \\
    0 & G^A(t',t'')
    \end{pmatrix}
\end{align}
where
\begin{subequations}
\begin{align}
    &G^R(t',t'') 
    = 
    -i \Theta(t'-t'')e^{-i(\epsilon_0-i\frac{\gamma}{2})(t'-t'')}
    \\
    &G^A(t',t'')
    =
    (G^R(t'',t'))^*
    \\\nonumber 
    &G^K(t',t'')
    =
    -i\kappa(1-2\bar{n})
    \int_0^t
    dT
    G^R(t',T)G^A(T,t'')
    \\
    &=
     -i(1-2\bar{n}) e^{-i \epsilon_0(t'-t'')}
     \left(e^{-\frac{\gamma}{2}|t'-t''|}-e^{-\frac{\gamma}{2}(t'+t'')}
     \right).
\end{align}
\end{subequations}
Mimicking the bosonic case, we can make the appropriate translation to the Grassmann variables to eliminate the source term and integrate over the quadratic action which gives unity (see Appendix~\ref{app_sec:Propegator_Fermions} for details). We then obtain
\begin{align}\label{eq:K_fermions}
    K(\psi', \psi; t)
    =
    e^{-(1-2\bar{n})(1-e^{-\gamma t}) \bar{\psi}_1' \psi_2' 
    +e^{(-i\epsilon_0-\frac{\kappa}{2}) t} \bar{\psi}'_1\psi_1
    -
    e^{(i\epsilon_0-\frac{\kappa}{2}) t} \bar{\psi}_2\psi'_2}
\end{align}
From Eq.~(\ref{eq:Fermion_Coh_Def}), by taking derivatives of the Grassmann variables we can use this propagator to obtain expectation values of observables, e.g. $\partial_{\bar{\psi}'_1} \langle \langle \psi' | e^{-i \hL t} | \hrho(0) \rangle \rangle |_{\psi' = 0} = \sqrt{2} \langle \hat{c}(t) \rangle $. Of course, we could have obtained this answer easily in this single-mode example; the application of this path integral is more useful when there are several fermionic modes. 
\section{Application:  dissipative non-linear cavity} \label{sec:Kerr}
We now provide an example of how our formalism can be extremely useful even in situations where there are true interactions, and the Lindbladian is not simply quadratic.  We consider a paradigmatic problem: a single bosonic mode with a Kerr (or Hubbard) type nonlinearity, and subject to thermal single-photon dissipation.  The master equation for this system is
\begin{align}\nonumber
		i \partial_t \hrho
		&=
		[\omega_0\ha^\dagger \ha + \frac{U}{2}\ha^\dagger \ha^\dagger \ha \ha, \hrho ] 
		\\ \label{eq:NonLinear_Oscillator_EOM}
		&+ i \kappa(\nth+1) \mathcal{D}[\ha]\hrho
		+
		i \kappa \nth \mathcal{D}[\ha^\dagger]\hrho
		\equiv 
		\mathcal{L} \hrho
\end{align}
with $U$ the strength of the Kerr non-linearity. Since $\hL$ is now evidently quartic in creation and annihilation superoperators, it would seem as though describing this problem exactly is not feasible. 

Surprisingly, well-known analytic expressions for this system exist, derived using a variety of different techniques (see e.g.~\cite{Mark_1973, 
Chaturvedi_1991_Kerr_1, Chaturvedi_1991_Kerr_2, Milburn_1986_Kerr_Liouville, Milburn_1986_Kerr_Liouville_Dissipation, Daniel_1989_Kerr_Coherence, Peinova_1990_Kerr_Exact}). While elegant, these solutions typically provide little direct intuition or insight into the underlying physics.  Further, using them for practical calculation is often complicated (e.g.~Ref.~\cite{Millburn_Kerr_Numerics_2008} for instance chose to study Eq.~(\ref{eq:NonLinear_Oscillator_EOM}) numerically as opposed to exploiting existing analytic expressions). A more recent work  Ref.~\cite{Alexander_Weak_Symm_PRL} demonstrated that the solvability of this interacting model is related to a special symmetry.  However, the results of that work still do not provide a simple prescription for calculating observable quantities, especially if one is interested in a phase-space picture of the dynamics.

In this section, we will use the equivalence between third-quantization and the Keldysh path integral to show that there exists simple physical picture of the interplay between the non-linearity and damping in this system.  We call this ``self-dephasing":  the fluctuating photon number in the oscillator causes it to dephase itself, i.e. degrade Fock space coherences. Further, using a novel gauge transformation within the path integral, this intuition will lead to simple expressions for physical observables of interest. Namely, we will demonstrate that we can obtain the time evolution of the Wigner function for an arbitrary initial state. We note that we will also implicitely make use of the results in Secs.~\ref{sec:Single_Osc_Phase_Space}-\ref{sec:Keldysh_Single_Osc} extensively. 

\subsection{Propagator via Keldysh path integral}
In our formulation of third quantization, the nonlinear term in Eq.~(\ref{eq:NonLinear_Oscillator_EOM}) takes the form
\begin{align}\label{eq:U_Third_Quantized}
    \frac{U}{2}
    \left(
    \haq^\dagger \hacl
    +
    \haq \hacl^\dagger
    \right)
    \left(
    \hacl^\dagger \hacl
    +
    \haq^\dagger \haq
    -2
    \right)
    |
    \hrho
    \rangle \rangle
\end{align}
which, using Table~\ref{tb:Quantization_Rules}, in phase space is equivalent to 
\begin{align} \nonumber
    &\frac{U}{2}
    \left(
    -\partial_{\alpha} \alpha
    +
    \partial_{\alpha^*} \alpha^*
    \right)
    \left(
    |\alpha|^2
    -
    \partial_{\alpha} \partial_{\alpha^*} - 2
    \right)
    W_{\hrho}(\alpha)
    \\ \label{eq:U_Phase_Space}
    &=
    \frac{U}{2}
    i \partial_{\phi}
    \left(
    |\alpha|^2
    -
    \partial_{\alpha} \partial_{\alpha^*} - 2
    \right)
    W_{\hrho}(\alpha)
\end{align}
where $\phi = \arg \alpha$. The first term in the bracket tells us that the frequency of the mode is amplitude-dependent, whereas the second stems from the quantum noise associated with inherent uncertainty associated with the amplitude $|\alpha|^2$. The last term is due the ordering convention: phase-space averages weighted by the Wigner function are equivalent to symmetrically-ordered operator averages \cite{gardiner_zoller_2010}. Our Hamiltonian on the other-hand is normal-ordered, hence this extra factor.

A path integral representation of the mixed phase-space kernel $K(\eta, \alpha; t)$ can immediately be read off from Eq.~(\ref{eq:L_Linear_Oscillator}) and Eq.~(\ref{eq:U_Phase_Space}), \begin{align}\label{eq:K_Non_Lin}
    &K(\eta, \alpha;t)
    =
    \langle 
    e^{\eta^* a_{\rm cl}(t) - \eta a^* _{\rm cl}(t)
    +a_{\rm q}^*(0) \alpha - a_{\rm q}(0)\alpha^*
    }
    \rangle_{S}
\end{align}
where the average is with respect to the Keldysh action
\begin{widetext}
\begin{align}  \label{eq:Keldysh_Action_U}
    S[a_{\rm cl}, a_{\rm q}]
    =
    \int_0^t dt'
    \begin{pmatrix}
    a_{\rm cl}^* & a_{\rm q}^*
    \end{pmatrix}
     \begin{pmatrix}
    0 & i\partial_{t'} - \left(\omega_0-U+ \frac{U}{2}n_{\rm cl}(t')+i\frac{\kappa}{2}\right) \\
    i\partial_{t'} - \left(\omega_0-U + \frac{U}{2} n_{\rm cl}(t')-i\frac{\kappa}{2}\right) & i \kappa(2\nth+1)
    \end{pmatrix}
    \begin{pmatrix}
        a_{\rm cl} \\ a_{\rm q}
    \end{pmatrix}
\end{align}.
\end{widetext}
Here we have at times dropped the temporal argument for notational compactness.  We have also defined $n_{\rm cl}(t') \equiv a_{{\rm cl}}^*(t') a_{{\rm cl}}(t')
    +
    a_{{\rm q}}^*(t') a_{{\rm q}}(t')$. The nomenclature is justified by noting that this term comes from the contribution $\hacl^\dagger \hacl + \haq^\dagger \haq = \{\hat{n}_{\rm sym} , \cdot\}$ in third quantization where $\hat{n}_{\rm sym} \equiv \frac{1}{2}\{\ha, \ha^\dagger\}$ is the symmetrized number operator. Despite the name, our formulation still takes into account all quantum fluctuations. Only by dropping the quantum noise term $a_{{\rm cl}}^*(t') a_{{\rm cl}}(t')
    +
    a_{{\rm q}}^*(t') a_{{\rm q}}(t') \to a_{{\rm cl}}^*(t') a_{{\rm cl}}(t')$ would the dynamics be rendered classical, since the phase-space equation of motion would be equivalent to that of a classical Fokker-Planck equation, see Eq.~(\ref{eq:U_Phase_Space}).
    
    It is worth emphasizing that Eq.~(\ref{eq:Keldysh_Action_U}) is the usual Keldysh action for this model written in a non-standard way in that we have singled-out $n_{\rm cl}(t')$. This way of writing the action is useful however since it makes it evident that the Kerr non-linearity serves as an amplitude-dependent frequency. By making the following gauge transformation within the path integral
    \begin{align}\label{eq:Gauge_Path_Integral}
        a_{\rm cl/ q}(t')
        \to 
        e^{
        i U t'
        -i\frac{U}{2}\int_0^{t'} dt'' n_{\rm cl}(t'')
        }
        a_{\rm cl/q}(t'),
    \end{align}
    we effectively move to a frame where the number-dependent phase accumulated by the fields has been taken into account. 

    This transformation has two effects: it both removes the non-linearity in the action $S \to S_0$ and introduces a multiplicative phase factor attached to $a_{\rm cl}(t)$
    \begin{align} \nonumber
    &K(\eta, \alpha;t)
    \\
    &=
    \langle 
    \exp
    \left( 
    \eta^* e^{i U t -i \frac{U}{2} \int_0^t dt' n_{\rm cl}(t') }a_{\rm cl}(t)
    + \alpha a_{\rm q}^*(0)
    -
    \rm{c.c.}
    \right)
    \rangle_{S_0}.
    \label{eq:KKernelAfterTrans}
\end{align}
Although a seemingly a more unwieldy expression than our starting one Eq.~(\ref{eq:K_Non_Lin}), we now show how this can not can not only be straightforwardly evaluated but has a simple physical interpretation.

\subsection{Self-dephasing through amplitude fluctuations}
To remove the phase factor $e^{-i \frac{U}{2}\int_0^t dt' n_{\rm cl}(t')}$ from the exponential in Eq.~(\ref{eq:KKernelAfterTrans}), we can use  the delta function representation of an angular variable $\delta(\theta -\theta') = \frac{1}{2\pi}\sum_{l= -\infty}^{\infty} e^{-i l (\theta-\theta')}$ and write
\begin{widetext}
\begin{subequations}
\begin{align}\label{eq:K_U_1}
    K(\eta, \alpha; t)
    &=
    \sum_{l = -\infty}^{\infty}
    \int_{-\pi}^\pi
    \frac{d \theta}{2\pi}
    e^{i l \theta+i U l t}
    \langle 
    \exp
    \left(
    - i \frac{U l}{2} \int_0^t
    dt'
    n_{\rm cl}(t')
    +
    \left[
    \eta^* e^{-i \theta} 
    a_{\rm cl}(t)
    +
    \alpha a_q^*(0) 
    -
    \rm{c.c.}
    \right]
    \right)
    \rangle_{S_0} 
    \\ \label{eq:K_U_2}
    &=
    \sum_{l = -\infty}^\infty
    \int_{-\pi}^\pi
    \frac{d \theta}{2\pi}
    e^{i l \theta+i U l t}
    \langle 
    \exp
    \left(
    \eta^* e^{-i \theta} 
    a_{\rm cl}(t)
    +
    \alpha a_q^*(0) 
    -
    \rm{c.c.}
    \right)
    \rangle_{S_{\frac{U l}{2}}}
\end{align}
\end{subequations}
In going from Eq.~(\ref{eq:K_U_1}) to Eq.(\ref{eq:K_U_2}), we have lumped the phase proportional to the integral of $n_{\rm cl}(t')$ in a new \textit{quadratic} action defined by
\begin{align}\label{eq:S_Ul/2}
    S_{\frac{U l}{2}}[a_{\rm cl}, a_{\rm q}]
    \equiv
    S_0[a_{\rm cl}, a_{\rm q}]
    -
    \frac{U l}{2}
    \int_0^t
    dt'
   (
    a_{\rm cl}^* a_{\rm cl}
    +
    a_{\rm q}^* a_{\rm q}
   ).
\end{align}
As shown in Appendix~ \ref{sec:Kerr}, the resulting Gaussian integral for each $l$ can then be performed and we arrive at 
\begin{align}\label{eq:K_U_3}
    K(\eta, \alpha;t)
    &=
    \sum_{l = -\infty}^\infty
    e^{-i (\omega_0 -U) l t + \frac{\kappa}{2} t}
    A_l(t)
    e^{- B_{{\rm{q}}, l}(t)|\eta|^2 }
    e^{- B_{{\rm{cl}}, l}(t)|\alpha|^2 }
    \int_{-\pi}^\pi
    \frac{d \theta}{2\pi}
    e^{i l \theta
    +
    (
\eta^* \alpha e^{-i \theta} - {\rm{c.c.}}) A_l(t)
    }
    \\ \label{eq:Prop_Kerr}
    &
    =
    \sum_{l=-\infty}^\infty
    e^{-i (\omega_0 -U) l t + \frac{\kappa}{2} t}
    A_l(t)
    e^{-B_{{\rm q},l}(t)|\eta|^2
    -B_{{\rm cl},l}(t)|\alpha|^2
    }
    e^{-il (\varphi-\phi)}
    \mathcal{J}_l\Bigl(2|\eta||\alpha|A_l(t)\Bigr)
\end{align}
\end{widetext}
where we have defined $\phi = \arg \alpha$, $\varphi = \arg \eta$, $\mathcal{J}_l$ as the $l$-th Bessel function of the first kind and
\begin{subequations}
\begin{align}
    A_{l}(t)
    &\equiv
    \frac{\Gamma_l}
    {
 \Gamma_l \cosh \frac{\Gamma_l}{2}t + \kappa \sinh \frac{\Gamma_l}{2}t
    },
    \\
    B_{{\rm q}, l}(t)
    &\equiv
     \frac{
     \left[ i U l + 2 \kappa(2\nth+1) \right]
    \sinh \frac{\Gamma_l}{2} t
     }
    {
 \Gamma_l \cosh \frac{\Gamma_l}{2}t + \kappa \sinh \frac{\Gamma_l}{2}t
    },
    \\
    B_{{\rm cl}, l}(t)
    &\equiv
     \frac{
     i U l  
    \sinh \frac{\Gamma_l}{2} t
     }
    {
 \Gamma_l \cosh \frac{\Gamma_l}{2}t + \kappa \sinh \frac{\Gamma_l}{2}t
    }
\end{align}
\end{subequations}
with $\Gamma_l  \equiv \sqrt{\kappa^2-U^2 l^2 + 2 i \kappa U l (2\nth+1)}$.

It is worth emphasizing that for $l \neq 0$ Eq.~(\ref{eq:S_Ul/2}) is not a valid Keldysh action, as it does not vanish when the quantum fields are set to zero \cite{Kamenev_Book_2011, Diehl_Keldysh_Review_2016}. We should not expect this to be the case however, since this action has been modified to include the effects of the fluctuating particle-dependent phase and is only physically meaningful when computing specific averages. For example, if we recall that the wavefunction in the quantum basis $\Lambda_{\hrho(t)}(\eta)$ is the generator of symmetrically-ordered correlation functions, we have 
\begin{align} \nonumber
    &\sqrt{2} 
    \langle \hat{a}(t) \rangle 
    = \partial_{\eta^*} \Lambda_{\hrho(t)}(\eta)|_{\eta = \eta^* = 0}
    \\ \nonumber
    &=
    \int \frac{d^2 \alpha}{2\pi}
    \langle 
    e^{i U t -i \frac{U}{2} \int_0^t dt' n_{\rm cl}(t')}
    e^{a_{\rm q}^*(0)\alpha - \rm{c.c.}} 
    a_{\rm cl}(t) 
    \rangle_{S_0}
    W_{\hrho(0)}(\alpha)
    \\
    &
    =e^{-i(\omega_0-U)t + \frac{\kappa}{2}t}
    \left(A_1(t)\right)^2
    \int \frac{d^2\alpha}{2\pi}
    \alpha
    e^{-B_{{\rm cl},1}(t) |\alpha|^2}
    W_{\hrho(0)}(\alpha).
\end{align}
We thus have a compact expression for the exact evolution of the average photon amplitude, starting with an {\it arbitrary initial state}. It can also be used to compute unequal-time correlation functions such as e.g. linear response coefficients \cite{Mark_Old_SPR_1984, Alexander_Weak_Symm_PRL}.

This result can be directly extended to more general averages.  The general prescription is that the relevant quadratic action that one needs to use has a form that {\it explicitly depends} on the chosen correlation function.  For example, for a correlation function like $\langle \{\ha^{\mu} (\ha^\dagger)^{\nu} \}_{\rm sym}  \rangle$ the corresponding action would be $S_{\frac{U l}{2}}$ with $l = \mu - \nu$.

\begin{figure}[t]
		\centering
		\includegraphics [width=0.475\textwidth]{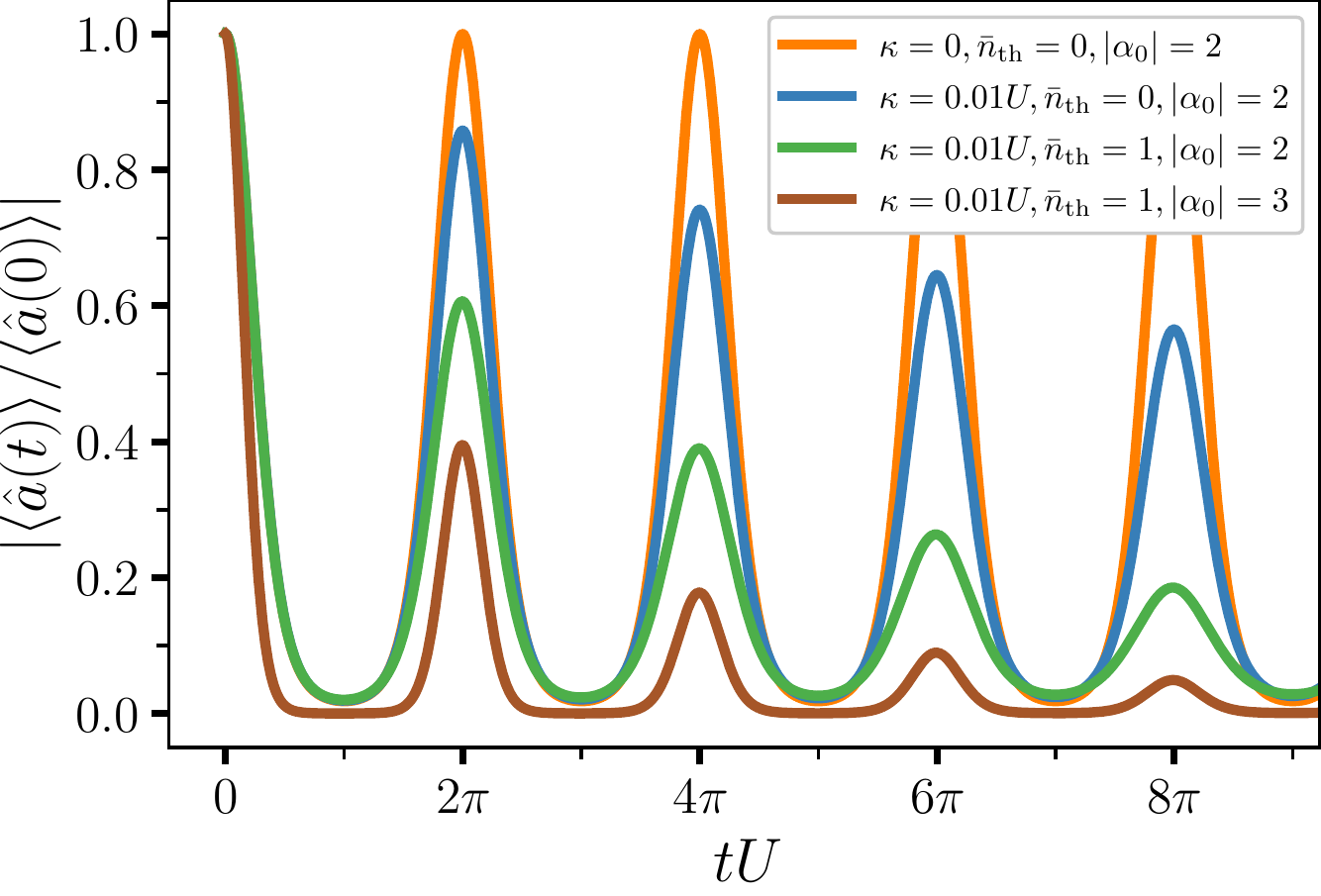}
		\caption{Scaled time-dependent average for a dissipative non-linear oscillator with an initial coherent state $\hrho(0) = |\frac{\alpha_0}{\sqrt{2}}\rangle\langle \frac{\alpha_0}{\sqrt{2}}|$. Increasing the thermal occupation $\nth$ or the magnitude $|\alpha_0|$ of the initial coherent state decreases the amplitude of the revivals at times $t U = 2\pi n$ for $n \in \mathbb{N}$. This is due to an increase in particle number fluctuations and thus an accompanying increase in dephasing. } 
		\label{fig:Avg}
	    \end{figure}

We see that our analytically exact results here directly validate the interpretation that the Kerr nonlineary generates a fluctuating frequency which causes dephasing.  In Fig.~\ref{fig:Avg} we plot a scaled version of
$\langle \hat{a}(t) \rangle$ for various parameters with $\hrho(0)$ a coherent state of varying magnitude.  We see that increasing either $\nth$ or the magnitude of the initial state tends to kill the revival of the average. This is in agreement with our intuition: increasing the noise or $|\alpha_0|$ causes an increase in particle-number fluctuations and consequently an increase in dephasing. Unlike the linear problem presented in Sec.~(\ref{sec:Harmonic_Oscillator}), the temperature changes both average values and in addition has an impact on the oscillation and decay rates.

\subsection{Time-dependent Wigner function}
Our third-quantization fueled approach for deriving exact expressions for this nonlinear dissipative cavity problem is useful even beyond calculating specific time-dependent averages or correlators.  As we now demonstrate, it allows us to analytically describe the time evolution of the full Wigner function describing the state.  From Eqs.~(\ref{eq:Prop_Cl_Cl_One})-(\ref{eq:Prop_Cl_Cl_Two}) we have access to this information by Fourier-transforming in the $\eta$ variable to obtain the propagator relating the Wigner function at time $t = 0$ to that at later times. We give this expression in Appendix~\ref{app_sec:Kerr_Prop}; it has the same functional form as Eq.~(\ref{eq:Prop_Kerr}). As an example, let us assume that the initial state is a coherent state $\hrho(0) = |\frac{\alpha_0}{\sqrt{2}} \rangle \langle \frac{\alpha_0}{\sqrt{2}} | \rightarrow W_{\hrho(0)}(\alpha) = 2 e^{-|\alpha-\alpha_0|^2}$. The time-evolved Wigner function can be computed using only Gaussian integrals. Leaving these details to Appendix~\ref{app_sec:Kerr_Prop}, we have 
\begin{align}\nonumber
    W_{\hrho(t)}(\alpha)
    =
    \sum_{l = - \infty}^\infty
    \Big[
    &P_l(t)
    e^{-Q_l(t)|\alpha|^2 - R_l(t)|\alpha_0|^2}
    e^{-i l (\phi-\Phi)}
    \\ \label{eq:Wigner_Initial_Coherent}
    &
    \times
    \mathcal{J}_l
    \Bigl(
    2 |\alpha| |\alpha_0| S_l(t) 
    \Bigr)
    \Big]
\end{align}
where $\Phi = \arg \alpha_0 $ and we have defined
\begin{align}
    P_l(t)
    &\equiv
    \frac{2 e^{-i(\omega_0 - U)l t + \frac{\kappa}{2} t} 
    \Gamma_l i^{-l}}
    {
    \Gamma_l \cosh \frac{\Gamma_l}{2} t
    +
    [i U l + \kappa(4\nth+1)] \sinh \frac{\Gamma_l}{2} t
    }
    \\
    Q_l(t)
    &\equiv
    \frac{
    \Gamma_l \cosh \frac{\Gamma_l}{2} t
    +
    [i U l + \kappa ] \sinh \frac{\Gamma_l}{2} t
    }
    {
    \Gamma_l \cosh \frac{\Gamma_l}{2} t
    +
    [i U l + \kappa(4\nth+1)] \sinh \frac{\Gamma_l}{2} t
    }
    \\
    R_l(t)
    &\equiv
    \frac{
    \Gamma_l \cosh \frac{\Gamma_l}{2} t
    -\kappa  \sinh \frac{\Gamma_l}{2} t
    }
    {
    \Gamma_l \cosh \frac{\Gamma_l}{2} t
    +
    [i U l + \kappa(4\nth+1)] \sinh \frac{\Gamma_l}{2} t
    }
    \\
    S_l(t)
    & \equiv
    \frac{i \Gamma_l}
     {
    \Gamma_l \cosh \frac{\Gamma_l}{2} t
    +
    [i U l + \kappa(4\nth+1)] \sinh \frac{\Gamma_l}{2} t
    }
\end{align}

To the best of our knowledge, this relatively compact expression was not previously known.  The closest analogue were results derived in Refs. \cite{Milburn_1986_Kerr_Liouville_Dissipation, Daniel_1989_Kerr_Coherence} that were more complicated, involving two infinite summations. Other equivalent exact solutions instead work with equally-intricate expressions involving the evolution of Fock states \cite{Mark_1973}.

In Fig.~\ref{fig:Wigner_Function} we plot this function for various choice of parameters. We stress that our choice of initial state was purely for illustrative purposes, and one can obtain the Wigner function of $\hrho(t)$ for an \textit{arbitrary} initial state with the propagator given in Eq.~(\ref{app_eq:Prop_Kerr_Wigner})
\\
\section{Conclusion}
By starting with a different set of superoperators for both bosons Eq.~(\ref{eq:Bosons_Sup_Def}) and fermions Eq.~(\ref{eq:Fermions_Sup_Def}), we have reinterpreted the standard approach to third quantization introduced by Prosen \cite{Prosen_Fermions_2008} and Prosen and Seligman \cite{Prosen_Seligman_Bosons_2010}. These new superoperators allows us to naturally relate third quantization, Keldysh field theory and the phase-space formulation of quantum mechanics. Further, our approach let us straightforwardly identify a symmetry of all quadratic Lindbladians, which can then be used to effectively gauge away fluctuations in these models. This leads to a simple and intuitive diagonalization procedure, and provides a quantum-inspired way of demonstrate that dynamics are not affected by noise in linear systems. Finally, we have shown that our formalism provides a simple picture of the dynamics in a paradigmatic dissipative non-linear cavity model, which can be used to provide straightforward exact expressions.

In future work, it would be worth exploring if third quantization can be extended to spins in a useful manner. Further, it is interesting to ask whether the set of non-Hermitian quasiparticles which were used to diagonalize the Lindbladian can be used to build a mean-field theory of interacting open quantum systems.

\section*{Acknowledgements}
    
    This work was supported by the Air Force Office of Scientific Research MURI program under Grant No. FA9550-19-1-0399, the Simons Foundation through a Simons Investigator award Grant No. 669487 and by the Air Force Office of Scientific Research under Award No. FA9550- 19-1-0362.

 \appendix

  \section{Phase-space representations of right and left eigenvectors of a single thermally-damped harmonic oscillator}\label{app_section:Phase_space_right_left_eigenmodes}
  We can easily derive the phase-space representations of the right and left eigenvectors using Table~\ref{tb:Quantization_Rules}. Starting with the right eigenvectors
  \begin{widetext}
  \begin{align}\nonumber
     &W_{\hat{r}_{\mu, \nu}}(\alpha)
     \equiv
     \langle \langle \halphacl| \hat{r}_{\mu, \nu} \rangle \rangle 
     =
     \frac{1}{\sqrt{\mu!\nu!}}
     \langle \langle \halphacl|
     (\haq^\dagger)^\mu
     (-\haq)^\nu
     |\hrho_{\rm ss} \rangle \rangle
     =
     \frac{(-1)^{\mu+\nu}}{\sqrt{\mu!\nu!}}
    \partial_{\alpha}^\mu 
     \partial_{\alpha^*}^\nu
     W_{\hrho_{\rm ss}}(\alpha)
     =
     \frac{2(-1)^{\mu+\nu}}{\sqrt{\mu!\nu!}}
    \partial_{\alpha}^\mu 
     \partial_{\alpha^*}^\nu
     \left(
     \frac{e^{-\frac{|\alpha|^2}{2\nth+1}}}{2\nth+1}
     \right)
     \\ \nonumber
     &=
     \frac{2 (-1)^\mu}{\sqrt{\mu! \nu!}}
     \sum_{j=0}^\mu
     \binom{\mu}{j}
     \partial_{\alpha}^j
     \left(
     \frac{\alpha}{2\nth+1}
     \right)^\nu
     \partial_{\alpha}^{\mu-j}
     \left(
      \frac{e^{-\frac{|\alpha|^2}{2\nth+1}}}{2\nth+1}
     \right)
     =
     \frac{2e^{-\frac{|\alpha|^2}{2\nth+1}}}{\sqrt{\mu! \nu!}}
     \sum_{j=0}^{\min(\mu, \nu)}
     \frac{\mu! \nu!}{j!(\mu-j)!(\nu-j)!}
     \frac{
     (-1)^j(\alpha^*)^{\mu-j}\alpha^{\nu-j}
     }
     {(2\nth+1)^{\mu+\nu-j+1}}
     \\
     &
     =
    \sqrt{\frac{\min(\mu,\nu)!}{\max(\mu, \nu)!}}
     \frac{2(-1)^{\min(\mu, \nu)}e^{-\frac{|\alpha|^2}{2\nth+1}}}{(2\nth+1)^{\max(\mu,\nu)+1}}
     e^{-i \phi(\mu-\nu)}
     |\alpha|^{|\mu-\nu|}
     L^{|\mu-\nu|}_{\min(\mu, \nu)}
     \left(\frac{|\alpha|^2}{2\nth+1} \right)
  \end{align}
  where $\phi = \arg \alpha$ and $L^{|\mu-\nu|}_{\min(\mu,\nu)}$ is an associated Laguerre polynomial. Using $[\hacl, \hacl^\dagger] = 0$, the left eigenvectors can be written as
  \begin{align}\nonumber
      &W^*_{\hat{l}_{\mu,\nu}}(\alpha)
      \equiv
      \frac{1}{\sqrt{\mu!\nu!}}
      \langle \langle \hat{0}_{\rm q}|
      \hacl^\mu (\hacl^\dagger)^\nu
      e^{(2\nth+1)\haq^\dagger \haq}
      |\halphacl\rangle \rangle
    =
      \frac{1}{\sqrt{\mu!\nu!}}
      \partial^{\mu}_{\rm \eta^*}
      \partial^\nu_{\rm \eta}
      \langle \langle \hat{0}_{\rm q}|
      e^{\eta^* \hacl + \eta \hacl^\dagger}
      e^{(2\nth+1)\haq^\dagger \haq}
      |\halphacl\rangle \rangle
      \Bigl\vert_{\eta = \eta^* = 0}.
  \end{align}
  With the Baker-Campbell-Hausdorff identity, we can move $e^{\eta^* \hacl + \eta \hacl^\dagger}$ past the factor of $e^{(2\nth+1)\haq^\dagger \haq}$ and use the defining property of the classical and quantum eigenvectors to obtain
    \begin{align}\nonumber
      &W^*_{\hat{l}_{\mu,\nu}}(\alpha)
      =
      \frac{1}{\sqrt{\mu!\nu!}}
      \partial^{\mu}_{\rm \eta^*}
      \partial^\nu_{\rm \eta}
      e^{-(2\nth+1)|\eta|^2+\eta^* \alpha + \eta \alpha^*}
      \Bigl\vert_{\eta = \eta^* = 0}
      =
      \frac{1}{\sqrt{\mu!\nu!}}
      \sum_{j=0}^{\min(\mu,\nu)}
      \frac{\mu! \nu!(-(2\nth+1))^j}{j!(\mu-j)!(\nu-j)!}
      \alpha^{\mu-j}(\alpha^*)^{\nu-j}
      \\
      &
    =\sqrt{\frac{\min(\mu,\nu)!}{\max(\mu, \nu)!}}
     \frac{(-1)^{\min(\mu,\nu)}}{(2\nth+1)^{-\min(\mu,\nu)}}
     e^{i \phi(\mu-\nu)}
     |\alpha|^{|\mu-\nu|}
     L^{|\mu-\nu|}_{\min(\mu, \nu)}
     \left(\frac{|\alpha|^2}{2\nth+1} \right).
  \end{align}
  \end{widetext}


  \section{Continuous-time representation of the bosonic path integral - Details}\label{app_sec:Propegator_Harmonic_Oscillator}
  In this Appendix, we provide the explicit computations which led us to the continuous-time representation of the propagator $K(\eta, \alpha ; t) = \langle \langle \hetaq | e^{-i \hL t} |\halphacl \rangle \rangle $ in Eq.~(\ref{eq:Continuum_Limit}). We start with the usual approach \cite{Altland_Simons_2010} and divide the propagator in $N-1$ parts $e^{-i \hL t} = e^{-i \hL \Delta t}\cdots e^{-i \hL \Delta t}$ with $\Delta t = t/(N-1)$. We then insert the resolution of the identity Eq.~(\ref{eq:Resolution_Identity_cl_q_j}) between each element $e^{- i \hL \Delta t}$ and $\langle \langle \hetaq|$ and $|\halphacl \rangle \rangle$. Expanding to linear order in $\Delta t$ and re-exponentiating gives 
\begin{align} \nonumber
   &\langle \langle \hat{a}_{{\rm q}, j} |
   e^{-i \hL \Delta t}
   | \hat{a}_{{\rm cl}, j-1} \rangle \rangle
   =
   \\\label{eq:Overlap_Path_Integral}
   &e^{\aqj^* a_{{\rm cl}, j-1} - \aqj a_{{\rm cl}, j-1}^*
   -i \mathcal{L}(\aqj, a_{{\rm cl}, j-1}) \Delta t}
   + \mathcal{O}([\Delta t]^2)
\end{align}
where $\mathcal{L}(\aqj, a_{{\rm cl}, j-1}) \equiv \langle \langle \hat{a}_{{\rm q}, j} | \hL | \hat{a}_{{\rm cl}, j-1} \rangle \rangle$. Using the defining property of the classical and quantum eigenvectors Eqs.~(\ref{eq:acl_eigen_property})-(\ref{eq:aq_eigen_property}), $\mathcal{L}(\aqj, a_{{\rm cl}, j-1})$ is obtained by simply replacing the classical and quantum superoperators that appear in $\hL$ by $a_{{\rm cl}, j-1}$ and $\aqj$ respectively. We thus have
  \begin{align} \nonumber
      K(\eta, \alpha; t)
      &=
      \int
      \prod_{j=1}^N
      \frac{d \aclj \aqj}{\pi^2}
      e^{
      i
      \tilde{\mathbbm{a}}^\dagger 
      \tilde{\mat{G}}^{-1}
      \tilde{\mathbbm{a}}
      +
      \tilde{\mathbbm{a}}^\dagger \tilde{\mat{J}}
      -
      \tilde{\mat{J}}^\dagger \tilde{\mathbbm{a}}
      }
      \\ \label{app:Discrete_Path_Integral}
      &+
      \mathcal{O}(\Delta t)
  \end{align}
  where the tilde indicates that we are working with discrete-time objects. We have defined the two $2N$ column vectors
  \begin{align}
      \tilde{\mathbbm{a}}
      =
      \begin{pmatrix}
        a_{\rm cl, 1}
        \\
        \vdots
        \\
        a_{{\rm cl}, N}
        \\
        a_{\rm q, 1}
        \\
        \vdots
        \\
        a_{{\rm q}, N}
      \end{pmatrix}
  \end{align}
  and 
  \begin{align}
      \tilde{\mat{J}}
      =
      \begin{pmatrix}
        0
        \\
        \vdots
        \\
        -\eta
        \\
        \alpha
        \\
        \vdots
        \\
        0
      \end{pmatrix},
  \end{align}
 the latter only having 2 non-zero entries at position $N$ and $N+1$. The discrete inverse matrix Green's function is given by
 \begin{align}
     \tilde{\mat{G}}^{-1}
     \equiv
     \begin{pmatrix}
      0 & [\tilde{\mat{G}}^{-1}]^A \\
      [\tilde{\mat{G}}^{-1}]^R & [\tilde{\mat{G}}^{-1}]^K
     \end{pmatrix}
 \end{align}
 where $[\tilde{\mat{G}}^{-1}]^R$ and $[\tilde{\mat{G}}^{-1}]^A$ are Hermitian conjugate $N\times N$ lower and upper-triangular matrices respectively, whose only non-vanishing elements are on the main off-diagonal 
 \begin{align}\label{app:G_R_inv}
     [\tilde{\mat{G}}^{-1}]^R
     &=
    \begin{pmatrix}
    i & 0  &  0        & 0    &  0 \\
      -i h& i      & 0     &  0   & 0 \\
      0&    -ih    & \ddots   &   0     & 0   \\
      0 &     0    &    \ddots       &     \ddots      &  0 \\
     0 &     0    &   0       &      -i h    & i
  \end{pmatrix}
    \\ \label{eq:app:G_A_inv}
    [\tilde{\mat{G}}^{-1}]^A
    &= 
    \left([\tilde{\mat{G}}^{-1}]^R\right)^\dagger
  \end{align}
  where $h \equiv 1-i(\omega_0-i\kappa/2)\Delta t$. The Keldysh component is diagonal and almost proportional to the identity, except for the first entry where it is zero
  \begin{align}\label{app:G_K_inv}
      \left(\left[\tilde{\mat{G}}^{-1}\right]^K\right)_{jj'}
      =
      i \kappa (2\nth+1) \Delta t
      (
      \delta_{jj'}
      -
      \delta_{j1} \delta_{j'1}
      ).
  \end{align}
  We can readily obtain the matrix Green's function since its inverse has a vanishing upper-left block
  \begin{align}
      \tilde{\mat{G}}
      =
      \begin{pmatrix}
      \tilde{\mat{G}}^K & \tilde{\mat{G}}^R \\
      \tilde{\mat{G}}^A & 0
      \end{pmatrix}
  \end{align} 
  where 
  \begin{align}\label{app:GR_Discreet}
      \left(\tilde{\mat{G}}^{R}\right)_{jj'}
      &=
      \begin{cases}
        0, & j' > j \\
        -i h^{j-j'} & j \geq j'
      \end{cases}
      \\ \label{app:GA_Discreet}
      \left(\tilde{\mat{G}}^{
      A}\right)_{jj'}
      &=
      \begin{cases}
        i (h^*)^{j'-j}, & j' \geq j \\
        0, & j > j'
      \end{cases}
  \end{align}
 \begin{align}\nonumber
      &\left(\tilde{\mat{G}}^K\right)_{jj'}
      =
      -i \kappa(2\nth+1)\sum_{k =2}^N
      \Delta t
      \left(\tilde{\mat{G}}^{R}\right)_{jk}
      \left(\tilde{\mat{G}}^{A}\right)_{kj'}
      \\ \nonumber
      =
      &\frac{-i (2\nth+1)(1-\delta_{j1}\delta_{j'1})}{1-\frac{\Delta t}{\kappa}(\omega_0^2+(\frac{\kappa}{2})^2)}
      h^{j}(h^{*})^{j'}
      \\ \label{app:GK_Discreet}
      &
      \times\left(
       h^{-\min(j,j')}(h^{*})^{-\min(j,j')}
      -
      1
      \right).
 \end{align}
The strange delta function terms $(1-\delta_{j1}\delta_{j'1})$ simply ensures that the Keldysh component vanishes when either argument is $1$, i.e. at the lower boundary. This is automatically satisfied in the continuum representation where $G^K(0,t'') = G^K(t',0) = 0$.

In Eq.~(\ref{app:Discrete_Path_Integral}), we can then make the displacement 
 \begin{align}
     \tilde{\mathbbm{a}}
     \to
     \tilde{\mathbbm{a}}
    + i
     \tilde{\mat{G}}
     \tilde{\mat{J}}
 \end{align}
and perform the integral over the various $\aclj$ and $\aqj$ using the well-known result for Gaussian integrals $\int \prod_{j=1}^N \frac{d \aclj \aqj}{\pi^2} e^{
      i
      \tilde{\mathbbm{a}}^\dagger 
      \tilde{\mat{G}}^{-1}
      \tilde{\mathbbm{a}}
      } = \det(i \tilde{\mat{G}})$ \cite{Altland_Simons_2010}. Since the lower-right block of $\tilde{\mat{G}}$ vanishes and because the retarded and advanced Green's functions are lower and upper triangular respectively, we have $\det(i \tilde{\mat{G}}) = \det(\tilde{\mat{G}}^R)\det(\tilde{\mat{G}}^A) = 1$. We are then left with
  \begin{align} \nonumber
      &K(\eta, \alpha; t)
      =
      \exp
      \left(
      -i
      \tilde{\mat{J}}^\dagger 
      \tilde{\mat{G}}
      \tilde{\mat{J}}
      \right)
      +
      \mathcal{O}(\Delta t)
      \\ \nonumber 
      &
      =
      \exp
      \left(
      -i\left[
      |\eta|^2 (\tilde{\mat{G}}^K)_{NN}
      - \eta^* \alpha (\tilde{\mat{G}}^R)_{N1}
      - \eta \alpha^* (\tilde{\mat{G}}^A)_{1N}
      \right]
      \right)
      \\
      & +
      \mathcal{O}(\Delta t)
  \end{align}
 Using Eqs.~(\ref{app:GR_Discreet})-(\ref{app:GK_Discreet}) and taking the $N \to \infty$ limit while keeping the length of the contour fixed $\Delta t (N-1) = t$, we recover Eq.~(\ref{eq:K_Solution}). As we have emphasized in the main text, Eqs.~(\ref{app:GR_Discreet})-(\ref{app:GA_Discreet}) indicate that the correct normalization of the Heaviside step function at time $t = 0$ as they appear in the retarded and advanced Green's function is $\Theta(0) =1$. This stems from the ordering of the classical and quantum variables within the path integral; in the discrete representation one always takes the classical field one time step before the quantum fields. 
  \section{Continuous-time representation of the fermionic path integral - Details}\label{app_sec:Propegator_Fermions}
  In this appendix, we demonstrate how to achieve the continuous-time representation of the path integral of Eq.~(\ref{eq:PI_fermions}). The procedure is nearly identical to the bosonic case presented in Appendix~\ref{app_sec:Propegator_Harmonic_Oscillator} and we will thus be rather brief in our derivation. 

  After writing the propagator $e^{-i \hL t} = e^{-i \Delta t \hL} \cdots e^{-i\Delta t \hL}$ as a product of $N-1$ terms $\Delta t = t/(N-1)$, using the resolution of the identity formed by the coherent states Eq.~(\ref{eq:Res_Identity_Fermions}), expanding to linear order in $\Delta t$, using the defining property of coherent states Eqs.~(\ref{eq:Femrionic_Coh_State_1})-(\ref{eq:Femrionic_Coh_State_2}) and exponentiating we arrive at a Gaussian integral whose discrete action is nearly identical to the bosonic version Eq.~(\ref{app:Discrete_Path_Integral}) with Grasmmann variables taking the place of complex variables. If we define the $4 N$ Grassmann variables to be integrated over with two indices $\bar{\psi}_{1/2, j}, \bar{\psi}_{1/2, j} $, then the main difference between the bosonic expression for the propagator is that the source term is
  \begin{align}
      e^{\bar\psi{_1}' \psi_{1,N} 
      -\bar{\psi}_{2,N} \psi_2'
      +
      \bar{\psi'_{1,1}} \psi_1
      -\bar{\psi}_{2,1}\psi'_{2,1}}
  \end{align}
  and the $2N\times 2N$ inverse Green's function reads
   \begin{align}
     \tilde{\mat{G}}^{-1}
     \equiv
     \begin{pmatrix}
      [\tilde{\mat{G}}^{-1}]^R & [\tilde{\mat{G}}^{-1}]^K \\
      0 & [\tilde{\mat{G}}^{-1}]^A
     \end{pmatrix}
 \end{align}
where $[\tilde{\mat{G}}^{-1}]^R, [\tilde{\mat{G}}^{-1}]^A$ and $[\tilde{\mat{G}}^{-1}]^K$ are exactly the same as in Eq.~(\ref{app:G_R_inv})-(\ref{app:G_K_inv}) with $\kappa \to \gamma$ and $1+2\nth \to 1-2\bar{n}$. The matrix Green's function is then 
\begin{align}
    \tilde{\mat{G}}
    =
    \begin{pmatrix}
        \tilde{\mat{G}}^R & \tilde{\mat{G}}^K \\
        0 & \tilde{\mat{G}}^A
    \end{pmatrix}
\end{align}
where once again the retarded, advanced and Keldysh Green's function are the same as the bosonic version in Eqs.~(\ref{app:GR_Discreet})-(\ref{app:GK_Discreet}) with the aforementioned substitutions to the decay frate and occupation factor. The continuous-time version of the path integral is then given by Eq.~(\ref{eq:PI_fermions}) in the main text.

 In the discrete representation, after eliminating the source term, one can integrate over the remaining Gaussian Grassmann integral using a well-known identity \cite{Altland_Simons_2010}, after which one concludes that this term is the identity. The final result Eq.~(\ref{eq:K_fermions}) then follows.  

  \section{Path integral approach to the non-linear Kerr oscillator}\label{app_sec:Kerr_Prop}
  In this appendix, we provide all the technical steps to obtain the results in Sec.~(\ref{sec:Kerr}). Let us first  
  show that Eq.~(\ref{eq:K_U_2}) and Eq.~(\ref{eq:K_U_3}) are equivalent by computing
  \begin{align}\label{app:To_Average_Over}
      \langle 
    \exp
    \left(
    \eta^* e^{-i \theta} 
    a_{\rm cl}(t')
    +
    \alpha a_q^*(0) 
    -
    \rm{c.c.}
    \right)
    \rangle_{S_{\frac{U l}{2}}}
  \end{align}
  where $S_{\frac{U l}{2}}$ is the non-standard quadratic action defined in Eq.~(\ref{eq:S_Ul/2}). It follows that to compute the functional integral, we must first find the matrix Green's function $\mat{G}_{\frac{Ul}{2}}(t',t'')$ corresponding to the action $S_{\frac{U l}{2}}$ which by definition satisfies 
  \begin{align}\label{app:G_Ul/2}
  \left(
    i \partial_{t'} 
    \mat{X}
    -
    \mat{D}_{\frac{Ul}{2}}
    \right)
    \mat{G}_{\frac{U l}{2}}(t', t'')
=
\delta(t'-t'')
\end{align}
where $\mat{X}$ is the first Pauli matrix
\begin{align}
    \mat{X}
    \equiv
    \begin{pmatrix}
        0 & 1 \\
        1 & 0
    \end{pmatrix}
\end{align}
and we have defined
\begin{align}
    \mat{D}_{\frac{U l}{2}}
    \equiv
    \begin{pmatrix}
       \frac{U l}{2} & \omega_0 +i \frac{\kappa}{2}\\ 
        \omega_0 -i \frac{\kappa}{2} & \frac{U l}{2} -i \kappa(2\nth+1)
    \end{pmatrix}.
\end{align}
The most general solution to this differential equation is 
\begin{align}\nonumber
    \mat{G}_{\frac{U l}{2}}(t', t'')
    =
    &-i \Theta(t'-t'')
    e^{-i t' \mat{X}
    \mat{D}_{\frac{U l}{2}}} 
    \mat{Y}_> 
    e^{i t'' \mat{D}_{\frac{U l}{2}} \mat{X}}
    \\ \label{app:General_Diff_Eq_Sol}
    &
    + i \Theta(t''-t')
    e^{-i t' \mat{X}
    \mat{D}_{\frac{U l}{2}}} 
    \mat{Y}_< 
    e^{i t'' \mat{D}_{\frac{U l}{2}} \mat{X}}
\end{align}
where $\mat{Y}_>$ and $\mat{Y}_<$ are two matrices that satisfy 
\begin{align}\label{app:Y_Constraint}
    \mat{Y}_> + \mat{Y}_< = \mat{X}.
\end{align}
We thus have four degrees of freedom remaining, as expected from a solution to a first order differential equation of a $2 \times 2$ matrix, which must be fixed by the appropriate boundary conditions.

Instead of going back to the discrete representation of the action to fix these boundary conditions, we can instead make use of Dyson's equation
\begin{align}\label{app:Dyson}
    \mat{G}_{\frac{U l}{2}}(t', t'')
    &=
    \mat{G}_{0}(t', t'')
    +
    \frac{U l}{2}
    \int_0^t
    dt'''
    \mat{G}_{0}(t', t''')
    \mat{G}_{\frac{Ul}{2}}(t''', t'')
\end{align}
in conjunction with the fact that we already know $\mat{G}_0(t',t'')$. Defining 
\begin{align}
    &\mat{Y}_+
    \equiv
    \begin{pmatrix}
        0 & 1  \\
        0 & 0
    \end{pmatrix},
    \\
    &\mat{Y}_-
    \equiv
    \begin{pmatrix}
        0 & 0  \\
        1 & 0
    \end{pmatrix},
\end{align}
one verifies that for $U = 0$, $\mat{Y}_> = \mat{Y}_+, \mat{Y}_< = \mat{Y}_-$. Since both of these matrices square to zero, from Eq.~(\ref{app:Dyson}) we have
\begin{align}
    \mat{Y}_- \mat{G}_{\frac{U l}{2}}(0, t'')
    =
    \mat{Y}_+ e^{i t \mat{X}\mat{D}_{0}} \mat{G}_{\frac{U l}{2}}(t, t'')
    =
    0.
\end{align}
Summing both these equations, using Eq.~(\ref{app:General_Diff_Eq_Sol}) and Eq.~(\ref{app:Y_Constraint}) we have an equation for $\mat{Y}_>$ or $\mat{Y}_<$ only. This yields
\begin{align}
    \mat{Y}_>
   &=
    \frac{
    e^{i t \mat{X}\mat{D}_\frac{U l}{2}}
    e^{-i t \mat{X}\mat{D}_0} 
    \mat{Y}_+ 
    }
    {\Tr
    \left( 
    \mat{Y}_+ \mat{Y}_-
    e^{i t \mat{X} \mat{D}_{\frac{U l}{2}}}
    e^{-i t \mat{X}\mat{D}_0}
    \right)
    }
    \\
    \mat{Y}_<
    &=
    \frac{\mat{Y}_- 
    e^{i t \mat{D}_0 \mat{X}} 
    e^{-i t \mat{D}_\frac{U l}{2}\mat{X}}}
    {\Tr
    \left( 
    \mat{Y}_+ \mat{Y}_-
    e^{i t \mat{X} \mat{D}_{\frac{U l}{2}}}
    e^{-i t \mat{X}\mat{D}_0}
    \right)
    }.
\end{align}
Computing the matrix exponential only requires diagonalizing a $2 \times 2$ matrix and thus can be done exactly. With $\mat{G}_{\frac{Ul}{2}}(t',t'')$ in hand, we can then eliminate the source term in Eq.~(\ref{app:To_Average_Over}) as usual by making the appropriate displacement. 

After performing the displacement, we must still compute the source-free functional integral, which is equal to the inverse of the functional determinant of the matrix Green's function \cite{Kamenev_Book_2011}
\begin{align}\label{app:Det}
    \frac{1}{
    {\rm{det}} 
    \left(i \mat{G}_{\frac{U l}{2}}\right)
    }
    =
    \langle 
    e^{-i \frac{U l}{2} \int_0^t dt' n_{\rm cl}(t')}
    \rangle_{S_0}
\end{align}
where recall $n_{\rm cl}(t') \equiv a^*_{\rm cl}(t')a_{\rm cl}(t') + a_{\rm q}^*(t')a_{\rm q}(t')$. Taking the derivative with respect to $U$ on both sides and using standard Gaussian integral identities we find
\begin{align}
    \frac{
    \frac{d}{d U}
    \langle 
    e^{-i \frac{U l}{2} \int_0^t dt' n_{\rm cl}(t')}
    \rangle_{S_0}
    }
    {
    \langle 
    e^{-i \frac{U l}{2} \int_0^t dt' n_{\rm cl}(t')}
    \rangle_{S_0}
    }
    =
    \frac{l}{2}
    \int_{0}^t
    dt'
    \Tr
    \left(
    \mat{G}_{\frac{U l}{2}}(t',t')
    \right)
\end{align}
which should be supplemented with the boundary condition $\det\left(i \mat{G}_{0}\right) = 1$. It should also be noted that although $\mat{G}(t', t'')_{\frac{U l}{2}}$ is discontinuous at $t' = t''$, its trace is not. With this differential equation and boundary condition, one verifies that we have
\begin{align}
    {
    \langle 
    e^{-i \frac{U l}{2} \int_0^t dt' n_{\rm cl}(t')}
    \rangle_{S_0}
    }
    =
    \frac{1}
    {\Tr
    \left( 
    \mat{Y}_+ \mat{Y}_-
    e^{i t \mat{X} \mat{D}_{\frac{U l}{2}}}
    e^{-i t \mat{X}\mat{D}_0}
    \right)
    }.
\end{align}
One then has all the required expressions to show that Eq.~(\ref{eq:K_U_2}) and Eq.~(\ref{eq:K_U_3}) are equal. 

 We can then obtain the propagator $\Xi(\beta, \alpha;t) = \langle \langle \hat{\beta}_{\rm cl} | e^{-i \hL t} | \halphacl \rangle \rangle$  relating the initial Wigner function to that at a later time by simply Fourier-transforming $K(\eta, \alpha ;t)$ in the $\eta$ variable. From Eq.~(\ref{eq:K_U_3}) in the Fourier transform can be computed using only Gaussian integrals. After a straightforward but tedious calculation, we arrive at
 \begin{widetext}
\begin{align} \label{app_eq:Prop_Kerr_Wigner}
    \Xi(\beta, \alpha; t)
    =
    \int
    \frac{2 d^2 \eta}{\pi}
    e^{\beta^* \eta - \eta^* \beta}
    K(\eta, \alpha; t)
    &
    =
    \sum_{l=-\infty}^\infty
    D_l(t)
    e^{-E_{-, l}(t)|\beta|^2
    -E_{+,l}(t)|\alpha|^2
    }
    e^{-il (\phi_{\beta}-\phi_\alpha)}
    \mathcal{J}_l\Bigl(2|\beta||\alpha|F_l(t)\Bigr)
\end{align}
\end{widetext}
with $\phi_{\alpha} = \arg \alpha$, $\phi_\beta = \arg \beta$ and we have defined
\begin{align}
    D_l(t)
    &\equiv
    \frac{2 e^{-i l (\omega_0 - U) t + \frac{\kappa}{2} t} 
    \Gamma_l i^{-l}}
    {
    [i U l + 2 \kappa(2\nth+1)] \sinh \frac{\Gamma_l}{2} t
    },
    \\
    E_{\pm ,l}
    &\equiv
    \frac
    {
    \Gamma_l \cosh \frac{\Gamma_l t}{2}
    \pm
    \kappa \sinh \frac{\Gamma_l t}{2} 
    }
    {
    [i U l + 2 \kappa(2\nth+1)] \sinh \frac{\Gamma_l}{2} t
    },
    \\
    F_l(t)
    &\equiv
    \frac{i \Gamma_l}
     {
    [i U l + 2 \kappa(2\nth+1)] \sinh \frac{\Gamma_l}{2} t
    }.
\end{align}
We can then obtain Eq.~(\ref{eq:Wigner_Initial_Coherent}), the Wigner function $W_{\hrho(t)}(\alpha)$ with $\hrho(0)$ a coherent state by using Eq.~(\ref{eq:Prop_Cl_Cl_Two}), Eq.~(\ref{app_eq:Prop_Kerr_Wigner}) and one more set of Gaussian integrals.
\bibliography{Third_Quant_Bib}

\begin{thebibliography}{45}%
\makeatletter
\providecommand \@ifxundefined [1]{%
 \@ifx{#1\undefined}
}%
\providecommand \@ifnum [1]{%
 \ifnum #1\expandafter \@firstoftwo
 \else \expandafter \@secondoftwo
 \fi
}%
\providecommand \@ifx [1]{%
 \ifx #1\expandafter \@firstoftwo
 \else \expandafter \@secondoftwo
 \fi
}%
\providecommand \natexlab [1]{#1}%
\providecommand \enquote  [1]{``#1''}%
\providecommand \bibnamefont  [1]{#1}%
\providecommand \bibfnamefont [1]{#1}%
\providecommand \citenamefont [1]{#1}%
\providecommand \href@noop [0]{\@secondoftwo}%
\providecommand \href [0]{\begingroup \@sanitize@url \@href}%
\providecommand \@href[1]{\@@startlink{#1}\@@href}%
\providecommand \@@href[1]{\endgroup#1\@@endlink}%
\providecommand \@sanitize@url [0]{\catcode `\\12\catcode `\$12\catcode
  `\&12\catcode `\#12\catcode `\^12\catcode `\_12\catcode `\%12\relax}%
\providecommand \@@startlink[1]{}%
\providecommand \@@endlink[0]{}%
\providecommand \url  [0]{\begingroup\@sanitize@url \@url }%
\providecommand \@url [1]{\endgroup\@href {#1}{\urlprefix }}%
\providecommand \urlprefix  [0]{URL }%
\providecommand \Eprint [0]{\href }%
\providecommand \doibase [0]{https://doi.org/}%
\providecommand \selectlanguage [0]{\@gobble}%
\providecommand \bibinfo  [0]{\@secondoftwo}%
\providecommand \bibfield  [0]{\@secondoftwo}%
\providecommand \translation [1]{[#1]}%
\providecommand \BibitemOpen [0]{}%
\providecommand \bibitemStop [0]{}%
\providecommand \bibitemNoStop [0]{.\EOS\space}%
\providecommand \EOS [0]{\spacefactor3000\relax}%
\providecommand \BibitemShut  [1]{\csname bibitem#1\endcsname}%
\let\auto@bib@innerbib\@empty
\bibitem [{\citenamefont {Prosen}(2008)}]{Prosen_Fermions_2008}%
  \BibitemOpen
  \bibfield  {author} {\bibinfo {author} {\bibfnamefont {T.}~\bibnamefont
  {Prosen}},\ }\bibfield  {title} {\bibinfo {title} {Third quantization: a
  general method to solve master equations for quadratic open fermi systems},\
  }\href {https://doi.org/10.1088/1367-2630/10/4/043026} {\bibfield  {journal}
  {\bibinfo  {journal} {New Journal of Physics}\ }\textbf {\bibinfo {volume}
  {10}},\ \bibinfo {pages} {043026} (\bibinfo {year} {2008})}\BibitemShut
  {NoStop}%
\bibitem [{\citenamefont {Prosen}\ and\ \citenamefont
  {Seligman}(2010)}]{Prosen_Seligman_Bosons_2010}%
  \BibitemOpen
  \bibfield  {author} {\bibinfo {author} {\bibfnamefont {T.}~\bibnamefont
  {Prosen}}\ and\ \bibinfo {author} {\bibfnamefont {T.~H.}\ \bibnamefont
  {Seligman}},\ }\bibfield  {title} {\bibinfo {title} {Quantization over boson
  operator spaces},\ }\href {https://doi.org/10.1088/1751-8113/43/39/392004}
  {\bibfield  {journal} {\bibinfo  {journal} {J. Phys. A: Math. Theor.}\
  }\textbf {\bibinfo {volume} {43}},\ \bibinfo {pages} {392004} (\bibinfo
  {year} {2010})}\BibitemShut {NoStop}%
\bibitem [{\citenamefont {Lieu}\ \emph {et~al.}(2020)\citenamefont {Lieu},
  \citenamefont {McGinley},\ and\ \citenamefont
  {Cooper}}]{Lieu_Ten_Fold_Lindblad_PRL}%
  \BibitemOpen
  \bibfield  {author} {\bibinfo {author} {\bibfnamefont {S.}~\bibnamefont
  {Lieu}}, \bibinfo {author} {\bibfnamefont {M.}~\bibnamefont {McGinley}},\
  and\ \bibinfo {author} {\bibfnamefont {N.~R.}\ \bibnamefont {Cooper}},\
  }\bibfield  {title} {\bibinfo {title} {Tenfold way for quadratic
  lindbladians},\ }\href {https://doi.org/10.1103/PhysRevLett.124.040401}
  {\bibfield  {journal} {\bibinfo  {journal} {Phys. Rev. Lett.}\ }\textbf
  {\bibinfo {volume} {124}},\ \bibinfo {pages} {040401} (\bibinfo {year}
  {2020})}\BibitemShut {NoStop}%
\bibitem [{\citenamefont {Zheng}\ \emph {et~al.}(2022)\citenamefont {Zheng},
  \citenamefont {Wang},\ and\ \citenamefont {Chen}}]{Shu_Boundary_Mode_arXiv}%
  \BibitemOpen
  \bibfield  {author} {\bibinfo {author} {\bibfnamefont {Z.-Y.}\ \bibnamefont
  {Zheng}}, \bibinfo {author} {\bibfnamefont {X.}~\bibnamefont {Wang}},\ and\
  \bibinfo {author} {\bibfnamefont {S.}~\bibnamefont {Chen}},\ }\href
  {https://doi.org/10.48550/ARXIV.2212.04785} {\bibinfo {title} {Exact solution
  of boundary-dissipated transverse field ising model: structure of liouvillian
  spectrum and dynamical duality}} (\bibinfo {year} {2022})\BibitemShut
  {NoStop}%
\bibitem [{\citenamefont {Costa}\ \emph {et~al.}(2022)\citenamefont {Costa},
  \citenamefont {Ribeiro}, \citenamefont {de~Luca}, \citenamefont {Prosen},\
  and\ \citenamefont {Sá}}]{Prosen_Random_Liouvillian_Fermions}%
  \BibitemOpen
  \bibfield  {author} {\bibinfo {author} {\bibfnamefont {J.}~\bibnamefont
  {Costa}}, \bibinfo {author} {\bibfnamefont {P.}~\bibnamefont {Ribeiro}},
  \bibinfo {author} {\bibfnamefont {A.}~\bibnamefont {de~Luca}}, \bibinfo
  {author} {\bibfnamefont {T.}~\bibnamefont {Prosen}},\ and\ \bibinfo {author}
  {\bibfnamefont {L.}~\bibnamefont {Sá}},\ }\href
  {https://doi.org/10.48550/ARXIV.2210.07959} {\bibinfo {title} {Spectral and
  steady-state properties of fermionic random quadratic liouvillians}}
  (\bibinfo {year} {2022})\BibitemShut {NoStop}%
\bibitem [{\citenamefont {Talkington}\ and\ \citenamefont
  {Claassen}(2022)}]{Dissipation_Flat_Bands_PRL}%
  \BibitemOpen
  \bibfield  {author} {\bibinfo {author} {\bibfnamefont {S.}~\bibnamefont
  {Talkington}}\ and\ \bibinfo {author} {\bibfnamefont {M.}~\bibnamefont
  {Claassen}},\ }\bibfield  {title} {\bibinfo {title} {Dissipation-induced flat
  bands},\ }\href {https://doi.org/10.1103/PhysRevB.106.L161109} {\bibfield
  {journal} {\bibinfo  {journal} {Phys. Rev. B}\ }\textbf {\bibinfo {volume}
  {106}},\ \bibinfo {pages} {L161109} (\bibinfo {year} {2022})}\BibitemShut
  {NoStop}%
\bibitem [{\citenamefont {Ribeiro}\ and\ \citenamefont
  {Prosen}(2019)}]{Prosen_Integrable_Spin_PRL}%
  \BibitemOpen
  \bibfield  {author} {\bibinfo {author} {\bibfnamefont {P.}~\bibnamefont
  {Ribeiro}}\ and\ \bibinfo {author} {\bibfnamefont {T.}~\bibnamefont
  {Prosen}},\ }\bibfield  {title} {\bibinfo {title} {Integrable quantum
  dynamics of open collective spin models},\ }\href
  {https://doi.org/10.1103/PhysRevLett.122.010401} {\bibfield  {journal}
  {\bibinfo  {journal} {Phys. Rev. Lett.}\ }\textbf {\bibinfo {volume} {122}},\
  \bibinfo {pages} {010401} (\bibinfo {year} {2019})}\BibitemShut {NoStop}%
\bibitem [{\citenamefont {Nakanishi}\ and\ \citenamefont
  {Sasamoto}(2022)}]{PT_Transition_PRA}%
  \BibitemOpen
  \bibfield  {author} {\bibinfo {author} {\bibfnamefont {Y.}~\bibnamefont
  {Nakanishi}}\ and\ \bibinfo {author} {\bibfnamefont {T.}~\bibnamefont
  {Sasamoto}},\ }\bibfield  {title} {\bibinfo {title} {$\mathcal{PT}$ phase
  transition in open quantum systems with lindblad dynamics},\ }\href
  {https://doi.org/10.1103/PhysRevA.105.022219} {\bibfield  {journal} {\bibinfo
   {journal} {Phys. Rev. A}\ }\textbf {\bibinfo {volume} {105}},\ \bibinfo
  {pages} {022219} (\bibinfo {year} {2022})}\BibitemShut {NoStop}%
\bibitem [{\citenamefont {Yang}\ \emph {et~al.}(2022)\citenamefont {Yang},
  \citenamefont {Jiang},\ and\ \citenamefont
  {Bergholtz}}]{Bergholtz_Exact_PRR}%
  \BibitemOpen
  \bibfield  {author} {\bibinfo {author} {\bibfnamefont {F.}~\bibnamefont
  {Yang}}, \bibinfo {author} {\bibfnamefont {Q.-D.}\ \bibnamefont {Jiang}},\
  and\ \bibinfo {author} {\bibfnamefont {E.~J.}\ \bibnamefont {Bergholtz}},\
  }\bibfield  {title} {\bibinfo {title} {Liouvillian skin effect in an exactly
  solvable model},\ }\href {https://doi.org/10.1103/PhysRevResearch.4.023160}
  {\bibfield  {journal} {\bibinfo  {journal} {Phys. Rev. Res.}\ }\textbf
  {\bibinfo {volume} {4}},\ \bibinfo {pages} {023160} (\bibinfo {year}
  {2022})}\BibitemShut {NoStop}%
\bibitem [{\citenamefont {Alaeian}\ and\ \citenamefont {Bu{\v
  c}a}(2022)}]{Buca_Multistability_Comm_Physics}%
  \BibitemOpen
  \bibfield  {author} {\bibinfo {author} {\bibfnamefont {H.}~\bibnamefont
  {Alaeian}}\ and\ \bibinfo {author} {\bibfnamefont {B.}~\bibnamefont {Bu{\v
  c}a}},\ }\bibfield  {title} {\bibinfo {title} {Exact multistability and
  dissipative time crystals in interacting fermionic lattices},\ }\href
  {https://doi.org/10.1038/s42005-022-01090-z} {\bibfield  {journal} {\bibinfo
  {journal} {Communications Physics}\ }\textbf {\bibinfo {volume} {5}},\
  \bibinfo {pages} {318} (\bibinfo {year} {2022})}\BibitemShut {NoStop}%
\bibitem [{\citenamefont {Shibata}\ and\ \citenamefont
  {Katsura}(2019)}]{Non-Hermitian_Kitaev_PRB}%
  \BibitemOpen
  \bibfield  {author} {\bibinfo {author} {\bibfnamefont {N.}~\bibnamefont
  {Shibata}}\ and\ \bibinfo {author} {\bibfnamefont {H.}~\bibnamefont
  {Katsura}},\ }\bibfield  {title} {\bibinfo {title} {Dissipative spin chain as
  a non-hermitian kitaev ladder},\ }\href
  {https://doi.org/10.1103/PhysRevB.99.174303} {\bibfield  {journal} {\bibinfo
  {journal} {Phys. Rev. B}\ }\textbf {\bibinfo {volume} {99}},\ \bibinfo
  {pages} {174303} (\bibinfo {year} {2019})}\BibitemShut {NoStop}%
\bibitem [{\citenamefont {Barthel}\ and\ \citenamefont
  {Zhang}(2021)}]{Barthel_Thomas_Third_Quant_2022}%
  \BibitemOpen
  \bibfield  {author} {\bibinfo {author} {\bibfnamefont {T.}~\bibnamefont
  {Barthel}}\ and\ \bibinfo {author} {\bibfnamefont {Y.}~\bibnamefont
  {Zhang}},\ }\href {https://doi.org/10.48550/ARXIV.2112.08344} {\bibinfo
  {title} {Solving quasi-free and quadratic lindblad master equations for open
  fermionic and bosonic systems}} (\bibinfo {year} {2021})\BibitemShut
  {NoStop}%
\bibitem [{\citenamefont {Briegel}\ and\ \citenamefont
  {Englert}(1993)}]{Englert_Damping_Basis_1993}%
  \BibitemOpen
  \bibfield  {author} {\bibinfo {author} {\bibfnamefont {H.-J.}\ \bibnamefont
  {Briegel}}\ and\ \bibinfo {author} {\bibfnamefont {B.-G.}\ \bibnamefont
  {Englert}},\ }\bibfield  {title} {\bibinfo {title} {Quantum optical master
  equations: The use of damping bases},\ }\href
  {https://doi.org/10.1103/PhysRevA.47.3311} {\bibfield  {journal} {\bibinfo
  {journal} {Phys. Rev. A}\ }\textbf {\bibinfo {volume} {47}},\ \bibinfo
  {pages} {3311} (\bibinfo {year} {1993})}\BibitemShut {NoStop}%
\bibitem [{\citenamefont {Honda}\ \emph {et~al.}(2010)\citenamefont {Honda},
  \citenamefont {Nakazato},\ and\ \citenamefont
  {Yoshida}}]{Honda_Spectral_Resolution_2010}%
  \BibitemOpen
  \bibfield  {author} {\bibinfo {author} {\bibfnamefont {D.}~\bibnamefont
  {Honda}}, \bibinfo {author} {\bibfnamefont {H.}~\bibnamefont {Nakazato}},\
  and\ \bibinfo {author} {\bibfnamefont {M.}~\bibnamefont {Yoshida}},\
  }\bibfield  {title} {\bibinfo {title} {Spectral resolution of the liouvillian
  of the lindblad master equation for a harmonic oscillator},\ }\href
  {https://doi.org/10.1063/1.3442363} {\bibfield  {journal} {\bibinfo
  {journal} {Journal of Mathematical Physics}\ }\textbf {\bibinfo {volume}
  {51}},\ \bibinfo {pages} {072107} (\bibinfo {year} {2010})}\BibitemShut
  {NoStop}%
\bibitem [{\citenamefont {Thompson}\ and\ \citenamefont
  {Kamenev}(2023)}]{Kamanev_arXiv_Field_Theory}%
  \BibitemOpen
  \bibfield  {author} {\bibinfo {author} {\bibfnamefont {F.}~\bibnamefont
  {Thompson}}\ and\ \bibinfo {author} {\bibfnamefont {A.}~\bibnamefont
  {Kamenev}},\ }\href {https://doi.org/10.48550/ARXIV.2301.02953} {\bibinfo
  {title} {Field theory of many-body lindbladian dynamics}} (\bibinfo {year}
  {2023})\BibitemShut {NoStop}%
\bibitem [{\citenamefont {Yurke}\ and\ \citenamefont
  {Stoler}(1986)}]{Yurke_Cat}%
  \BibitemOpen
  \bibfield  {author} {\bibinfo {author} {\bibfnamefont {B.}~\bibnamefont
  {Yurke}}\ and\ \bibinfo {author} {\bibfnamefont {D.}~\bibnamefont {Stoler}},\
  }\bibfield  {title} {\bibinfo {title} {Generating quantum mechanical
  superpositions of macroscopically distinguishable states via amplitude
  dispersion},\ }\href {https://doi.org/10.1103/PhysRevLett.57.13} {\bibfield
  {journal} {\bibinfo  {journal} {Phys. Rev. Lett.}\ }\textbf {\bibinfo
  {volume} {57}},\ \bibinfo {pages} {13} (\bibinfo {year} {1986})}\BibitemShut
  {NoStop}%
\bibitem [{\citenamefont {Albert}(2018)}]{Victor_Thesis}%
  \BibitemOpen
  \bibfield  {author} {\bibinfo {author} {\bibfnamefont {V.~V.}\ \bibnamefont
  {Albert}},\ }\href@noop {} {\bibinfo {title} {Lindbladians with multiple
  steady states: theory and applications}} (\bibinfo {year} {2018}),\ \Eprint
  {https://arxiv.org/abs/1802.00010} {arXiv:1802.00010 [quant-ph]} \BibitemShut
  {NoStop}%
\bibitem [{Note1()}]{Note1}%
  \BibitemOpen
  \bibinfo {note} {Technically, for the infinite-dimensional spaces under
  consideration, both $\protect \hat {X}$ and $\protect \hat {Y}$ must be trace
  class to be part of this Hilbert space of operators. This is cumbersome,
  since operators of interest such as $\protect \hat {a}$ and $\protect \hat
  {a}^\dagger $ do not satisfy this property. We do not concern ourselves with
  these technical details. We will thus use the language and notation that one
  uses for a normal Hilbert space.}\BibitemShut {Stop}%
\bibitem [{\citenamefont {Kamenev}(2011)}]{Kamenev_Book_2011}%
  \BibitemOpen
  \bibfield  {author} {\bibinfo {author} {\bibfnamefont {A.}~\bibnamefont
  {Kamenev}},\ }\href {https://doi.org/10.1017/CBO9781139003667} {\emph
  {\bibinfo {title} {Field Theory of Non-Equilibrium Systems}}}\ (\bibinfo
  {publisher} {Cambridge University Press},\ \bibinfo {year}
  {2011})\BibitemShut {NoStop}%
\bibitem [{\citenamefont {Sieberer}\ \emph {et~al.}(2016)\citenamefont
  {Sieberer}, \citenamefont {Buchhold},\ and\ \citenamefont
  {Diehl}}]{Diehl_Keldysh_Review_2016}%
  \BibitemOpen
  \bibfield  {author} {\bibinfo {author} {\bibfnamefont {L.~M.}\ \bibnamefont
  {Sieberer}}, \bibinfo {author} {\bibfnamefont {M.}~\bibnamefont {Buchhold}},\
  and\ \bibinfo {author} {\bibfnamefont {S.}~\bibnamefont {Diehl}},\ }\bibfield
   {title} {\bibinfo {title} {Keldysh field theory for driven open quantum
  systems},\ }\href@noop {} {\bibfield  {journal} {\bibinfo  {journal} {Rep.
  Prog. Phys.}\ }\textbf {\bibinfo {volume} {79}},\ \bibinfo {pages} {096001}
  (\bibinfo {year} {2016})}\BibitemShut {NoStop}%
\bibitem [{\citenamefont {Clerk}\ \emph {et~al.}(2010)\citenamefont {Clerk},
  \citenamefont {Devoret}, \citenamefont {Girvin}, \citenamefont {Marquardt},\
  and\ \citenamefont {Schoelkopf}}]{RMP_Clerk}%
  \BibitemOpen
  \bibfield  {author} {\bibinfo {author} {\bibfnamefont {A.~A.}\ \bibnamefont
  {Clerk}}, \bibinfo {author} {\bibfnamefont {M.~H.}\ \bibnamefont {Devoret}},
  \bibinfo {author} {\bibfnamefont {S.~M.}\ \bibnamefont {Girvin}}, \bibinfo
  {author} {\bibfnamefont {F.}~\bibnamefont {Marquardt}},\ and\ \bibinfo
  {author} {\bibfnamefont {R.~J.}\ \bibnamefont {Schoelkopf}},\ }\bibfield
  {title} {\bibinfo {title} {Introduction to quantum noise, measurement, and
  amplification},\ }\href {https://doi.org/10.1103/RevModPhys.82.1155}
  {\bibfield  {journal} {\bibinfo  {journal} {Rev. Mod. Phys.}\ }\textbf
  {\bibinfo {volume} {82}},\ \bibinfo {pages} {1155} (\bibinfo {year}
  {2010})}\BibitemShut {NoStop}%
\bibitem [{Note2()}]{Note2}%
  \BibitemOpen
  \bibinfo {note} {One should be careful when working with the inverse of
  $\protect \mathbfcal {\protect \hat {V}}$, since as we we show in Sec.~\ref
  {sec:Single_Osc_Phase_Space} it is unbounded. One can always write down
  expression which involve only $\protect \mathbfcal {\protect \hat {V}}$
  however, e.g. $\protect \mathbfcal {\protect \hat {L}}\protect \mathbfcal
  {\protect \hat {V}}|\protect \hat {\rho }' \rangle \rangle = \protect
  \mathbfcal {\protect \hat {V}}\left [(\omega _0-i \protect \frac {\kappa
  }{2})\protect \bm {\protect \hat {a}}_{\protect \rm q}^\dagger \protect \bm
  {\protect \hat {a}}_{\protect \rm cl}+ (\omega _0+i \protect \frac {\kappa
  }{2})\protect \bm {\protect \hat {a}}_{\protect \rm q}\protect \bm {\protect
  \hat {a}}_{\protect \rm cl}^\dagger )\right ] |\protect \hat {\rho }' \rangle
  \rangle $}\BibitemShut {NoStop}%
\bibitem [{\citenamefont {Bishop}\ and\ \citenamefont
  {Vourdas}(1994)}]{Displaced_Squeezed_Ops_1994_PRA}%
  \BibitemOpen
  \bibfield  {author} {\bibinfo {author} {\bibfnamefont {R.~F.}\ \bibnamefont
  {Bishop}}\ and\ \bibinfo {author} {\bibfnamefont {A.}~\bibnamefont
  {Vourdas}},\ }\bibfield  {title} {\bibinfo {title} {Displaced and squeezed
  parity operator: Its role in classical mappings of quantum theories},\ }\href
  {https://doi.org/10.1103/PhysRevA.50.4488} {\bibfield  {journal} {\bibinfo
  {journal} {Phys. Rev. A}\ }\textbf {\bibinfo {volume} {50}},\ \bibinfo
  {pages} {4488} (\bibinfo {year} {1994})}\BibitemShut {NoStop}%
\bibitem [{\citenamefont {Royer}(1977)}]{Royer_1977_Wigner_Disaplced}%
  \BibitemOpen
  \bibfield  {author} {\bibinfo {author} {\bibfnamefont {A.}~\bibnamefont
  {Royer}},\ }\bibfield  {title} {\bibinfo {title} {Wigner function as the
  expectation value of a parity operator},\ }\href
  {https://doi.org/10.1103/PhysRevA.15.449} {\bibfield  {journal} {\bibinfo
  {journal} {Phys. Rev. A}\ }\textbf {\bibinfo {volume} {15}},\ \bibinfo
  {pages} {449} (\bibinfo {year} {1977})}\BibitemShut {NoStop}%
\bibitem [{\citenamefont {Gardiner}\ and\ \citenamefont
  {Zoller}(2010)}]{gardiner_zoller_2010}%
  \BibitemOpen
  \bibfield  {author} {\bibinfo {author} {\bibfnamefont {C.~W.}\ \bibnamefont
  {Gardiner}}\ and\ \bibinfo {author} {\bibfnamefont {P.}~\bibnamefont
  {Zoller}},\ }\href@noop {} {\emph {\bibinfo {title} {Quantum Noise: A
  handbook of Markovian and non-Markovian Quantum stochastic methods with
  applications to Quantum Optics}}}\ (\bibinfo  {publisher} {Springer},\
  \bibinfo {year} {2010})\BibitemShut {NoStop}%
\bibitem [{\citenamefont {Curtright}\ \emph {et~al.}(2014)\citenamefont
  {Curtright}, \citenamefont {Fairlie},\ and\ \citenamefont
  {Zachos}}]{Quantization_Phase_Space_Treatise}%
  \BibitemOpen
  \bibfield  {author} {\bibinfo {author} {\bibfnamefont {T.}~\bibnamefont
  {Curtright}}, \bibinfo {author} {\bibfnamefont {D.}~\bibnamefont {Fairlie}},\
  and\ \bibinfo {author} {\bibfnamefont {C.}~\bibnamefont {Zachos}},\
  }\href@noop {} {\emph {\bibinfo {title} {A concise treatise on Quantum
  Mechanics in phase space}}}\ (\bibinfo  {publisher} {World Scientific},\
  \bibinfo {year} {2014})\BibitemShut {NoStop}%
\bibitem [{\citenamefont {Fano}(1957)}]{Fano_1957_RMP}%
  \BibitemOpen
  \bibfield  {author} {\bibinfo {author} {\bibfnamefont {U.}~\bibnamefont
  {Fano}},\ }\bibfield  {title} {\bibinfo {title} {Description of states in
  quantum mechanics by density matrix and operator techniques},\ }\href@noop {}
  {\bibfield  {journal} {\bibinfo  {journal} {Rev. Mod. Phys.}\ }\textbf
  {\bibinfo {volume} {29}},\ \bibinfo {pages} {74} (\bibinfo {year}
  {1957})}\BibitemShut {NoStop}%
\bibitem [{\citenamefont {Wootters}(1987)}]{Phase_Point_Wooter_1987}%
  \BibitemOpen
  \bibfield  {author} {\bibinfo {author} {\bibfnamefont {W.~K.}\ \bibnamefont
  {Wootters}},\ }\bibfield  {title} {\bibinfo {title} {A wigner-function
  formulation of finite-state quantum mechanics},\ }\href@noop {} {\bibfield
  {journal} {\bibinfo  {journal} {Ann. Phys.}\ }\textbf {\bibinfo {volume}
  {176}},\ \bibinfo {pages} {1} (\bibinfo {year} {1987})}\BibitemShut {NoStop}%
\bibitem [{\citenamefont {Risken}(1996)}]{Fokker_Planck_Textbook}%
  \BibitemOpen
  \bibfield  {author} {\bibinfo {author} {\bibfnamefont {H.}~\bibnamefont
  {Risken}},\ }\href@noop {} {\emph {\bibinfo {title} {The Fokker-Planck
  Equation: Methods of Solutions and Applications}}}\ (\bibinfo  {publisher}
  {Springer},\ \bibinfo {year} {1996})\BibitemShut {NoStop}%
\bibitem [{\citenamefont {Feynman}\ and\ \citenamefont
  {Vernon}(1963)}]{Feynman_Vernon_1962}%
  \BibitemOpen
  \bibfield  {author} {\bibinfo {author} {\bibfnamefont {R.}~\bibnamefont
  {Feynman}}\ and\ \bibinfo {author} {\bibfnamefont {F.}~\bibnamefont
  {Vernon}},\ }\bibfield  {title} {\bibinfo {title} {The theory of a general
  quantum system interacting with a linear dissipative system},\ }\href
  {https://doi.org/https://doi.org/10.1016/0003-4916(63)90068-X} {\bibfield
  {journal} {\bibinfo  {journal} {Annals of Physics}\ }\textbf {\bibinfo
  {volume} {24}},\ \bibinfo {pages} {118} (\bibinfo {year} {1963})}\BibitemShut
  {NoStop}%
\bibitem [{\citenamefont {Sharan}(1979)}]{Star_product_PRD}%
  \BibitemOpen
  \bibfield  {author} {\bibinfo {author} {\bibfnamefont {P.}~\bibnamefont
  {Sharan}},\ }\bibfield  {title} {\bibinfo {title} {Star-product
  representation of path integrals},\ }\href
  {https://doi.org/10.1103/PhysRevD.20.414} {\bibfield  {journal} {\bibinfo
  {journal} {Phys. Rev. D}\ }\textbf {\bibinfo {volume} {20}},\ \bibinfo
  {pages} {414} (\bibinfo {year} {1979})}\BibitemShut {NoStop}%
\bibitem [{\citenamefont {Marinov}(1991)}]{Marinov_1991}%
  \BibitemOpen
  \bibfield  {author} {\bibinfo {author} {\bibfnamefont {M.}~\bibnamefont
  {Marinov}},\ }\bibfield  {title} {\bibinfo {title} {A new type of phase-space
  path integral},\ }\href
  {https://doi.org/https://doi.org/10.1016/0375-9601(91)90352-9} {\bibfield
  {journal} {\bibinfo  {journal} {Physics Letters A}\ }\textbf {\bibinfo
  {volume} {153}},\ \bibinfo {pages} {5} (\bibinfo {year} {1991})}\BibitemShut
  {NoStop}%
\bibitem [{\citenamefont {Altland}\ and\ \citenamefont
  {Simons}(2010)}]{Altland_Simons_2010}%
  \BibitemOpen
  \bibfield  {author} {\bibinfo {author} {\bibfnamefont {A.}~\bibnamefont
  {Altland}}\ and\ \bibinfo {author} {\bibfnamefont {B.~D.}\ \bibnamefont
  {Simons}},\ }\href@noop {} {\emph {\bibinfo {title} {Condensed Matter Field
  Theory}}}\ (\bibinfo  {publisher} {Cambridge University Press},\ \bibinfo
  {year} {2010})\BibitemShut {NoStop}%
\bibitem [{\citenamefont {Bergholtz}\ \emph {et~al.}(2021)\citenamefont
  {Bergholtz}, \citenamefont {Budich},\ and\ \citenamefont
  {Kunst}}]{Bergholtz_Kunst_RMP}%
  \BibitemOpen
  \bibfield  {author} {\bibinfo {author} {\bibfnamefont {E.~J.}\ \bibnamefont
  {Bergholtz}}, \bibinfo {author} {\bibfnamefont {J.~C.}\ \bibnamefont
  {Budich}},\ and\ \bibinfo {author} {\bibfnamefont {F.~K.}\ \bibnamefont
  {Kunst}},\ }\bibfield  {title} {\bibinfo {title} {Exceptional topology of
  non-hermitian systems},\ }\href@noop {} {\bibfield  {journal} {\bibinfo
  {journal} {Rev. Mod. Phys.}\ }\textbf {\bibinfo {volume} {93}},\ \bibinfo
  {pages} {015005} (\bibinfo {year} {2021})}\BibitemShut {NoStop}%
\bibitem [{\citenamefont {Ashida}\ \emph {et~al.}(2020)\citenamefont {Ashida},
  \citenamefont {Gong},\ and\ \citenamefont {Ueda}}]{Ueda_NH_Review}%
  \BibitemOpen
  \bibfield  {author} {\bibinfo {author} {\bibfnamefont {Y.}~\bibnamefont
  {Ashida}}, \bibinfo {author} {\bibfnamefont {Z.}~\bibnamefont {Gong}},\ and\
  \bibinfo {author} {\bibfnamefont {M.}~\bibnamefont {Ueda}},\ }\bibfield
  {title} {\bibinfo {title} {{Non-Hermitian} physics},\ }\href@noop {}
  {\bibfield  {journal} {\bibinfo  {journal} {Adv. Phys.}\ }\textbf {\bibinfo
  {volume} {69}},\ \bibinfo {pages} {249} (\bibinfo {year} {2020})}\BibitemShut
  {NoStop}%
\bibitem [{\citenamefont {Dykman}(1973)}]{Mark_1973}%
  \BibitemOpen
  \bibfield  {author} {\bibinfo {author} {\bibfnamefont {M.~I.}\ \bibnamefont
  {Dykman}},\ }\bibfield  {title} {\bibinfo {title} {Quantum theory of spectral
  distribution of isolated nonlinear vibration modes near the combination
  frequency},\ }\href@noop {} {\bibfield  {journal} {\bibinfo  {journal}
  {Soviet Physics - Solid State}\ }\textbf {\bibinfo {volume} {15}},\ \bibinfo
  {pages} {735} (\bibinfo {year} {1973})}\BibitemShut {NoStop}%
\bibitem [{\citenamefont {Chaturvedi}\ and\ \citenamefont
  {Srinivasan}(1991{\natexlab{a}})}]{Chaturvedi_1991_Kerr_1}%
  \BibitemOpen
  \bibfield  {author} {\bibinfo {author} {\bibfnamefont {S.}~\bibnamefont
  {Chaturvedi}}\ and\ \bibinfo {author} {\bibfnamefont {V.}~\bibnamefont
  {Srinivasan}, \bibfnamefont {V}},\ }\bibfield  {title} {\bibinfo {title}
  {Class of exactly solvable master equations describing coupled nonlinear
  oscillators},\ }\href@noop {} {\bibfield  {journal} {\bibinfo  {journal}
  {Phys. Rev. A}\ }\textbf {\bibinfo {volume} {43}},\ \bibinfo {pages} {4054}
  (\bibinfo {year} {1991}{\natexlab{a}})}\BibitemShut {NoStop}%
\bibitem [{\citenamefont {Chaturvedi}\ and\ \citenamefont
  {Srinivasan}(1991{\natexlab{b}})}]{Chaturvedi_1991_Kerr_2}%
  \BibitemOpen
  \bibfield  {author} {\bibinfo {author} {\bibfnamefont {S.}~\bibnamefont
  {Chaturvedi}}\ and\ \bibinfo {author} {\bibfnamefont {V.}~\bibnamefont
  {Srinivasan}},\ }\bibfield  {title} {\bibinfo {title} {Solution of the master
  equation for an attenuated or amplified nonlinear oscillator with an
  arbitrary initial condition},\ }\href@noop {} {\bibfield  {journal} {\bibinfo
   {journal} {J. Mod. Opt.}\ }\textbf {\bibinfo {volume} {38}},\ \bibinfo
  {pages} {777} (\bibinfo {year} {1991}{\natexlab{b}})}\BibitemShut {NoStop}%
\bibitem [{\citenamefont {Milburn}(1986)}]{Milburn_1986_Kerr_Liouville}%
  \BibitemOpen
  \bibfield  {author} {\bibinfo {author} {\bibfnamefont {G.~J.}\ \bibnamefont
  {Milburn}},\ }\bibfield  {title} {\bibinfo {title} {Quantum and classical
  liouville dynamics of the anharmonic oscillator},\ }\href@noop {} {\bibfield
  {journal} {\bibinfo  {journal} {Phys. Rev. A Gen. Phys.}\ }\textbf {\bibinfo
  {volume} {33}},\ \bibinfo {pages} {674} (\bibinfo {year} {1986})}\BibitemShut
  {NoStop}%
\bibitem [{\citenamefont {Milburn}\ and\ \citenamefont
  {Holmes}(1986)}]{Milburn_1986_Kerr_Liouville_Dissipation}%
  \BibitemOpen
  \bibfield  {author} {\bibinfo {author} {\bibfnamefont {G.~J.}\ \bibnamefont
  {Milburn}}\ and\ \bibinfo {author} {\bibfnamefont {C.~A.}\ \bibnamefont
  {Holmes}},\ }\bibfield  {title} {\bibinfo {title} {Dissipative quantum and
  classical liouville mechanics of the anharmonic oscillator},\ }\href@noop {}
  {\bibfield  {journal} {\bibinfo  {journal} {Phys. Rev. Lett.}\ }\textbf
  {\bibinfo {volume} {56}},\ \bibinfo {pages} {2237} (\bibinfo {year}
  {1986})}\BibitemShut {NoStop}%
\bibitem [{\citenamefont {Daniel}\ and\ \citenamefont
  {Milburn}(1989)}]{Daniel_1989_Kerr_Coherence}%
  \BibitemOpen
  \bibfield  {author} {\bibinfo {author} {\bibfnamefont {D.~J.}\ \bibnamefont
  {Daniel}}\ and\ \bibinfo {author} {\bibfnamefont {G.~J.}\ \bibnamefont
  {Milburn}},\ }\bibfield  {title} {\bibinfo {title} {Destruction of quantum
  coherence in a nonlinear oscillator via attenuation and amplification},\
  }\href@noop {} {\bibfield  {journal} {\bibinfo  {journal} {Phys. Rev. A Gen.
  Phys.}\ }\textbf {\bibinfo {volume} {39}},\ \bibinfo {pages} {4628} (\bibinfo
  {year} {1989})}\BibitemShut {NoStop}%
\bibitem [{\citenamefont {Peinov{\'a}}\ and\ \citenamefont
  {Luk}(1990)}]{Peinova_1990_Kerr_Exact}%
  \BibitemOpen
  \bibfield  {author} {\bibinfo {author} {\bibfnamefont {V.}~\bibnamefont
  {Peinov{\'a}}, \bibfnamefont {V}}\ and\ \bibinfo {author} {\bibfnamefont
  {A.}~\bibnamefont {Luk}},\ }\bibfield  {title} {\bibinfo {title} {Exact
  quantum statistics of a nonlinear dissipative oscillator evolving from an
  arbitrary state},\ }\href@noop {} {\bibfield  {journal} {\bibinfo  {journal}
  {Phys. Rev. A}\ }\textbf {\bibinfo {volume} {41}},\ \bibinfo {pages} {414}
  (\bibinfo {year} {1990})}\BibitemShut {NoStop}%
\bibitem [{\citenamefont {Stobi{\'n}ska}\ \emph {et~al.}(2008)\citenamefont
  {Stobi{\'n}ska}, \citenamefont {Milburn},\ and\ \citenamefont
  {W{\'o}dkiewicz}}]{Millburn_Kerr_Numerics_2008}%
  \BibitemOpen
  \bibfield  {author} {\bibinfo {author} {\bibfnamefont {M.}~\bibnamefont
  {Stobi{\'n}ska}}, \bibinfo {author} {\bibfnamefont {G.~J.}\ \bibnamefont
  {Milburn}},\ and\ \bibinfo {author} {\bibfnamefont {K.}~\bibnamefont
  {W{\'o}dkiewicz}},\ }\bibfield  {title} {\bibinfo {title} {Wigner function
  evolution of quantum states in the presence of self-kerr interaction},\
  }\href@noop {} {\bibfield  {journal} {\bibinfo  {journal} {Phys. Rev. A}\
  }\textbf {\bibinfo {volume} {78}},\ \bibinfo {pages} {013810} (\bibinfo
  {year} {2008})}\BibitemShut {NoStop}%
\bibitem [{\citenamefont {McDonald}\ and\ \citenamefont
  {Clerk}(2022)}]{Alexander_Weak_Symm_PRL}%
  \BibitemOpen
  \bibfield  {author} {\bibinfo {author} {\bibfnamefont {A.}~\bibnamefont
  {McDonald}}\ and\ \bibinfo {author} {\bibfnamefont {A.~A.}\ \bibnamefont
  {Clerk}},\ }\bibfield  {title} {\bibinfo {title} {Exact solutions of
  interacting dissipative systems via weak symmetries},\ }\href
  {https://doi.org/10.1103/PhysRevLett.128.033602} {\bibfield  {journal}
  {\bibinfo  {journal} {Phys. Rev. Lett.}\ }\textbf {\bibinfo {volume} {128}},\
  \bibinfo {pages} {033602} (\bibinfo {year} {2022})}\BibitemShut {NoStop}%
\bibitem [{\citenamefont {Dykman}\ and\ \citenamefont
  {Krivoglaz}(1984)}]{Mark_Old_SPR_1984}%
  \BibitemOpen
  \bibfield  {author} {\bibinfo {author} {\bibfnamefont {M.~I.}\ \bibnamefont
  {Dykman}}\ and\ \bibinfo {author} {\bibfnamefont {M.~A.}\ \bibnamefont
  {Krivoglaz}},\ }\bibfield  {title} {\bibinfo {title} {Theory of nonlinear
  oscillator interacting with a medium},\ }\href@noop {} {\bibfield  {journal}
  {\bibinfo  {journal} {Soviet Physics Reviews(vol 5)}\ ,\ \bibinfo {pages} {pp
  265–441}} (\bibinfo {year} {1984})}\BibitemShut {NoStop}%
\end{thebibliography}%
\end{document}